\documentclass[11pt,a4paper]{article}

\usepackage{jheppub}
\usepackage{amsthm}

\usepackage{amsmath,amssymb,amsfonts,mathtools}
\usepackage{mathrsfs}
\usepackage{bm}
\usepackage{bbm}
\usepackage{physics}
\usepackage{slashed}
\usepackage{graphicx}
\usepackage{hyperref}
\usepackage{xcolor}
\usepackage{float}
\usepackage[section]{placeins}

\usepackage{microtype}
\usepackage{bm}

\title{(Pseudo-)Dirac Gravitinos}

\author[ ]{Karim Benakli}
\author[ ]{and Arno Goudeau}

\affiliation[ ]{Sorbonne Universit\'e, CNRS, Laboratoire de Physique Th\'eorique et Hautes \'Energies, LPTHE, F-75005 Paris, France}

\emailAdd{kbenakli@lpthe.jussieu.fr}
\emailAdd{goudeau@lpthe.jussieu.fr}

\abstract{
We discuss low-energy effective theories with a Dirac gravitino. Our main benchmark is
Scherk--Schwarz supersymmetry breaking with anti-periodic boundary conditions
for fermions on \(S^1/\mathbb Z_2\). In this case two Majorana
spin-\(\tfrac32\) modes are degenerate, so the low-energy theory is naturally
organized around a Dirac gravitino and an associated \(R\)-symmetry selection
rule. We show, in minimal supergravity, that this Dirac spin-\(\tfrac32\)
mixing has no local Dirac-gaugino-type \(\mathcal N=1\) superspace
realization: a superspace description requires projection onto the transverse
superspin-\(\tfrac32\) sector. 

We then analyze some consequences for matter couplings and radiative masses.
The ordinary gravitino supercurrent coupling gives the standard universal
Scherk--Schwarz scalar threshold; in the minimal benchmark this contribution is
negative, and should therefore be regarded as one calculable contribution to
the scalar mass matrix, not as a complete scalar-spectrum prediction. Dirac
fermion masses can be generated if an additional companion-channel matter
current is present. The existence and normalization of this current are not
fixed by the Dirac gravitino alone and depend on the ultraviolet completion.
We finally discuss small \(R\)-breaking deformations, the resulting
pseudo-Dirac regime, and comment on possible applications to singlet fermions,
modulini, and radiative Dirac neutrino masses.
}

\begin{document}
\maketitle
\flushbottom

\section{Introduction}
\label{sec:introduction}

Scherk--Schwarz boundary conditions provide one of the simplest ways of
breaking supersymmetry in a higher-dimensional theory
\cite{Scherk:1978ta,Scherk:1979zr,Cremmer:1979uq,Fayet:1985kt}.
In field-theory applications the Scherk--Schwarz twist is often treated as a
continuous parameter \(\omega\). In string-inspired constructions, however, the
pure Scherk--Schwarz choice can be more restricted. In particular,
anti-periodic boundary conditions for fermions,
\begin{equation}
\omega=\frac12 ,
\end{equation}
are a distinguished possibility in perturbative string constructions
\cite{Rohm:1983aq,Antoniadis:1988jn,Banks:1988yz,Ferrara:1988jx,
Kounnas:1989dk,Antoniadis:1990ew,Ferrara:1994kg,Benakli:1995ut}.
Thus the anti-periodic point may be the natural pure Scherk--Schwarz option.
In this work we restrict ourselves to this anti-periodic Scherk--Schwarz case
and to its low-energy field-theory description.

In a five-dimensional field theory on \(S^1/\mathbb Z_2\), \(\omega\) may be
treated as a continuous Scherk--Schwarz parameter. In this paper we use the
pure anti-periodic point as our benchmark. Boundary gravitino mass terms can
shift the effective boundary conditions away from the pure half-twist point,
and may then be encoded either as jump conditions or as an effective
\(\omega\neq\tfrac12\)
\cite{Bagger:2001qi,Bagger:2001ep,Gherghetta:2001sa,Meissner:2002dg,
Biggio:2002rb,Rattazzi:2003rj,Benakli:2007zza,Benakli:2014daa}. More general
localized supersymmetry breaking need not reduce to a shifted Scherk--Schwarz angle.

The anti-periodic point has an important consequence for the four-dimensional
spin-\(\tfrac32\) spectrum. At \(\omega=\tfrac12\), the light
spin-\(\tfrac32\) sector is not the single Majorana gravitino of ordinary
\(\mathcal N=1\) supergravity. Rather, it consists of two degenerate
spin-\(\tfrac32\) modes which are naturally organized as a Dirac gravitino. 
This Dirac organization is protected
by a residual \(U(1)_R\): the off-diagonal mass is allowed, while diagonal
Majorana masses are forbidden. The same selection rule applies to
four-dimensional matter fermions coupled to this sector. In the exact
anti-periodic limit, gravitational mediation therefore cannot generate
Majorana masses for matter fermions unless this symmetry is broken.

This observation is the starting point of the present paper. The
anti-periodic Scherk--Schwarz spectrum, its Dirac spin-\(\tfrac32\)
organization, and the corresponding gravity-mediated scalar threshold are
known results \cite{Antoniadis:1997ic,Gherghetta:2001sa,Rattazzi:2003rj,Antoniadis:2015chx}.
Here we address a related question which, to our knowledge, has not been
isolated systematically: what matter-sector masses can still be generated when
the natural Scherk--Schwarz limit is the \(U(1)_R\)-symmetric point where
Majorana fermion masses vanish? More precisely, can one obtain radiative Dirac
fermion masses, and what scalar terms necessarily accompany the additional
matter couplings required to generate them?

A useful distinction will be important throughout the paper. Degeneracy of two
spin-\(\tfrac32\) fields is only a kinematic statement. A Dirac interpretation
requires a symmetry that distinguishes the two fields and forbids diagonal
Majorana entries. In the constructions considered below this role is played by
the residual \(U(1)_R\). The same symmetry also organizes the matter-sector
operators:
\begin{equation}
m_f^{\rm Maj}=0
\qquad
\text{in the exact }U(1)_R\text{ limit},
\end{equation}
while Dirac fermion masses require an \(R\)-neutral partner channel.


We first recall how the same Dirac-gravitino structure appears in
four-dimensional \(\mathcal N=2\) gauged supergravity. This discussion is
closely related to the early analyses of extended-supergravity breaking and
massive gravitini
\cite{Freedman:1976xh,Ferrara:1976fu,Deser:1977uq}. We use the standard
matter-coupled \(\mathcal N=2\) framework with vector and hypermultiplets,
fermion shifts, moment maps, special K\"ahler geometry and
quaternionic-K\"ahler geometry
\cite{deWit:1984rvr,Andrianopoli:1996cm,Andrianopoli:1996ve,Freedman:2012zz}.

The purpose of this part is limited. We do not aim to construct a fully
stabilized \(\mathcal N=2\) vacuum, nor to use the full \(\mathcal N=2\)
matter sector for phenomenology. Rather, we use \(\mathcal N=2\) gauged
supergravity to identify the local data which realize a massive
spin-\(\tfrac32\) Dirac pair protected by a residual \(U(1)_R\). The relevant
object is the aligned gravitino shift
\begin{equation}
P_\Lambda^a L^\Lambda ,
\end{equation}
where \(P_\Lambda^a\) are the moment maps and \(L^\Lambda\) are the
covariantly holomorphic sections. When this vector selects a single direction
in \(SU(2)_R\) space, the two gravitini can be organized with opposite
\(U(1)_R\) charges, and the spin-\(\tfrac32\) mass matrix takes the Dirac form.

We illustrate this mechanism with two simple representatives: a vector-sector
Fayet--Iliopoulos gauging and a hypermultiplet gauging. These examples are not
complete model-building constructions. Their role is to show how the
gravitino mass, charge and Pauli couplings are tied to the same gauging data,
and how the aligned shift can arise either from constant or from
field-dependent moment maps. Once this Dirac spin-\(\tfrac32\) structure has
been identified, the phenomenological analysis will be formulated in
\(\mathcal N=1\) language, appropriate for chiral matter and visible-sector
applications. Questions of full moduli stabilization and of the detailed
classification of \(\mathcal N=2\) vacua, including STU-type examples
\cite{Cremmer:1984hj,Duff:1995sm,Behrndt:1996hu}, are not needed for the
low-energy EFT analysis below.


We then compare this four-dimensional description with the higher-dimensional
Scherk--Schwarz realization on \(S^1/\mathbb Z_2\). In the field-theory
description, the Scherk--Schwarz breaking is implemented by boundary conditions
periodic up to an \(SU(2)_R\) rotation. At the anti-periodic point the two
lightest spin-\(\tfrac32\) levels have
\begin{equation}
m_0=m_{-1}=\frac{1}{2R}.
\end{equation}
The low-energy spin-\(\tfrac32\) sector must therefore be kept as a pair and
is naturally organized as a Dirac gravitino. The orbifold also fixes boundary
selection rules. On a given boundary the even gravitino couples directly,
whereas the odd gravitino vanishes and can enter only through normal-derivative
couplings, bulk propagation, or the second-supersymmetry structure. This
distinction will be important for the matter couplings.

The anti-periodic point will be our reference background for the analysis of
fields coupled to the spin-\(\tfrac32\) sector. It is the point at which the
residual \(U(1)_R\) is exact, the light spin-\(\tfrac32\) sector is Dirac, and
Majorana masses for four-dimensional fermions are forbidden. In what follows,
the ``matter sector'' includes the localized chiral and gauge multiplets, and
therefore both matter fermions and gauginos. The anti-periodic point is then
the natural limit in which to ask what gravitationally induced masses remain in
this sector. In the minimal setup considered here, these are threshold effects:
the ordinary gravitino coupling gives the known scalar threshold, whereas
fermion masses require additional \(R\)-compatible couplings.

The situation is analogous to Dirac-gaugino model building
\cite{Fayet:1978qc,Hall:1990hq,Fox:2002bu,Nelson:2002ca,
Antoniadis:2006uj,Antoniadis:2006eb,Kribs:2007ac,Belanger:2009wf,
Benakli:2008pg,Benakli:2009mk,Benakli:2010gi,Benakli:2011kz,
Benakli:2011vb,Benakli:2012cy,Benakli:2013msa,Dudas:2013gga,
Benakli:2014cia,Benakli:2015ioa,Abel:2011dc,Gherghetta:2011na,
Davies:2011mp,Martin:2015eca,Bertuzzo:2014bwa,Ardu:2025qym,
Carpenter:2015mna,Csaki:2013fla,Carpenter:2010as,Choi:2008ub,
Benakli:2018vqz,Benakli:2018vjk,Benakli:2018ldd,Benakli:2019yaq}.
There the Dirac mass requires, besides the gauge multiplet, an additional
adjoint chiral multiplet and an \(R\)-symmetric mass operator. The same
structural logic applies here. A Dirac mass for a four-dimensional fermion
requires a second fermion with the appropriate \(U(1)_R\) charge and a matter
current coupled to the companion spin-\(\tfrac32\) channel. The existence and
normalization of this channel are therefore additional dynamical input, not
consequences of the universal ordinary gravitino coupling. Our purpose is to
identify which matter-current structures are required, what is fixed by the
Scherk--Schwarz geometry, and what remains UV-dependent.

This leads naturally to the \(\mathcal N=1\) description. Boundary matter, and
more generally many visible-sector models, are written in terms of
\(\mathcal N=1\) chiral and vector multiplets, with K\"ahler potential \(K\),
superpotential \(W\), and gauge kinetic functions \(f\). This language is
appropriate for the scalar geometry and for the matter sector. The
spin-\(\tfrac32\) sector is more subtle. After choosing one
\(\mathcal N=1\) subalgebra inside the parent \(\mathcal N=2\) theory, the
ordinary supergravity multiplet is accompanied by an additional
spin-\(\tfrac32\) multiplet. In a genuine \(\mathcal N=2\) or five-dimensional
completion this companion field is the second gravitino. In a more general
low-energy EFT it is a companion spin-\(\tfrac32\) field whose couplings must
be specified.

There is, however, an important difference with the Dirac-gaugino case. For
gauginos the relevant \(R\)-symmetric mass operator admits a local
\(\mathcal N=1\) superspace description. For spin-\(\tfrac32\), no analogous
local chiral rank-one building block exists. The Dirac spin-\(\tfrac32\)
mixing can be represented only after projection onto the irreducible
transverse superspin-\(\tfrac32\) sector, and is therefore non-local in
superspace. This is a property of the superspace representation, not of the
component interaction restricted to physical spin-\(\tfrac32\) modes.

The matter couplings have to be specified at three different levels. In a
generic four-dimensional EFT, the ordinary gravitino has the universal coupling
to the supercurrent of the localized chiral and gauge multiplets. By contrast,
the coupling of the companion spin-\(\tfrac32\) field is an additional EFT
datum, constrained by Lorentz invariance, gauge invariance, dimensional
analysis, and the residual \(U(1)_R\).

The Scherk--Schwarz benchmark adds kinematical selection rules. On a given
boundary the even spin-\(\tfrac32\) wavefunction is nonzero and couples
directly to the boundary supercurrent, whereas the odd wavefunction vanishes.
The odd, or wrong-parity, spin-\(\tfrac32\) field can enter only through
normal-derivative couplings, bulk propagation, or the matter current associated
with the second supersymmetry of the extended multiplet structure. Thus the
orbifold fixes which kernels are allowed, but it does not by itself fix the
existence or normalization of the companion matter current.

Finally, in a complete \(\mathcal N=2\) or five-dimensional supergravity
model, the companion field may be identified with the second gravitino and its
matter current must be derived from the underlying matter sector. In that case
the normalization is no longer arbitrary, but it is still model-dependent: it
depends on how the localized or bulk matter fields are embedded into the
extended-supergravity construction.

This separation is essential for the loop effects. The scalar sector contains
a universal contribution. The operator \(\phi^\dagger\phi\) is neutral under
the residual \(U(1)_R\), and the ordinary supercurrent coupling is always
present. In the anti-periodic Scherk--Schwarz benchmark this gives the standard
gravity-mediated scalar threshold,
\begin{equation}
m_{\phi,{\rm univ}}^2
\sim
\frac{1}{16\pi^2}\,
\frac{m_{3/2}^4}{M_P^2},
\end{equation}
with a negative sign in the minimal setup, using the convention
\(V\supset m_\phi^2|\phi|^2\). This sign refers to the calculable universal
Scherk--Schwarz contribution alone; it is not, by itself, a prediction for the
physical scalar spectrum in a stabilized completion. The exact coefficient for
a canonically normalized boundary scalar was obtained in Scherk--Schwarz
soft-mass computations
\cite{Antoniadis:1997ic,Gherghetta:2001sa,Rattazzi:2003rj,Antoniadis:2015chx}.
In this paper we use that scalar result in a form adapted to the same-boundary
kernel normalization needed for the loop analysis.

The mixed scalar terms have a different status. They are not fixed by the
ordinary supergravity coupling and are absent unless the matter sector contains
the additional structure associated with the companion spin-\(\tfrac32\)
channel. Depending on the topology and on the matter mass insertions, such
terms can generate diagonal or off-diagonal entries in the scalar mass matrix.
The stability question is therefore a question about the full scalar mass
matrix, not about the ordinary Scherk--Schwarz scalar threshold alone.

The fermion sector is more restrictive. In the exact \(U(1)_R\)-symmetric
limit, Majorana masses are absent. A Dirac fermion mass requires three
ingredients: an \(R\)-neutral fermion bilinear, a partner matter channel, and a
companion spin-\(\tfrac32\) matter current. If these ingredients are absent,
the one-loop Dirac mass is absent even though the scalar threshold remains. If
they are present, the tensor structure of the amplitude can be computed, but
the absolute size depends on the normalization of the companion-channel
coupling. With conservative supergravity normalization, two explicit
spin-\(\tfrac32\)--matter vertices give the scaling
\begin{equation}
m_f^{\rm Dirac}
\sim
\frac{1}{16\pi^2}\,
\alpha_f\,
\frac{m_{3/2}^3}{M_P^2}.
\end{equation}
A larger estimate proportional to \(m_{3/2}^2/M_P\) is possible only as an
additional EFT assumption about the companion-channel normalization; it is not
fixed by the Scherk--Schwarz geometry or by the boundary kernels alone.

The Scherk--Schwarz benchmark also distinguishes local and non-local
thresholds. Same-boundary amplitudes contain local pieces and are sensitive to
boundary counterterms. Opposite-boundary amplitudes are finite
boundary-to-boundary thresholds, with ultraviolet behavior controlled by the
separation along the interval. This distinction is useful for identifying
which scalar entries are local-threshold dependent and which are genuinely
non-local effects.

Finally, we also discuss the regime near, but not exactly at, the
anti-periodic point. Writing
\begin{equation}
\omega=\frac12-\epsilon_\omega ,
\end{equation}
localized \(R\)-breaking in the gravitino sector produces a pseudo-Dirac
splitting
\begin{equation}
\Delta m_{3/2}=\frac{2\epsilon_\omega}{R}.
\end{equation}
If direct localized \(R\)-breaking matter operators and pseudo-Goldstino
mixings are absent or sequestered, the induced Majorana masses for localized
fermions are transmitted only through gravitational loops. For gauginos one
finds parametrically
\begin{equation}
M_{1/2}^{\rm grav}
\sim
\epsilon_\omega\,
\frac{1}{16\pi^2}\,
\frac{m_{3/2}^3}{M_P^2}.
\end{equation}
The universal scalar mass is unchanged at first order in \(\epsilon_\omega\).
This gives a pseudo-Dirac regime distinct from the \(R\)-symmetric
companion-channel mechanism for radiative Dirac masses.

Possible applications include Dirac gaugino sectors, neutral singlet sectors
such as modulini, and radiative Dirac neutrino constructions. We do not try to
build a complete phenomenological model. Our purpose is to isolate the
structural ingredients of the mechanism: the symmetry protecting the Dirac
gravitino, the matter couplings required for Dirac fermion masses, the
Scherk--Schwarz kernels entering the thresholds, and the constraints coming
from the scalar sector.

A complete ultraviolet embedding must stabilize the radion and other moduli
without generating \(R\)-breaking effects that erase the pseudo-Dirac
hierarchy. It must also explain the origin and normalization of the companion
matter current. We do not construct such a completion here; instead, scalar
stability, companion-channel normalization, and \(R\)-breaking are treated as
constraints on possible completions.

The paper is organized as follows. Sections~\ref{sec:N2_to_N0_dirac} and
\ref{sec:SS_origin} collect the standard supergravity and Scherk--Schwarz
material needed to make the rest of the discussion self-contained and to fix
our conventions. In Section~\ref{sec:N2_to_N0_dirac} we recall how an aligned
gravitino shift in broken \(\mathcal N=2\) gauged supergravity gives a Dirac
gravitino protected by a residual \(U(1)_R\), and we discuss vector and
hypermultiplet realizations. In Section~\ref{sec:SS_origin} we review the
Scherk--Schwarz realization on \(S^1/\mathbb Z_2\), the special role of
\(\omega=\tfrac12\), and the effect of localized boundary deformations.

The new analysis begins in Section~\ref{sec:N1_language_and_obstruction},
where we formulate the low-energy system in \(\mathcal N=1\) language and
explain the obstruction to a local Dirac-gaugino-type superspace bilinear for
the spin-\(\tfrac32\) mixing. Section~\ref{sec:coupling_to_matter} identifies
the matter couplings relevant for the ordinary and companion spin-\(\tfrac32\)
channels. Sections~\ref{sec:scalar_masses} and
\ref{sec:fermion_masses} contain the scalar and fermion one-loop analyses,
including the distinction between the universal scalar threshold and the
conditional Dirac fermion masses. Section~\ref{sec:pheno} summarizes 
the phenomenological implications of the
exact \(U(1)_R\) limit. Section~\ref{sec:small_R_breaking} discusses small
\(R\)-breaking deformations away from \(\omega=\tfrac12\), the resulting
pseudo-Dirac regime, and the constraints from induced Majorana masses.
Section~\ref{sec:conclusion} contains our conclusions.

Several technical points are collected in appendices.
Appendix~\ref{app:projectors} reviews the superspin projectors used in the
\(\mathcal N=1\) superspace description. Appendix~\ref{app:SS_kernels} derives
the mixed-representation Green functions and boundary kernels on the interval.
Appendix~\ref{app:scalar_cubic_trace} explains the off-shell boundary origin
of the universal scalar normalization. It separates the full Scherk--Schwarz
scalar threshold from the local cubic--cubic tensor reduction and shows how the
coefficient \(\mathcal C_A\) used in Section~\ref{sec:scalar_masses} is
obtained from the complete locally supersymmetric boundary coupling rather
than from an isolated cubic graph.

\section{Dirac gravitino from broken \texorpdfstring{$\mathcal N=2$}{N=2} supergravity}
\label{sec:N2_to_N0_dirac}

In this section we explain how a Dirac gravitino arises in four-dimensional
\(\mathcal N=2\) gauged supergravity when supersymmetry is completely broken,
\begin{equation}
\mathcal N=2 \;\longrightarrow\; \mathcal N=0,
\end{equation}
while a continuous \(U(1)_R\subset SU(2)_R\) remains unbroken. Our purpose is
not to review the full formalism of gauged \(\mathcal N=2\) supergravity, but
to isolate the ingredients that control the spin-\(\tfrac32\) sector and the
low-energy interpretation used later in the paper.\footnote{For the component
formulation of \(4D\), \(\mathcal N=2\) supergravity coupled to matter
multiplets, see \cite{deWit:1984rvr}. For reviews and conventions on gauged
\(\mathcal N=2\) supergravity, including fermion shifts, special K\"ahler
geometry, quaternionic-K\"ahler geometry and moment maps, see for example
\cite{Andrianopoli:1996cm,Andrianopoli:1996ve,Freedman:2012zz}.}

Four-dimensional \(\mathcal N=2\) supergravity contains two Weyl gravitini,
\(\psi_{\mu A}\), \(A=1,2\), transforming as a doublet under \(SU(2)_R\).
After gauging, the supersymmetry variations acquire fermion shifts. In
particular, the gravitini are controlled by a symmetric \(2\times2\) shift
matrix \(S_{AB}\). The central point is that if the gravitino shift selects a
single direction in \(SU(2)_R\) space and the remaining vacuum data are
compatible with that direction, then a residual \(U(1)_R\) can remain
unbroken. In a basis adapted to this \(U(1)_R\), the two gravitini have
opposite charges and may be organized into one Dirac gravitino.

By a \emph{Dirac gravitino} we mean more than the statement that two Weyl
gravitini have equal masses. Degeneracy alone is not enough, since two
equal-mass Weyl fields can always be recombined by a change of basis. The
Dirac interpretation becomes intrinsic only when a symmetry distinguishes the
two fields and forbids diagonal Majorana masses. Here that symmetry is the
residual \(U(1)_R\). In a basis adapted to it, the spin-\(\tfrac32\) mass term
takes the form
\begin{equation}
\mathcal L_m^{(3/2)}
=
-\,m_{3/2}\big(\psi^{1\mu}\sigma_{\mu\nu}\psi^{2\nu}
+\text{h.c.}\big),
\qquad
m_{3/2}>0,
\label{eq:sec2_DiracRS_new}
\end{equation}
with no \(\psi^1\psi^1\) or \(\psi^2\psi^2\) terms. Equivalently, the two Weyl
gravitini carry opposite \(U(1)_R\) charges,
\begin{equation}
R(\psi^1_\mu)=+1,
\qquad
R(\psi^2_\mu)=-1,
\label{eq:sec2_Rcharges_new}
\end{equation}
so that the off-diagonal mass term is neutral whereas diagonal Majorana masses
are forbidden.

The rest of this section makes this statement precise. We first recall the
minimal \(\mathcal N=2\) ingredients entering the gravitino shift and the
scalar potential. We then illustrate the mechanism in two standard settings:
a vector-sector Fayet--Iliopoulos realization and a hypermultiplet-gauging
realization. The first is algebraically transparent and exhibits a no-scale
Minkowski background with a Dirac gravitino. The second shows how the same
aligned structure appears in genuine hypermultiplet gauging, where the scalar
potential contains both the universal negative gravitino contribution and
positive Killing-vector terms. In that case, the simple axionic shift gauging
displayed below should be viewed as the elementary building block; a fully
stabilized Minkowski vacuum generally requires additional gauging data or
additional sectors.

\subsection{Minimal \texorpdfstring{$\mathcal N=2$}{N=2} ingredients, gravitino shift, and vacuum conditions}

The gravity multiplet contains
\begin{equation}
(g_{\mu\nu},\psi_{\mu A},A_\mu^0),
\end{equation}
namely the metric, the two gravitini, and the graviphoton. Vector multiplets
contain one vector, two gaugini, and one complex scalar, while hypermultiplets
contain two hyperini and four real scalars. The vector-multiplet scalars
\(z^i\) span a special-K\"ahler manifold, whereas the hypermultiplet scalars
\(q^u\) span a quaternionic-K\"ahler manifold.

On the vector side one introduces a holomorphic symplectic section
\(X^\Lambda(z)\), \(\Lambda=0,\dots,n_V\), together with the special-K\"ahler
potential \(K(z,\bar z)\). The covariantly holomorphic section is
\begin{equation}
L^\Lambda \equiv e^{K/2}X^\Lambda .
\label{eq:sec2_Lambda_new}
\end{equation}
On the hypermultiplet side the relevant data are the quaternionic-K\"ahler
metric \(h_{uv}\), the Killing vectors \(k_\Lambda^u(q)\) associated with the
gauged isometries, and the corresponding triplets of moment maps
\(P_\Lambda^a(q)\), with
\begin{equation}
a=1,2,3
\end{equation}
an adjoint \(SU(2)_R\) index.

Once isometries are gauged by the vectors \(A_\mu^\Lambda\), the gravitino
variation takes the schematic form
\begin{equation}
\delta\psi_{\mu A}
=
D_\mu\epsilon_A
+
i\,S_{AB}\gamma_\mu\epsilon^B
+\cdots ,
\label{eq:sec2_gravitino_variation_new}
\end{equation}
with
\begin{equation}
S_{AB}
=
\frac{i}{2}\,(\sigma^a)_{AB}\,P_\Lambda^a\,L^\Lambda .
\label{eq:sec2_SAB_new}
\end{equation}
This is the key formula for the spin-\(\tfrac32\) discussion. The overall
normalization of \(m_{3/2}\) follows from the convention chosen for
\(S_{AB}\); with \eqref{eq:sec2_SAB_new}, the common gravitino mass is
\(\frac12 |P_\Lambda^aL^\Lambda|\) when the shift is aligned along a single
\(SU(2)_R\) direction.

The three quantities \(P_\Lambda^aL^\Lambda\) define a vector in \(SU(2)_R\)
space. If, at the vacuum, this vector is aligned along a single direction,
\begin{equation}
\Big(P_\Lambda^aL^\Lambda\Big)\Big|_{\rm vac}
=
\delta^{aa_0}\,\mathcal P\Big|_{\rm vac},
\qquad
a_0\in\{1,2,3\},
\label{eq:sec2_alignment_new}
\end{equation}
then the gravitino shift is aligned along a single \(SU(2)_R\) generator. In
the situations considered below, this aligned generator is the origin of the
residual \(U(1)_R\subset SU(2)_R\) relevant for the spin-\(\tfrac32\) sector.

This alignment immediately implies degeneracy of the two gravitini. In the
presence of the residual \(U(1)_R\), one can then choose a basis of definite
charge in which the degenerate pair is naturally described as a Dirac
gravitino. Concretely,
\begin{equation}
S_{AB}\Big|_{\rm vac}
=
\frac{i}{2}\,\mathcal P\,(\sigma^{a_0})_{AB},
\end{equation}
so that the two eigenvalues have the same modulus. With the convention
\eqref{eq:sec2_SAB_new}, the common gravitino mass is
\begin{equation}
m_{3/2}
=
\frac12\,\big|\mathcal P\big|_{\rm vac}
=
\frac12\,\big|P_\Lambda^{a_0}L^\Lambda\big|_{\rm vac}.
\label{eq:sec2_mD_new}
\end{equation}
In a basis of definite charge under the residual \(U(1)_R\), the mass matrix
then takes the off-diagonal Dirac form, while diagonal Majorana entries are
forbidden by \(U(1)_R\).

For the Minkowski vacua relevant to this paper, this Dirac interpretation is
tied to complete supersymmetry breaking. If one local supersymmetry remained
unbroken, one gravitino combination would have to remain massless in order to
gauge the surviving \(\mathcal N=1\) algebra. A genuine massive Dirac
gravitino in flat space therefore corresponds to
\begin{equation}
\mathcal N=2 \;\longrightarrow\; \mathcal N=0 .
\end{equation}

For a background \((z^i_\ast,q^u_\ast)\), the vacuum conditions of interest
are
\begin{equation}
\partial_i V\big|_\ast = 0,
\qquad
\partial_u V\big|_\ast = 0,
\qquad
V\big|_\ast = 0,
\qquad
m_{3/2}\big|_\ast \neq 0,
\label{eq:sec2_vacuum_conditions_new}
\end{equation}
together with the absence of any Killing spinor. The condition \(V|_\ast=0\)
gives a Minkowski vacuum, while \(m_{3/2}\neq0\) means that both gravitini are
lifted into the massive spin-\(\tfrac32\) sector.

The full scalar potential will not be needed in detail, but its characteristic
structure is important. For hypermultiplet gauging one has
\begin{equation}
V
=
\Big(4\,h_{uv}\,k_\Lambda^u k_\Sigma^v\Big)\,\bar L^\Lambda L^\Sigma
+
\Big(U^{\Lambda\Sigma}-3\,\bar L^\Lambda L^\Sigma\Big)
P_\Lambda^aP_\Sigma^a,
\label{eq:sec2_scalar_potential_hyper_new}
\end{equation}
where
\begin{equation}
U^{\Lambda\Sigma}\equiv g^{i\bar j}f_i^\Lambda f_{\bar j}^\Sigma,
\qquad
f_i^\Lambda\equiv D_iL^\Lambda .
\end{equation}
The first term is positive semidefinite and comes from the gauged Killing
vectors. The second contains, in particular, the universal negative gravitino
contribution proportional to
\(-3\,\bar L^\Lambda L^\Sigma P_\Lambda^aP_\Sigma^a\). A Minkowski vacuum may
arise when all terms in \eqref{eq:sec2_scalar_potential_hyper_new} balance at
a stationary point.

\subsection{A first explicit realization: vector-sector FI gauging}
\label{subsec:vector_FI_realization}

A particularly transparent realization is provided by a pure vector-multiplet
model with constant Fayet--Iliopoulos gauging, namely constant moment maps in
the \(\mathcal N=2\) gauged-supergravity language
\cite{deWit:1984rvr,Andrianopoli:1996cm,Freedman:2012zz}. A standard example
is the STU model \cite{Cremmer:1984hj,Duff:1995sm,Behrndt:1996hu}, with cubic
prepotential
\begin{equation}
F(X)=\frac{X^1X^2X^3}{X^0}.
\label{eq:sec2_STU_prepotential_new}
\end{equation}
Introducing the special coordinates
\begin{equation}
S=\frac{X^1}{X^0},
\qquad
T=\frac{X^2}{X^0},
\qquad
U=\frac{X^3}{X^0},
\label{eq:sec2_STU_coords_new}
\end{equation}
and fixing the projective gauge \(X^0=1\), one has
\begin{equation}
X^\Lambda=(1,S,T,U).
\end{equation}
The corresponding special-K\"ahler potential is
\begin{equation}
K_{\rm SK}
=
-\log\!\Big[-i(S-\bar S)(T-\bar T)(U-\bar U)\Big],
\label{eq:sec2_KSTU_new}
\end{equation}
so that
\begin{equation}
L^\Lambda=e^{K_{\rm SK}/2}X^\Lambda,
\qquad
L^0=
\Big[-i(S-\bar S)(T-\bar T)(U-\bar U)\Big]^{-1/2}.
\label{eq:sec2_L_sections_new}
\end{equation}
On the physical slice
\begin{equation}
S=-is,\qquad T=-it,\qquad U=-iu,\qquad s,t,u>0,
\label{eq:sec2_STU_physical_slice_new}
\end{equation}
one has
\begin{equation}
-i(S-\bar S)(T-\bar T)(U-\bar U)=8stu,
\qquad
L^0=\frac{1}{2\sqrt{2stu}}.
\label{eq:sec2_L0_physical_slice_new}
\end{equation}

We choose a constant FI gauging aligned along a fixed \(SU(2)_R\) direction,
\begin{equation}
P_\Lambda^a=\xi\,\delta_\Lambda^0\,\delta^{a2},
\qquad
\Lambda=0,1,2,3,
\qquad
a=1,2,3.
\label{eq:sec2_FI_choice_new}
\end{equation}
Only the graviphoton \(A_\mu^0\) participates in the FI gauging, and the
moment-map triplet is aligned along a fixed \(SU(2)_R\) direction. The
alignment is therefore manifest. The parameter \(\xi\) includes both the gauge
coupling and the normalization of the chosen \(U(1)_R\) generator. In this
sense \(\xi\) is the single gauging parameter controlling the \(R\)-symmetry
gauging in this example.

The gravitino covariant derivative contains
\begin{equation}
D_\mu\psi_{\nu A}
=
\nabla_\mu\psi_{\nu A}
+\cdots
+\frac{i}{2}\,\xi\,A_\mu^0\,(\sigma^2)_A{}^{B}\psi_{\nu B}.
\label{eq:sec2_FI_covD_new}
\end{equation}
After diagonalizing the selected \(U(1)_R\subset SU(2)_R\), the two Weyl
gravitini carry opposite charges with respect to \(A_\mu^0\). In the vector
normalization used in \eqref{eq:sec2_FI_covD_new}, the corresponding charge
coefficient is
\begin{equation}
e_{\rm eff}=\frac{\xi}{2}.
\label{eq:sec2_FI_charge_new}
\end{equation}
If one instead works with canonically normalized vector fields, this
coefficient must be rescaled by the appropriate gauge-kinetic factor.

The same gauging fixes the gravitino shift,
\begin{equation}
S_{AB}
=
\frac{i}{2}\,\xi\,(\sigma^2)_{AB}\,L^0.
\label{eq:sec2_SAB_STU_new}
\end{equation}
With the convention \eqref{eq:sec2_SAB_new}, the common gravitino mass is
\begin{equation}
m_{3/2}=\frac{|\xi|}{2}\,|L^0|.
\label{eq:sec2_mD_STU_new}
\end{equation}
On the physical slice this becomes
\begin{equation}
m_{3/2}=\frac{|\xi|}{4\sqrt{2stu}}.
\label{eq:sec2_mD_physical_slice_new}
\end{equation}
Thus the gravitino charge coefficient and the gravitino mass are controlled by
the same gauging datum. In the vector normalization used above,
\begin{equation}
\frac{m_{3/2}}{|e_{\rm eff}|}=|L^0|,
\qquad
\frac{m_{3/2}^2}{e_{\rm eff}^2}=|L^0|^2.
\label{eq:sec2_mass_charge_ratio_FI_new}
\end{equation}
On the STU physical slice,
\begin{equation}
\frac{m_{3/2}^2}{e_{\rm eff}^2}
=
\frac{1}{8stu}.
\label{eq:sec2_mass_charge_ratio_FI_STU_new}
\end{equation}
Thus the mass and the charge coefficient are not independent. Their numerical
ratio is moduli-dependent because the no-scale model does not fix the vacuum
values of \(S,T,U\).

The Pauli coupling to the dressed graviphoton is fixed by the same special
geometry data. The anti-self-dual graviphoton field strength is
\begin{equation}
T^-_{\mu\nu}
=
2i\,(\Im\mathcal N)_{\Lambda\Sigma}L^\Sigma F^{-\Lambda}_{\mu\nu}.
\label{eq:sec2_Tminus_general_FI_new}
\end{equation}
If only \(F_{\mu\nu}^{0}\) is turned on,
\begin{equation}
T^-_{\mu\nu}
=
2i\,(\Im\mathcal N)_{00}L^0\,F^{-0}_{\mu\nu}.
\label{eq:sec2_Tminus_FI_A0_new}
\end{equation}
For the STU model on the physical slice one finds
\begin{equation}
(\Im\mathcal N)_{00}=stu,
\end{equation}
hence
\begin{equation}
T^-_{\mu\nu}
=
i\sqrt{\frac{stu}{2}}\,F^{-0}_{\mu\nu}.
\label{eq:sec2_Tminus_STU_physical_new}
\end{equation}
The Pauli coupling is therefore not an independent parameter either: it is
fixed by the same gauged \(\mathcal N=2\) structure.

For pure vector-multiplet gauging the scalar potential is
\begin{equation}
V=
\big(U^{\Lambda\Sigma}-3\bar L^\Lambda L^\Sigma\big)\,
P_\Lambda^aP_\Sigma^a.
\label{eq:sec2_V_vectors_new}
\end{equation}
With \eqref{eq:sec2_FI_choice_new},
\begin{equation}
V=\xi^2\big(U^{00}-3|L^0|^2\big).
\label{eq:sec2_V_STU_new}
\end{equation}
For STU, the no-scale identity gives
\begin{equation}
U^{00}=3|L^0|^2,
\label{eq:sec2_STU_noscale_new}
\end{equation}
and hence
\begin{equation}
V\equiv0
\label{eq:sec2_STU_Minkowski_new}
\end{equation}
along the no-scale moduli space. Supersymmetry is nevertheless broken whenever
\(\xi L^0\neq0\), since the fermion shifts are nonzero.

This example isolates the Dirac-gravitino mechanism in a minimal setting. The
gauging is explicit, the residual \(U(1)_R\) is manifest, the two gravitini are
degenerate, and the scalar potential vanishes along the no-scale moduli space
by the identity \eqref{eq:sec2_STU_noscale_new}. The limitation is also clear:
the moduli are not stabilized, so the mass and coupling ratios are not fixed
numbers. For our purposes this is sufficient, since the aim is to identify the
supergravity origin of the Dirac-gravitino structure, not to construct a fully
stabilized vacuum.

\subsection{A second realization: hypermultiplet gauging}
\label{subsec:hyper_realization}

A second realization is obtained by gauging an isometry in the hypermultiplet
sector. This case is useful for two reasons. First, it shows that the same
aligned \(U(1)_R\) structure is not restricted to constant FI terms. Second, it
displays the positive Killing-vector contribution in the scalar potential,
which is absent in the pure FI example and is one of the ingredients that can
participate in a Minkowski balance.

For concreteness, consider one vector multiplet and one hypermultiplet,
\begin{equation}
n_V=1,
\qquad
n_H=1.
\end{equation}
On the vector side choose the special-K\"ahler geometry
\begin{equation}
F(X)=-\,iX^0X^1.
\label{eq:sec2_one_vector_prepotential_new}
\end{equation}
With
\begin{equation}
z=\frac{X^1}{X^0},
\label{eq:sec2_one_vector_z_new}
\end{equation}
and \(X^0=1\), one has
\begin{equation}
X^\Lambda=(1,z),
\qquad
\Lambda=0,1,
\end{equation}
and
\begin{equation}
K_{\rm SK}=-\log\big(2(z+\bar z)\big),
\qquad \Re z>0.
\label{eq:sec2_one_vector_geometry_new}
\end{equation}
Thus
\begin{equation}
L^\Lambda=e^{K_{\rm SK}/2}X^\Lambda,
\qquad
L^0=\frac{1}{\sqrt{2(z+\bar z)}},
\qquad
L^1=\frac{z}{\sqrt{2(z+\bar z)}}.
\label{eq:sec2_one_vector_Lambda_new}
\end{equation}

On the hypermultiplet side take the universal hypermultiplet
\cite{Bagger:1983tt,Cecotti:1988qn,Ferrara:1989ik}, with real scalars
\((\phi,\sigma,\zeta,\tilde\zeta)\). Its metric may be written as
\begin{equation}
ds^2
=
d\phi^2
+\frac14 e^{4\phi}
\Big(d\sigma+\frac12(\zeta\,d\tilde\zeta-\tilde\zeta\,d\zeta)\Big)^2
+\frac14 e^{2\phi}\big(d\zeta^2+d\tilde\zeta^2\big).
\label{eq:sec2_UH_metric_new}
\end{equation}
It admits the Heisenberg isometries
\begin{equation}
k_\sigma=\partial_\sigma,
\qquad
k_\zeta=\partial_\zeta-\frac12\tilde\zeta\,\partial_\sigma,
\qquad
k_{\tilde\zeta}=\partial_{\tilde\zeta}+\frac12\zeta\,\partial_\sigma .
\label{eq:sec2_UH_isometries_new}
\end{equation}

We first gauge the axionic shift generated by \(k_\sigma\), a standard
hypermultiplet-isometry gauging in \(\mathcal N=2\) supergravity
\cite{Andrianopoli:1996cm,Freedman:2012zz},
\begin{equation}
k_\Lambda=g_\Lambda\,k_\sigma,
\qquad
\Lambda=0,1,
\label{eq:sec2_sigma_shift_gauging_new}
\end{equation}
with real constants \(g_\Lambda\). These constants are the gauging data. They
determine the vector combination \(g_\Lambda A_\mu^\Lambda\) and, after
canonical normalization of the vectors, the corresponding coupling of the
gravitini to that gauged direction.

On the truncation locus
\begin{equation}
\zeta=\tilde\zeta=0,
\label{eq:sec2_UH_truncation_new}
\end{equation}
the metric reduces to
\begin{equation}
ds^2
=
d\phi^2+\frac14 e^{4\phi}d\sigma^2
+\frac14 e^{2\phi}(d\zeta^2+d\tilde\zeta^2),
\label{eq:sec2_UH_metric_truncated_new}
\end{equation}
and
\begin{equation}
k_\Lambda^u\partial_u=g_\Lambda\,\partial_\sigma .
\label{eq:sec2_killing_sigma_only_new}
\end{equation}
The covariant derivative is
\begin{equation}
D_\mu \sigma=\partial_\mu \sigma + g_\Lambda A_\mu^\Lambda .
\label{eq:sec2_cov_sigma_new}
\end{equation}
The hypermultiplet kinetic term then contains the Stückelberg coupling
\begin{equation}
-\,h_{uv}D_\mu q^uD^\mu q^v
\supset
-\frac14 e^{4\phi}
\big(\partial_\mu\sigma+g_\Lambda A_\mu^\Lambda\big)^2.
\label{eq:sec2_stueckelberg_term_new}
\end{equation}
In unitary gauge the axion \(\sigma\) is eaten and the vector mass matrix is
\begin{equation}
\mathcal L_{\rm mass}^{(V)}
=
-\frac14 e^{4\phi_\ast}\,
g_\Lambda g_\Sigma\,A_\mu^\Lambda A^{\Sigma\mu}.
\label{eq:sec2_vector_mass_matrix_new}
\end{equation}
Thus the vector combination \(g_\Lambda A_\mu^\Lambda\) becomes massive.

The moment maps are aligned on the same locus:
\begin{equation}
P_\Lambda^1=0,
\qquad
P_\Lambda^2=0,
\qquad
P_\Lambda^3=-\frac14\,g_\Lambda e^{2\phi},
\label{eq:sec2_UH_moment_maps_truncated_new}
\end{equation}
up to the overall sign convention of the quaternionic moment-map equation.
Therefore
\begin{equation}
\big(P_\Lambda^aL^\Lambda\big)
=
\delta^{a3}\,\mathcal P,
\qquad
\mathcal P
=
-\frac14\,e^{2\phi}\,g_\Lambda L^\Lambda .
\label{eq:sec2_alignment_hyper_explicit_new}
\end{equation}
The gravitino shift is
\begin{equation}
S_{AB}
=
-\frac{i}{8}\,e^{2\phi}\,
(g_\Lambda L^\Lambda)\,(\sigma^3)_{AB},
\qquad
(\zeta=\tilde\zeta=0),
\label{eq:sec2_SAB_hyper_explicit_new}
\end{equation}
and the common gravitino mass is
\begin{equation}
m_{3/2}
=
\frac18\,e^{2\phi_\ast}\,
\big|g_\Lambda L^\Lambda\big|_\ast .
\label{eq:sec2_mD_hyper_general_new}
\end{equation}
Using \eqref{eq:sec2_one_vector_Lambda_new},
\begin{equation}
m_{3/2}
=
\frac18\,e^{2\phi_\ast}\,
\left|
\frac{g_0+z_\ast g_1}{\sqrt{2(z_\ast+\bar z_\ast)}}
\right|.
\label{eq:sec2_mD_hyper_explicit_new}
\end{equation}
Here and below, a subscript \(\ast\) denotes evaluation at the vacuum.

The same gauging data \(g_\Lambda\) determine the gauged vector combination,
enter its Stückelberg mass, and fix the gravitino shift. The numerical
mass-to-charge ratio is field-dependent because it also depends on the vacuum
values of the vector- and hypermultiplet scalars.

The scalar potential has the form
\begin{equation}
V
=
\Big(4h_{uv}k_\Lambda^u k_\Sigma^v\Big)\bar L^\Lambda L^\Sigma
+
\Big(U^{\Lambda\Sigma}-3\bar L^\Lambda L^\Sigma\Big)P_\Lambda^aP_\Sigma^a.
\label{eq:sec2_V_hyper_explicit_new}
\end{equation}
For the pure \(\sigma\)-shift gauging,
\begin{equation}
4h_{uv}k_\Lambda^u k_\Sigma^v
=
4h_{\sigma\sigma}\,g_\Lambda g_\Sigma
=
e^{4\phi}\,g_\Lambda g_\Sigma,
\qquad
(\zeta=\tilde\zeta=0),
\label{eq:sec2_k2_hyper_new}
\end{equation}
since \(h_{\sigma\sigma}=\frac14 e^{4\phi}\). This is the positive
Killing-vector contribution. The second line of
\eqref{eq:sec2_V_hyper_explicit_new} contains the universal negative term
proportional to
\(-3\bar L^\Lambda L^\Sigma P_\Lambda^aP_\Sigma^a\), as well as the
\(U^{\Lambda\Sigma}P_\Lambda^aP_\Sigma^a\) term. In a complete gauging, the
Minkowski condition is a balance among all these contributions.

For the simple pure \(\sigma\)-shift gauging displayed above, an isolated
Minkowski vacuum is not obtained by the ansatz alone. The gauging removes one
axionic degree of freedom and gives a Stückelberg mass to one vector
combination, but it does not generically stabilize all remaining scalar
directions. In particular, \(z\) and \(\phi\) remain unfixed without additional
gauging data or additional sectors. A complete construction would have to
supplement this block so as to obtain an isolated stationary point with
\begin{equation}
V\big|_\ast=0,
\qquad
\partial_iV\big|_\ast=\partial_uV\big|_\ast=0 .
\label{eq:sec2_Minkowski_condition_hyper_new}
\end{equation}

This limitation does not affect the role of the example in the present
analysis. We use it only to display the ingredients that will be needed later.
A genuine hypermultiplet gauging can align the gravitino shift, preserve a
\(U(1)_R\) in the spin-\(\tfrac32\) sector, relate the gravitino mass to the
same gauging data that determine the coupling to the gauged vector
combination, and add the positive \(k^2\) contribution to the scalar
potential. Thus the Dirac-gravitino structure is already visible in the
aligned hypermultiplet gauging, even before completing the construction to a
fully stabilized vacuum.

















\section{Scherk--Schwarz breaking on \texorpdfstring{$S^1/\mathbb Z_2$}{S1/Z2} and the Dirac gravitino}
\label{sec:SS_origin}

A particularly useful realization of the spin-\(\tfrac32\) structure discussed
above is provided by Scherk--Schwarz supersymmetry breaking on
\(S^1/\mathbb Z_2\)
\cite{Scherk:1978ta,Scherk:1979zr,Cremmer:1979uq}. Since the mechanism is
standard, we focus on the points needed below: the relation with an
\(SU(2)_R\) rotation in the covering theory, the orbifold parities at the two
fixed points, the signed mass spectrum on the interval, the special role of the
anti-periodic point, and the way in which localized boundary terms can move the
system away from the pure half-twist benchmark.

Before the orbifold projection is imposed, five-dimensional minimal
supergravity has eight real supercharges and corresponds, in four-dimensional
language, to \(\mathcal N=2\) supersymmetry. The two four-dimensional Weyl
gravitini \(\psi_{\mu 1}\) and \(\psi_{\mu 2}\) form a doublet under the
\(SU(2)_R\) automorphism group of the five-dimensional theory. In the
conventions of Ref.~\cite{Benakli:2007zza}, the Scherk--Schwarz twist on the
covering circle is written as the \(SU(2)_R\) rotation
\begin{equation}
\begin{pmatrix}
\psi_{\mu1}(y+2\pi R)\\[1mm]
\psi_{\mu2}(y+2\pi R)
\end{pmatrix}
=
\begin{pmatrix}
\cos(2\pi\omega) & \sin(2\pi\omega)\\[1mm]
-\sin(2\pi\omega) & \cos(2\pi\omega)
\end{pmatrix}
\begin{pmatrix}
\psi_{\mu1}(y)\\[1mm]
\psi_{\mu2}(y)
\end{pmatrix}.
\label{eq:SS_twist_rotation}
\end{equation}
Equivalently, in the covering-space description, the twist may be represented
by a background \(R\)-symmetry connection along the compact direction,
\begin{equation}
D_y=\partial_y-iA_y^{(R)},
\qquad
A_y^{(R)}=\frac{\omega}{R}\,T_R ,
\label{eq:SS_background_connection}
\end{equation}
where \(T_R\) is the \(SU(2)_R\) generator corresponding to the rotation in
\eqref{eq:SS_twist_rotation}. This is the setting in which the analogy with
four-dimensional \(\mathcal N=2\) gauged supergravity is most direct. Locally,
the Scherk--Schwarz background is described by an aligned gravitino shift,
\begin{equation}
S_{AB}
=
\frac{i}{2}(\sigma^a)_{AB}P_\Lambda^aL^\Lambda ,
\label{eq:SS_SAB_match_new}
\end{equation}
with
\begin{equation}
\big(P_\Lambda^aL^\Lambda\big)\Big|_{\rm vac}
=
\delta^{aa_0}\,\mathcal P\Big|_{\rm vac}.
\label{eq:SS_alignment_match_new}
\end{equation}
Thus the aligned-gauging language of
Section~\ref{sec:N2_to_N0_dirac} captures the local bulk
spin-\(\tfrac32\) data of the Scherk--Schwarz construction.

The orbifold projection changes the global interpretation of this statement.
On \(S^1/\mathbb Z_2\) the physical space is an interval. The modes are
standing waves rather than momentum eigenstates on the circle, and the
massless graviphoton associated with translations along the fifth direction is
projected out. Therefore the Kaluza--Klein momentum of the covering circle is
not a four-dimensional gauge charge of the orbifold theory. Moreover, the
fixed planes preserve only \(\mathcal N=1\) supersymmetry. Boundary
interactions are therefore not required, in general, to preserve the full
\(SU(2)_R\) structure of the bulk covering theory.

The interval spectrum is most cleanly described in terms of the parities at the
two fixed points. At \(y=0\) we choose
\begin{equation}
\psi_{\mu1}(-y)=+\psi_{\mu1}(y),
\qquad
\psi_{\mu2}(-y)=-\psi_{\mu2}(y),
\label{eq:SS_parity_zero}
\end{equation}
or equivalently
\begin{equation}
\partial_y\psi_{\mu1}\big|_{0}=0,
\qquad
\psi_{\mu2}\big|_{0}=0 .
\label{eq:SS_bc_zero}
\end{equation}
At \(y=\pi R\), the even and odd combinations are rotated by the
Scherk--Schwarz angle. Following Ref.~\cite{Benakli:2007zza}, define
\begin{align}
\psi_{\mu+}
&=
\cos(\pi\omega)\,\psi_{\mu1}
-
\sin(\pi\omega)\,\psi_{\mu2},
\label{eq:SS_rotated_plus}
\\[1mm]
\psi_{\mu-}
&=
\sin(\pi\omega)\,\psi_{\mu1}
+
\cos(\pi\omega)\,\psi_{\mu2}.
\label{eq:SS_rotated_minus}
\end{align}
The parity assignment at the second fixed point is
\begin{equation}
\psi_{\mu+}(\pi R-y)=+\psi_{\mu+}(\pi R+y),
\qquad
\psi_{\mu-}(\pi R-y)=-\psi_{\mu-}(\pi R+y),
\label{eq:SS_parity_pi}
\end{equation}
or equivalently
\begin{equation}
\partial_y\psi_{\mu+}\big|_{\pi R}=0,
\qquad
\psi_{\mu-}\big|_{\pi R}=0 .
\label{eq:SS_bc_pi}
\end{equation}

For a mode of signed four-dimensional mass \(m\), the conditions at \(y=0\)
are solved by
\begin{equation}
\psi_{\mu1}(x,y)
=
\psi_\mu^{(m)}(x)\cos(my),
\qquad
\psi_{\mu2}(x,y)
=
-\,\psi_\mu^{(m)}(x)\sin(my),
\label{eq:SS_signed_mode_functions}
\end{equation}
up to a convention-dependent sign in the second equation. Imposing
\(\psi_{\mu-}|_{\pi R}=0\) gives
\begin{equation}
\sin(\pi\omega)\cos(m\pi R)
-
\cos(\pi\omega)\sin(m\pi R)
=0,
\end{equation}
or
\begin{equation}
\sin(\pi\omega-m\pi R)=0 .
\end{equation}
Thus the signed masses are
\begin{equation}
m_n=\frac{n+\omega}{R},
\qquad n\in\mathbb Z .
\label{eq:SS_signed_masses}
\end{equation}
The physical masses are \(|m_n|\).

Since the spectrum depends on \(|n+\omega|\), the pure Scherk--Schwarz spectra
at \(\omega\) and \(1-\omega\) are equivalent after the relabelling
\(n\to -n-1\). We therefore take the fundamental range to be
\begin{equation}
0\leq \omega\leq \frac12 .
\label{eq:SS_fundamental_range}
\end{equation}
In this range the lightest spin-\(\tfrac32\) mode is the signed mode \(n=0\),
with physical mass
\begin{equation}
m_{\rm light}=\frac{\omega}{R}.
\label{eq:SS_light_mass_generic}
\end{equation}
For \(0<\omega<\tfrac12\), the next state has mass
\((1-\omega)/R\) and is separated by a finite gap. At energies below that
threshold, the appropriate effective description is an ordinary broken
\(\mathcal N=1\) supergravity with one massive Majorana gravitino. There is no
Dirac organization of the light spin-\(\tfrac32\) sector in this generic
case.

The endpoints are special. At \(\omega=0\), the mode \(n=0\) is massless and
gives the usual \(\mathcal N=1\) Majorana gravitino. The massive modes
\(n\) and \(-n\), \(n\geq1\), have equal physical masses and may be reorganized
into Dirac pairs at the level of the free spectrum. This is the standard
massive Kaluza--Klein pairing, not the low-energy Dirac-gravitino structure
studied in this paper.

The anti-periodic point,
\begin{equation}
\omega=\frac12,
\qquad
M_0=M_\pi=0 ,
\label{eq:SS_pure_half_benchmark_new}
\end{equation}
is qualitatively different. The signed masses are
\begin{equation}
m_n=\frac{n+\tfrac12}{R}.
\end{equation}
The modes \(n\) and \(-n-1\) have opposite signed masses and equal physical
masses. In particular, the two lightest signed modes are
\begin{equation}
m_0=+\frac{1}{2R},
\qquad
m_{-1}=-\frac{1}{2R}.
\label{eq:SS_two_half_modes}
\end{equation}
Thus the light spin-\(\tfrac32\) sector contains two degenerate Majorana
modes. After a phase redefinition of the negative-mass mode, they may be
combined into a Dirac basis. If we denote the two degenerate Weyl
spin-\(\tfrac32\) fields by \(\eta_\mu\) and \(\zeta_\mu\), the Dirac basis may
be chosen as
\begin{equation}
\psi_\mu=\frac{1}{\sqrt2}\big(\eta_\mu+i\zeta_\mu\big),
\qquad
\chi_\mu=\frac{1}{\sqrt2}\big(\eta_\mu-i\zeta_\mu\big).
\label{eq:SS_Dirac_basis_half}
\end{equation}
In this basis the mass term takes the Dirac form
\begin{equation}
\mathcal L_{m}^{(3/2)}
=
-\,m_{3/2}\,
\psi^\mu\sigma_{\mu\nu}\chi^\nu
+\text{h.c.},
\qquad
m_{3/2}=\frac{1}{2R}.
\label{eq:SS_Dirac_mass_half}
\end{equation}
This is the Dirac gravitino of the pure half-twist benchmark.

The same conclusion can be expressed in terms of the parities. At
\(\omega=\tfrac12\), the rotated fields at \(y=\pi R\) are
\begin{equation}
\psi_{\mu+}=-\psi_{\mu2},
\qquad
\psi_{\mu-}=\psi_{\mu1}.
\end{equation}
Thus the parity matrix at \(y=\pi R\) is opposite to the one at \(y=0\). The
two parity matrices are then simultaneously preserved by the \(U(1)\) generated
by the diagonal \(SU(2)_R\) generator which acts as
\begin{equation}
\psi_{\mu1}\to e^{i\alpha}\psi_{\mu1},
\qquad
\psi_{\mu2}\to e^{-i\alpha}\psi_{\mu2}.
\label{eq:SS_half_UR_action_bulk}
\end{equation}
This continuous symmetry is present for the pure boundary-condition problem at
\(\omega=\tfrac12\). In the Dirac basis of the degenerate light sector it acts
as a phase rotation,
\begin{equation}
\psi_\mu\to e^{i\alpha}\psi_\mu,
\qquad
\chi_\mu\to e^{-i\alpha}\chi_\mu .
\label{eq:SS_half_UR_action_Dirac}
\end{equation}
The off-diagonal mass term in \eqref{eq:SS_Dirac_mass_half} is neutral under
this \(U(1)_R\), while diagonal Majorana mass terms
\(\psi^\mu\sigma_{\mu\nu}\psi^\nu\) and
\(\chi^\mu\sigma_{\mu\nu}\chi^\nu\) are charged and are therefore forbidden.

This \(U(1)_R\) statement has a precise scope. It is a symmetry of the pure
orbifold boundary-condition problem at the half-twist point, and of any
completion which preserves the corresponding selection rule. It should not be
confused with a surviving Kaluza--Klein gauge symmetry: the massless
graviphoton associated with translations along the circle is projected out on
\(S^1/\mathbb Z_2\). It also need not be respected by arbitrary localized
boundary interactions, since the fixed planes preserve only
\(\mathcal N=1\) supersymmetry. Localized \(R\)-breaking or
supersymmetry-breaking terms can therefore split the Dirac pair or move the
system away from the pure half-twist. Small departures from this limit are
analyzed in Section~\ref{sec:small_R_breaking}.

A further point is important for later use. The condition
\(\omega=\tfrac12\) is the pure-bulk half-twist only when no additional
localized gravitino masses are present. If boundary sectors generate localized
gravitino mass parameters \(M_0\) and \(M_\pi\), the matching condition for an
unbroken Killing spinor is shifted. With \(M_b/M_\star\) denoting the
dimensionless localized mass parameters, one may write
\begin{equation}
\omega\pi
+\arctan\!\left(\frac{M_0}{M_\star}\right)
+\arctan\!\left(\frac{M_\pi}{M_\star}\right)
=n\pi,
\qquad n\in\mathbb Z .
\label{eq:SS_shifted_killing_condition}
\end{equation}
Equivalently, defining boundary angles by
\begin{equation}
\tan\Theta_b=\frac{M_b}{M_\star},
\qquad b=0,\pi,
\label{eq:SS_Theta_definition}
\end{equation}
the condition becomes
\begin{equation}
\omega\pi+\Theta_0+\Theta_\pi=n\pi .
\label{eq:SS_shifted_killing_condition_theta}
\end{equation}
Thus localized Scherk--Schwarz-type gravitino masses shift the effective
boundary conditions away from the pure-bulk point
\cite{Bagger:2001qi,Bagger:2001ep,Gherghetta:2001sa,Meissner:2002dg,
Biggio:2002rb,Rattazzi:2003rj,Benakli:2007zza}. In particular, if one starts
from the Dirac point \(\omega=\tfrac12\), nonzero boundary masses move the
system away from the exact half-twist unless their boundary angles satisfy a
compensating relation.

Localized gravitino mass terms and genuine boundary \(F\)-term breaking have
different status. Localized gravitino mass terms of the Scherk--Schwarz type
modify the boundary conditions of the bulk gravitini. They can be described
equivalently as boundary jump conditions, or, after a field redefinition, as
part of a generalized Scherk--Schwarz twist. This equivalence applies only to
the gravitino boundary conditions themselves. A brane \(F\)-term source
contains additional Goldstino data, because it represents spontaneous
supersymmetry breaking localized on the boundary. The resulting
gravitino--Goldstino system need not be reproducible by a single
\(SU(2)_R\) rotation of the bulk gravitino doublet. Thus such a source does
not in general amount to replacing \(\omega\) by an effective twist angle. It
introduces localized supersymmetry-breaking data beyond the pure
Scherk--Schwarz boundary condition. This distinction, including the appearance
of pseudo-Goldstino degrees of freedom when genuine brane \(F\)-terms are
present, was analyzed explicitly in Ref.~\cite{Benakli:2007zza}.

The orbifold parity structure also controls the coupling to boundary matter.
At the \(y=0\) fixed plane the field \(\psi_{\mu1}\) is even and has a
non-vanishing boundary value, whereas \(\psi_{\mu2}\) is odd and vanishes
there. Thus the direct boundary supercurrent coupling of a chiral multiplet
localized at \(y=0\) selects the even gravitino component. At the half-twist
point the mode functions used later in the mixed-representation kernels may be
chosen as
\begin{equation}
f_n^{(+)}(y)
=
\sqrt{\frac{2}{\pi R}}\,
\cos\!\left(\frac{n+\tfrac12}{R}y\right),
\qquad
f_n^{(-)}(y)
=
\sqrt{\frac{2}{\pi R}}\,
\sin\!\left(\frac{n+\tfrac12}{R}y\right),
\qquad n\ge0 .
\label{eq:SS_half_twist_profiles_new}
\end{equation}
On the \(y=0\) fixed plane,
\begin{equation}
f_n^{(+)}(0)\neq0,
\qquad
f_n^{(-)}(0)=0 .
\label{eq:SS_boundary_values_new}
\end{equation}
The ordinary boundary coupling is therefore to the even tower.

The odd tower is not absent from the physics. Since
\begin{equation}
\partial_y f_n^{(-)}(y)
=
\sqrt{\frac{2}{\pi R}}\,
\frac{n+\tfrac12}{R}
\cos\!\left(\frac{n+\tfrac12}{R}y\right),
\label{eq:SS_derivative_odd_new}
\end{equation}
its normal derivative is nonzero on the boundary. This is the
higher-dimensional origin of the statement that the odd gravitino can enter
through normal-derivative couplings or through the matter current generated by
the second supersymmetry of the extended multiplet structure. It does not mean
that the odd gravitino couples to the same boson--fermion pair as the even
one. Rather, the companion channel has its own matter-current structure. This
distinction is one of the ingredients underlying the effective assumptions made
in Section~\ref{sec:coupling_to_matter}: the Scherk--Schwarz geometry fixes the
even/odd kernels, but it does not determine the existence or normalization of
the companion matter current.

For the purposes of this paper, the main lesson is the following. The pure
anti-periodic Scherk--Schwarz point on \(S^1/\mathbb Z_2\),
\begin{equation}
\omega=\frac12,
\qquad
M_0=M_\pi=0,
\end{equation}
provides a standard benchmark in which
\begin{enumerate}
\item the free infrared spin-\(\tfrac32\) spectrum contains two degenerate
Majorana modes which can be organized as a Dirac gravitino with
\(m_{3/2}=1/(2R)\);
\item the pure boundary conditions preserve an effective \(U(1)_R\) acting on
the degenerate light pair, so that the off-diagonal Dirac mass is allowed
whereas diagonal Majorana deformations are forbidden;
\item this \(U(1)_R\) is a property of the pure half-twist benchmark, or of
completions preserving the same selection rule. It is not a surviving
Kaluza--Klein gauge symmetry of the orbifold theory;
\item boundary physics is only \(\mathcal N=1\) and can in general generate
localized deformations which move the theory away from the pure half-twist.
The small-deformation regime is analyzed in
Section~\ref{sec:small_R_breaking};
\item the direct boundary coupling selects the even gravitino wavefunction,
while the odd gravitino can enter through normal-derivative couplings or
through the companion matter current;
\item the geometry fixes the even/odd wavefunctions and the corresponding
boundary kernels, but not the companion matter current required for mixed
scalar terms or radiative Dirac fermion masses.
\end{enumerate}

This Scherk--Schwarz construction will serve as our explicit ultraviolet
benchmark. It realizes the same local Dirac spin-\(\tfrac32\) mass structure
that, in Section~\ref{sec:N2_to_N0_dirac}, was described in four-dimensional
\(\mathcal N=2\) gauged-supergravity language, while making explicit how the
orbifold projection and boundary \(\mathcal N=1\) physics restrict the
low-energy interpretation. In the limit where localized deformations away from
\(\omega=\tfrac12\) are negligible, it supplies the kinematical data used later
in the loop analysis: the half-twist mass \(m_{3/2}=1/(2R)\), the even/odd
parity selection rules, and the distinction between direct boundary couplings
and companion-channel structures.

\section{Low-energy EFT in \texorpdfstring{$\mathcal N=1$}{N=1} language}
\label{sec:N1_language_and_obstruction}

Sections~\ref{sec:N2_to_N0_dirac} and \ref{sec:SS_origin} established the
physical origin of the Dirac-gravitino structure relevant for this paper. The
starting point is a broken \(\mathcal N=2\) system, either described
intrinsically in four-dimensional gauged supergravity or realized explicitly by
Scherk--Schwarz breaking on \(S^1/\mathbb Z_2\). In the benchmark cases of
interest, a residual \(U(1)_R\) acts on the light spin-\(\tfrac32\) sector. At
the anti-periodic Scherk--Schwarz point this structure is visible directly: the
light spin-\(\tfrac32\) sector is not a single Majorana gravitino but a
degenerate pair of Weyl gravitino modes organized as a Dirac gravitino.

The next question is how to describe this structure in a language suitable for
phenomenology. The natural language for the matter sector is
\(\mathcal N=1\) supersymmetry. Visible sectors are usually written in terms of
chiral and vector multiplets, with K\"ahler potential \(K\), superpotential
\(W\), and gauge kinetic functions \(f\). In explicit boundary constructions,
such as Scherk--Schwarz compactifications, the matter sector localized on a
fixed plane is indeed organized as a softly broken \(\mathcal N=1\) theory.

The subtlety concerns the spin-\(\tfrac32\) sector. Once one chooses an
\(\mathcal N=1\) subalgebra inside the parent \(\mathcal N=2\) theory, one
gravitino belongs to the ordinary \(\mathcal N=1\) supergravity multiplet. The
second spin-\(\tfrac32\) field is a separate degree of freedom. In the strict
\(\mathcal N=2\) or five-dimensional Scherk--Schwarz realization it is the
second gravitino, but this need not be true in a more general low-energy EFT.
For the purposes of the effective theory, what is required is a companion
spin-\(\tfrac32\) field with the appropriate \(U(1)_R\) charge and couplings.
It may be the second gravitino of an extended-supergravity completion, or it
may be treated more generally as an independent companion spin-\(\tfrac32\)
field whose low-energy couplings have to be specified. After the Dirac mixing
is turned on, the physical eigenstates form a massive spin-\(\tfrac32\) system.

This distinction matters. If the companion field is the second gravitino of an
underlying \(\mathcal N=2\) or five-dimensional supergravity, its normalization
and part of its coupling structure are constrained by the UV theory. If instead
it is an independent low-energy spin-\(\tfrac32\) field, the existence and
normalization of its matter current are EFT data. In both cases, however, the
same structural question arises: can the Dirac mixing between the ordinary
gravitino and the companion spin-\(\tfrac32\) field be represented by a local
\(\mathcal N=1\) superspace bilinear, in close analogy with the familiar
Dirac-gaugino operator? The answer is no.

This negative result is compatible with, but stronger than, what is seen
directly in the Scherk--Schwarz construction. In the pure half-twist
compactification it is already clear that the massive spin-\(\tfrac32\) sector
is not obtained by keeping only a single four-dimensional \(\mathcal N=1\)
supergravity multiplet: the light mode is defined together with the
higher-dimensional KK completion and the orbifold boundary conditions. The
superspace obstruction derived below goes beyond this specific KK fact. It
shows that even in a purely four-dimensional low-energy description, and even
if the companion spin-\(\tfrac32\) field is not assumed to be literally the
second gravitino of a Scherk--Schwarz tower, the desired Dirac
spin-\(\tfrac32\) mixing is not generated by a standard local
\(\mathcal N=1\) superspace operator of Dirac-gaugino type. To isolate the
physical spin-\(\tfrac32\) sector, one must project onto the transverse
superspin-\(\tfrac32\) component; that projection is non-local in superspace.

Concretely, after choosing one \(\mathcal N=1\) subalgebra, the gravity sector
of an \(\mathcal N=2\) theory decomposes schematically as
\begin{equation}
(g_{\mu\nu},\psi_{\mu 1},\psi_{\mu 2},A_\mu^0)
\quad\longrightarrow\quad
\underbrace{(g_{\mu\nu},\psi_{\mu 1})}_{\text{\(\mathcal N=1\) gravity multiplet}}
\;\oplus\;
\underbrace{(\psi_{\mu 2},A_\mu^0)}_{\text{\(\mathcal N=1\) superspin-1 multiplet before mixing}} .
\label{eq:sec4_gravity_decomp}
\end{equation}
In a genuine \(\mathcal N=2\) completion, the companion spin-\(\tfrac32\)
field is \(\psi_{\mu2}\). In a more general EFT, we denote the companion field
by \(\chi_\mu\) and only require that it carry the appropriate \(U(1)_R\)
charge to pair with the ordinary gravitino. Similarly, an \(\mathcal N=2\)
vector multiplet decomposes into one \(\mathcal N=1\) vector multiplet plus one
\(\mathcal N=1\) chiral multiplet, while a hypermultiplet decomposes into two
\(\mathcal N=1\) chiral multiplets.

This rewriting is useful for the scalar sector. For example, in the
universal-hypermultiplet parametrization one may choose complex coordinates, of
the schematic form
\begin{equation}
S \sim e^{-2\phi}+i\sigma+\cdots,
\qquad
C \sim \zeta+i\tilde\zeta,
\label{eq:sec4_UH_chirals}
\end{equation}
in which the scalar metric is described by a K\"ahler potential of the form
\begin{equation}
K_{\rm UH}
=
-\log\!\big(S+\bar S-2C\bar C\big),
\label{eq:sec4_KUH}
\end{equation}
up to convention-dependent field redefinitions and normalizations
\cite{Cecotti:1988qn,Ferrara:1989ik}. For the purposes of this section, only
the existence of such an \(\mathcal N=1\) K\"ahler description matters.
Combined with the special-K\"ahler contribution \(K_{\rm SK}\), the scalar
kinetic terms take the standard \(\mathcal N=1\) sigma-model form with total
K\"ahler potential
\begin{equation}
K=K_{\rm SK}+K_{\rm UH}.
\label{eq:sec4_totalK}
\end{equation}
Likewise, gauged \(\mathcal N=2\) isometries act on suitable
\(\mathcal N=1\) chiral coordinates as holomorphic Killing vectors. From the
\(\mathcal N=1\) viewpoint the corresponding scalar couplings may be described
by the usual \((K,W,f)\) data whenever an \(\mathcal N=1\) truncation or
effective description is available.

In an \(\mathcal N=1\) effective description of the scalar sector, one often
parameterizes the mass scale associated with the ordinary gravitino sector as
\begin{equation}
m_{3/2}\sim e^{K/2}|W_{\rm eff}|.
\label{eq:sec4_weff}
\end{equation}
This formula should be understood only as a scalar-sector parametrization of
the supersymmetry-breaking scale in an \(\mathcal N=1\) description. By itself
it would correspond to the usual Majorana gravitino mass of ordinary
\(\mathcal N=1\) supergravity. In the exact \(U(1)_R\)-symmetric Dirac limit
studied here, such diagonal Majorana masses are forbidden. The physical
spin-\(\tfrac32\) mass is instead the off-diagonal Dirac mixing between the
ordinary gravitino and the companion spin-\(\tfrac32\) field.
This statement should not be confused with a local \(\mathcal N=1\) superspace
origin for the Dirac spin-\(\tfrac32\) mixing itself. In no-scale examples one
recovers the familiar pattern
\begin{equation}
V_F
=
e^K\big(K^{i\bar j}D_iW D_{\bar j}\bar W - 3|W|^2\big)
\equiv 0,
\qquad
F^i\neq0,
\label{eq:sec4_noscale}
\end{equation}
namely Minkowski space, broken supersymmetry, and nonzero gravitino mass.

Thus \(\mathcal N=1\) language is the correct language for the scalar geometry
and for visible matter. The issue is narrower and more structural. It concerns
the spin-\(\tfrac32\) mixing between the ordinary \(\mathcal N=1\) gravitino
and a companion spin-\(\tfrac32\) field. We now show that this mixing cannot be
written as a local \(\mathcal N=1\) superspace bilinear analogous to the
Dirac-gaugino operator. The obstruction is not merely that a naive component
mass term breaks gauge invariance. Rather, within the standard local
\(4D,\mathcal N=1\) superspace framework, the spin-\(\tfrac32\) sector does not
provide the local chiral rank-one building blocks from which such a
Dirac-gaugino-type operator could be constructed.

\subsection{Linearized superspace setup and the failure of the natural local bilinear}
\label{subsec:linearized_setup_and_local_failure}

The issue already appears at the quadratic level. In components, the desired
spin-\(\tfrac32\) structure is a Dirac-type mixing between the ordinary
gravitino \(\psi_\mu\) and a companion spin-\(\tfrac32\) field \(\chi_\mu\),
\begin{equation}
\mathcal L_{\rm Dirac}^{(3/2)}
=
-\,m_{3/2}\,
\psi^\mu\sigma_{\mu\nu}\chi^\nu
+\text{h.c.},
\label{eq:sec4_component_target}
\end{equation}
or equivalently its four-component Rarita--Schwinger form. The question is
whether such a term can arise from a \emph{local} \(\mathcal N=1\) superspace
bilinear, in direct analogy with the familiar Dirac-gaugino operator.

For this purpose it is sufficient to use the standard linearized superspace
descriptions of the two relevant massless multiplets.\footnote{In this section
we use van der Waerden notation, since the linearized \(4D,\mathcal N=1\)
superspace gauge structure is most transparent in dotted and undotted indices.
See for example \cite{Wess:1992cp,Gates:1983nr,Buchbinder:1998twe}.}
The ordinary graviton--gravitino system is described, in old-minimal
linearized supergravity, by a real vector superfield \(H_{\alpha\dot\alpha}\)
together with a chiral compensator \(\sigma\),
\begin{equation}
\bar D_{\dot\alpha}\sigma=0 .
\end{equation}
We use the linearized gauge transformations
\begin{equation}
\delta H_{\alpha\dot\alpha}
=
\bar D_{\dot\alpha}L_\alpha
-
D_\alpha\bar L_{\dot\alpha},
\qquad
\delta\sigma
=
-\frac{1}{12}\bar D^2D^\alpha L_\alpha ,
\label{eq:sec4_OMgauge}
\end{equation}
where \(L_\alpha\) is an unconstrained spinor gauge parameter. A convenient
quadratic old-minimal action may then be written in projector form as
\begin{equation}
S_{\rm om}[H,\sigma]
=
\int d^8z\,
H^a\Box\left(-\frac13 \Pi^L_0+\frac12 \Pi^T_{3/2}\right)H_a
+
S_{\rm comp}[H,\sigma] ,
\label{eq:sec4_OMaction}
\end{equation}
where \(S_{\rm comp}[H,\sigma]\) denotes the chiral-compensator sector required
by old-minimal supergravity. Its explicit form depends on the normalization
chosen for the compensator and will not be needed below. Only the transverse
superspin-\(\tfrac32\) part of \eqref{eq:sec4_OMaction} will be used in the
following. Its rôle is to identify the physical graviton--gravitino sector as
the \(\Pi^T_{3/2}H_a\) component of the real vector prepotential.

The companion spin-\(\tfrac32\) field can be described, before the Dirac
mixing is turned on, by the linearized \(\mathcal N=1\) massless gravitino
multiplet. This terminology refers to the representation, not necessarily to
its ultraviolet origin. In an \(\mathcal N=2\) completion the field is the
second gravitino; in a more general EFT it is simply the companion
spin-\(\tfrac32\) field. A standard superspace description uses a spinor
prepotential \(\Psi_\alpha\) together with a chiral compensator \(\Phi\). We
follow the conventional linearized superspace description of massless
superspin multiplets; see, for example,
Refs.~\cite{Gates:1983nr,Buchbinder:1998twe,Hutomo:2017phd}. For our purposes
only the gauge structure and the field content are needed, so we write the
action schematically as
\begin{equation}
S_{3/2}[\Psi,\Phi]
=
S_b[\Psi]
-\frac12\int d^8z
\Big\{
\Phi\bar\Phi
+
(\Phi+\bar\Phi)
\big(D^\alpha\Psi_\alpha+\bar D_{\dot\alpha}\bar\Psi^{\dot\alpha}\big)
\Big\},
\label{eq:sec4_S32}
\end{equation}
where \(S_b[\Psi]\) denotes the standard kinetic term. The detailed
normalization of \(S_b[\Psi]\) will not be used below. The only properties
needed are that \(\Psi_\alpha\) contains the physical Rarita--Schwinger mode
and that the multiplet has its own gauge redundancy.

To avoid confusing the genuine obstruction with a trivial gauge-exact
contribution, it is useful to strip off the pure-gradient part of the
\(\Psi_\alpha\) gauge transformation. If
\begin{equation}
\delta\Psi_\alpha=D_\alpha\Omega+\Lambda_\alpha,
\qquad
\bar D_{\dot\alpha}\Omega=0,
\label{eq:sec4_Psigauge}
\end{equation}
one may introduce a chiral superfield \(Y\), \(\bar D_{\dot\alpha}Y=0\), and
define
\begin{equation}
\widehat\Psi_\alpha\equiv \Psi_\alpha-D_\alpha Y.
\label{eq:sec4_shiftedPsi}
\end{equation}
Choosing
\begin{equation}
\delta Y=\Omega
\end{equation}
removes the pure-gradient piece and leaves
\begin{equation}
\delta\widehat\Psi_\alpha=\Lambda_\alpha .
\label{eq:sec4_shiftedPsiGauge}
\end{equation}
This step is only diagnostic: it ensures that the failure of the mixing term
below cannot be blamed on the trivial \(D_\alpha\Omega\) part.

With \(H_{\alpha\dot\alpha}\) and \(\widehat\Psi_\alpha\) in hand, the most
obvious real vector superfield candidate built from the companion prepotential
is
\begin{equation}
J_{\alpha\dot\alpha}
=
i\left(
\bar D_{\dot\alpha}\widehat\Psi_\alpha
-
D_\alpha\bar{\widehat\Psi}_{\dot\alpha}
\right).
\label{eq:sec4_current}
\end{equation}
The factor of \(i\) makes \(J_{\alpha\dot\alpha}\) real. This superfield has
the correct index structure to couple to \(H_{\alpha\dot\alpha}\), so the
natural local bilinear to try is
\begin{equation}
S_{\rm mix}^{\rm local}
=
\mu\int d^8z\;
H^{\alpha\dot\alpha}J_{\alpha\dot\alpha},
\label{eq:sec4_localmix}
\end{equation}
with \(\mu\) a mass parameter.

This term has the right superficial properties: it is real, quadratic, Lorentz
invariant, and local in superspace. However, it is not gauge invariant under
the old-minimal transformation \eqref{eq:sec4_OMgauge}. Indeed,
\begin{equation}
\delta S_{\rm mix}^{\rm local}
=
\mu\int d^8z\;
\big(
\bar D_{\dot\alpha}L_\alpha
-
D_\alpha\bar L_{\dot\alpha}
\big)
J^{\alpha\dot\alpha}.
\label{eq:sec4_variation1}
\end{equation}
After integrating by parts in superspace, this becomes
\begin{equation}
\delta S_{\rm mix}^{\rm local}
=
\mu\int d^8z\;
\Big[
\bar L_{\dot\alpha}D_\alpha J^{\alpha\dot\alpha}
-
L_\alpha\bar D_{\dot\alpha}J^{\alpha\dot\alpha}
\Big],
\label{eq:sec4_variation2}
\end{equation}
up to the standard signs from integrating spinor derivatives by parts in
superspace. Thus gauge invariance of this simple local bilinear would require
\begin{equation}
D^\alpha J_{\alpha\dot\alpha}=0,
\qquad
\bar D^{\dot\alpha}J_{\alpha\dot\alpha}=0 .
\label{eq:sec4_needed_constraints}
\end{equation}
One could try to add compensator-dependent local terms. Such terms can affect
lower-superspin components, but they do not change the basic index-structure
problem emphasized below.

For the vector superfield \eqref{eq:sec4_current} one finds
\begin{equation}
D^\alpha J_{\alpha\dot\alpha}
=
i\left(
D^\alpha\bar D_{\dot\alpha}\widehat\Psi_\alpha
-
D^2\bar{\widehat\Psi}_{\dot\alpha}
\right),
\label{eq:sec4_DJ}
\end{equation}
and similarly
\begin{equation}
\bar D^{\dot\alpha}J_{\alpha\dot\alpha}
=
i\left(
\bar D^2\widehat\Psi_\alpha
-
\bar D^{\dot\alpha}D_\alpha\bar{\widehat\Psi}_{\dot\alpha}
\right).
\label{eq:sec4_DbarJ}
\end{equation}
These expressions do not vanish identically. The problematic structures are
\begin{equation}
D^\alpha\bar D_{\dot\alpha}\widehat\Psi_\alpha,
\qquad
D^2\bar{\widehat\Psi}_{\dot\alpha},
\qquad
\bar D^2\widehat\Psi_\alpha,
\qquad
\bar D^{\dot\alpha}D_\alpha\bar{\widehat\Psi}_{\dot\alpha},
\label{eq:sec4_badterms}
\end{equation}
which are neither simple chiral field strengths nor removable gauge-exact
pieces.

The point is stronger than the statement that a mass term breaks a gauge
symmetry. What fails is the existence of a \emph{local superspace realization}
of the Dirac mixing analogous to the Dirac-gaugino operator. In the
Dirac-gaugino case the relevant local holomorphic building block is the
rank-one chiral field strength \(W_\alpha\). The analogous question here is
whether the spin-\(\tfrac32\) sector furnishes a local gauge-invariant chiral
superfield with a single undotted spinor index, built from the standard
\(\mathcal N=1\) prepotentials, whose physical fermionic content is the
Rarita--Schwinger mode or its irreducible field strength. In the standard
off-shell formulations of linearized \(4D,\mathcal N=1\) supergravity, the
local gauge-invariant chiral curvature carrying the physical gravitino is not
such a rank-one spinor superfield. In old-minimal language this is seen
directly from the super-Weyl tensor \(W_{\alpha\beta\gamma}\). The same
index-structure obstruction persists in new-minimal language, where the
compensator sector is different but the physical supergravity curvature is not
replaced by a \(W_\alpha\)-type object.

The argument above was phrased in old-minimal linearized supergravity, where
the compensator is chiral. This is sufficient for the matter effective
description used in the rest of the paper, but the obstruction is not an
artifact of old-minimal auxiliary fields. New-minimal supergravity uses a real
linear compensator and has a manifest local \(U(1)_R\) gauge symmetry. However,
it does not change the gauge-invariant local curvature that contains the
physical graviton--gravitino degrees of freedom. The linearized supergravity
prepotential is still \(H_{\alpha\dot\alpha}\), and its gauge-invariant chiral
supercurvature is still the super-Weyl tensor
\begin{equation}
W_{\alpha\beta\gamma}(H),
\end{equation}
which carries three undotted spinor indices
\cite{Gates:1983nr,Buchbinder:1998twe}. The new-minimal compensator supplies
the \(R\)-multiplet structure and the auxiliary \(U(1)_R\) gauge field, but it
does not supply a local gauge-invariant chiral rank-one spinor superfield
whose physical fermionic component is the Rarita--Schwinger mode.

Thus the auxiliary sector can change the form of lower-spin and \(R\)-current
couplings, but it cannot turn the spin-\(\tfrac32\) gauge curvature into a
rank-one chiral field strength analogous to the gauge multiplet \(W_\alpha\).
The available local chiral curvatures have the wrong index structure for a
Dirac-gaugino-like bilinear. The corresponding gauge-invariant chiral field
strength of the massless gravitino multiplet is not a rank-one spinor
superfield: it carries more than one undotted spinor index. The superfield
representation of the spin-\(\tfrac32\) sector therefore does not provide the
rank-one chiral ingredients needed to imitate the Dirac-gaugino construction.

The rôle of the Stückelberg field \(Y\) is only diagnostic. After removing the
trivial gauge-exact part of \(\Psi_\alpha\), the obstruction remains unchanged.
The problem lies in the coupling to the supergravity multiplet itself and in
the index structure of the available local chiral supercurvatures. In this
precise sense, the obstruction is structural.

\subsection{Projected non-local coupling, physical reduction, and the vector sector}
\label{subsec:projected_nonlocal}

Once the local obstruction is understood, the correct strategy is also clear.
The vector superfield \(J_{\alpha\dot\alpha}\) in \eqref{eq:sec4_current}
contains not only the physical spin-\(\tfrac32\) mode of
\(\widehat\Psi_\alpha\), but also lower-superspin components. Similarly, the
real prepotential \(H_{\alpha\dot\alpha}\) contains several superspin sectors,
of which only the transverse superspin-\(\tfrac32\) component describes the
physical graviton--gravitino system. The natural remedy is therefore not to
couple the raw prepotentials, but only their irreducible superspin-\(\tfrac32\)
parts.

We accordingly consider
\begin{equation}
S_{\rm mix}^{\rm nl}
=
\mu
\int d^8z\;
H^{\alpha\dot\alpha}
\big(\Pi^T_{3/2}J\big)_{\alpha\dot\alpha},
\label{eq:sec4_nonlocalmix}
\end{equation}
with
\begin{equation}
J_{\alpha\dot\alpha}
=
\bar D_{\dot\alpha}\widehat\Psi_\alpha
-
D_\alpha\bar{\widehat\Psi}_{\dot\alpha},
\label{eq:sec4_current_again}
\end{equation}
and \(\Pi^T_{3/2}\) the projector onto the transverse superspin-\(\tfrac32\)
sector of a real vector superfield. Using the self-adjointness of the
projector with respect to the full superspace measure, one may equivalently
write
\begin{equation}
S_{\rm mix}^{\rm nl}
=
\mu
\int d^8z\;
\big(\Pi^T_{3/2}H\big)^{\alpha\dot\alpha}
J_{\alpha\dot\alpha}.
\label{eq:sec4_nonlocalmix_sym}
\end{equation}

The explicit superspace form of \(\Pi^T_{3/2}\) is collected in
Appendix~\ref{app:projectors}. For the present discussion only three of its
properties matter. It is idempotent,
\begin{equation}
(\Pi^T_{3/2})^2=\Pi^T_{3/2},
\label{eq:sec4_projector_idempotent}
\end{equation}
it projects onto the transverse superspin-\(\tfrac32\) sector,
\begin{equation}
\partial^{\alpha\dot\alpha}\big(\Pi^T_{3/2}V\big)_{\alpha\dot\alpha}=0,
\label{eq:sec4_projector_transverse}
\end{equation}
and it annihilates the pure-gauge and lower-superspin components of a real
vector superfield. This last property is what makes
\eqref{eq:sec4_nonlocalmix} viable: the gauge-variant contamination that
obstructed the local bilinear is projected out before the mixing is formed.

The price is non-locality. The projector contains inverse powers of \(\Box\),
so \eqref{eq:sec4_nonlocalmix} is not a local superspace operator. This is not
an accidental inconvenience but the superspace manifestation of the structural
obstruction explained above. Within the standard local \(\mathcal N=1\)
formalism, one cannot simultaneously isolate the irreducible spin-\(\tfrac32\)
content and retain a local Dirac-gaugino-type bilinear. 
This obstruction does not forbid a local component description of the physical
massive spin-\(\tfrac32\) modes. Its implication is narrower: the Dirac
gravitino mass cannot be generated by a standard local \(\mathcal N=1\)
superspace operator of the same type as the Dirac-gaugino operator. Thus the
local UV completion, when it exists, must either be formulated in the parent
\(\mathcal N=2\) or five-dimensional theory, or in components after the
physical spin-\(\tfrac32\) sector has been isolated.

The physical meaning becomes transparent after reduction to the irreducible
Rarita--Schwinger subspace. The real prepotential \(H_{\alpha\dot\alpha}\)
contains, in Wess--Zumino gauge, the massless graviton, the ordinary gravitino,
and the old-minimal auxiliaries. The spinor prepotential \(\Psi_\alpha\)
contains a vector-spinor \(\chi_\mu\), together with lower-spin fields and a
spin-1 field belonging to the massless superspin-1 multiplet. The rôle of
\(\Pi^T_{3/2}\) is to remove the lower-superspin components and retain only the
irreducible transverse, \(\gamma\)-traceless spin-\(\tfrac32\) content. On the
physical subspace
\begin{equation}
\partial^\mu\psi_\mu=0,
\qquad
\gamma^\mu\psi_\mu=0,
\qquad
\partial^\mu\chi_\mu=0,
\qquad
\gamma^\mu\chi_\mu=0,
\label{eq:sec4_physicalsubspace}
\end{equation}
the projector acts trivially, and the fermionic sector reduces schematically
to
\begin{equation}
\mathcal L_{\rm ferm}^{(2)}
\sim
\bar\psi_\mu\slashed{\partial}\psi^\mu
+
\bar\chi_\mu\slashed{\partial}\chi^\mu
+
\mu\,\bar\psi_\mu\chi^\mu
+\text{h.c.}
\label{eq:sec4_fermreduction}
\end{equation}
Thus the projected superspace coupling reproduces the desired Dirac mixing on
the physical spin-\(\tfrac32\) subspace while keeping the graviton massless.

This conclusion agrees with the Scherk--Schwarz intuition but is not limited
to it. In the half-twist compactification, the need to go beyond a single
local \(\mathcal N=1\) supergravity multiplet is already visible from the KK
description: the light massive gravitino at \(\omega=\tfrac12\) is part of the
compactified higher-dimensional spin-\(\tfrac32\) system, and the KK tower
supplies the local higher-dimensional completion. The superspace result above
is a four-dimensional formulation of the same lesson in
representation-theoretic terms. It says that, even when the companion
spin-\(\tfrac32\) field is treated as a general low-energy field rather than
explicitly as a Scherk--Schwarz gravitino mode, the local \(\mathcal N=1\)
superspace description is insufficient. The correct quadratic mixing is local
only after reduction to the physical spin-\(\tfrac32\) component; in superspace
the required projection is non-local.

A separate comment is needed for the vector sector. The projected coupling
\eqref{eq:sec4_nonlocalmix} is designed to act on the irreducible
superspin-\(\tfrac32\) sector only. It therefore does not by itself determine
the vector mass spectrum. In particular, it does \emph{not} by itself generate
a Proca mass for the spin-1 field sitting in the companion superspin-1
multiplet, nor does it give a mass to the graviphoton. The fate of the vector
depends on the ultraviolet realization.

In the four-dimensional \(\mathcal N=2\) FI realization discussed above, the
graviphoton gauges the residual \(U(1)_R\) but remains massless, since no
hypermultiplet Stückelberg mechanism is present. By contrast, in the
hypermultiplet-gauged realization, the vector combination participating in the
gauging can become massive by eating the corresponding axion. Finally, in the
Scherk--Schwarz realization on \(S^1/\mathbb Z_2\), the graviphoton zero mode
is projected out by orbifold parity, so the low-energy four-dimensional theory
does not contain a massless graviphoton at all; the relevant vector partners
belong instead to the Kaluza--Klein tower. Thus
\eqref{eq:sec4_nonlocalmix} should be viewed as a description of the fermionic
Dirac mixing in the spin-\(\tfrac32\) sector, while the vector sector is
inherited from the specific four- or five-dimensional completion.

The practical conclusion is straightforward. \(\mathcal N=1\) language remains
the correct language for matter, scalar geometry, and the effective
\((K,W,f)\) data. What fails is only the expectation that the Dirac gravitino
should arise from a standard local holomorphic superspace operator exactly
analogous to the Dirac-gaugino case. Within the standard local framework, that
construction is obstructed. A viable superspace description appears only after
projection to the physical superspin-\(\tfrac32\) sector and is therefore
intrinsically non-local.

For the remainder of the paper, the low-energy EFT should therefore be
formulated in a hybrid way: \(\mathcal N=1\) language for visible matter and
scalar couplings, but the spin-\(\tfrac32\) sector written directly in terms of
the physical Dirac gravitino, the companion spin-\(\tfrac32\) field, and the
residual \(U(1)_R\). The companion field need not be assumed to be a gravitino
unless a specific \(\mathcal N=2\) or five-dimensional completion is imposed.
This is the standpoint from which the next section formulates the couplings to
matter, and from which the later sections analyze the radiatively induced
masses.
















\section{Coupling the Dirac gravitino to chiral matter}
\label{sec:coupling_to_matter}

We now specify the couplings between the spin-\(\tfrac32\) sector and
boundary matter. Three logically distinct descriptions must be kept separate.

First, there is the four-dimensional EFT. At this level one writes the
operators allowed by Lorentz invariance, gauge invariance, dimensional
analysis, and the residual \(U(1)_R\). This determines which couplings are
allowed, but not which of them are generated in a given ultraviolet
completion.

Second, there is the minimal low-energy setup used in the loop analysis. The
matter sector is written in ordinary \(\mathcal N=1\) language. The
spin-\(\tfrac32\) sector contains the ordinary gravitino \(\psi_\mu\) and a
companion spin-\(\tfrac32\) field \(\chi_\mu\), with an off-diagonal Dirac
mass. The ordinary gravitino coupling to the matter supercurrent is universal.
It is sufficient for the universal scalar threshold. By contrast, mixed scalar
terms and Dirac fermion masses require an additional matter current coupled to
\(\chi_\mu\). The existence and normalization of that current are dynamical
input.

Third, there is the pure anti-periodic Scherk--Schwarz benchmark on
\(S^1/\mathbb Z_2\),
\begin{equation}
\omega=\frac12,
\qquad
M_0=M_\pi=0 .
\label{eq:sec5_pure_SS_benchmark}
\end{equation}
In this construction the spin-\(\tfrac32\) wavefunctions and boundary kernels
are fixed. On a given fixed plane, the component with nonzero boundary value
has a direct boundary coupling, whereas the opposite-parity component vanishes
there and can enter only through normal-derivative couplings, bulk propagation,
or the matter current associated with the second-supersymmetry structure. This
is a kinematical statement fixed by the orbifold parities and the half-twist
boundary conditions. It does not determine the existence or normalization of
the companion matter current.

Throughout this section \(M_P\) denotes the reduced four-dimensional Planck
mass, \(M_P\equiv M_4\). When the five-dimensional Planck scale appears
explicitly, we write \(M_5\), with
\begin{equation}
M_4^2=\pi R\,M_5^3 .
\label{eq:sec5_M4_M5_relation}
\end{equation}

The purpose of this section is to state the four-dimensional couplings and
\(U(1)_R\) selection rules, to identify the assumptions used in the loop
analysis, and to collect the boundary kernels fixed by the pure
anti-periodic Scherk--Schwarz benchmark.

\subsection{Four-dimensional organization of the fields and couplings}
\label{subsec:4d_organization_fields_couplings_revised}

The boundary matter sector consists of standard \(\mathcal N=1\) chiral and
vector multiplets,
\begin{equation}
Q^i=(\phi^i,\psi^i,F^i),
\qquad
V^a=(A_\mu^a,\lambda^a,D^a),
\end{equation}
with dynamics described by a Kähler potential \(K\), a superpotential \(W\),
and a gauge kinetic function \(f\). The spin-\(\tfrac32\) sector contains two
Weyl spin-\(\tfrac32\) fields, denoted by \(\psi_\mu\) and \(\chi_\mu\). In
the exact \(U(1)_R\)-symmetric limit their mass term is
\begin{equation}
{\cal L}_{m}^{(3/2)}
=
-\,m_{3/2}\big(\psi^\mu\sigma_{\mu\nu}\chi^\nu+\text{h.c.}\big).
\label{eq:sec5_Dirac_mass_revised}
\end{equation}
Diagonal Majorana terms,
\begin{equation}
\psi^\mu\sigma_{\mu\nu}\psi^\nu,
\qquad
\chi^\mu\sigma_{\mu\nu}\chi^\nu ,
\end{equation}
violate the residual \(U(1)_R\) and are absent in the exact
\(U(1)_R\)-symmetric limit.

The coupling of the ordinary gravitino is fixed by local supersymmetry
\cite{Wess:1992cp,Freedman:2012zz}:
\begin{equation}
{\cal L}_{\psi J}
=
-\frac{1}{2M_P}\,\psi_\mu J^\mu_{\rm SUSY}
+\text{h.c.}
\label{eq:sec5_psicurrent_revised}
\end{equation}
For a chiral multiplet, the derivative part of the supercurrent is
\begin{equation}
J^\mu_{\rm chiral}
=
\sqrt2\,K_{i\bar j}\,
(\partial_\nu\bar\phi^{\bar j})\,\sigma^\nu\bar\sigma^\mu\psi^i
+\cdots ,
\label{eq:sec5_chiralcurrent_revised}
\end{equation}
and for a gauge multiplet,
\begin{equation}
J^\mu_{\rm gauge}
=
\sigma^{\nu\rho}\sigma^\mu\bar\lambda^a F^a_{\nu\rho}
+\cdots .
\label{eq:sec5_gaugecurrent_revised}
\end{equation}
This coupling is universal. It is present whenever the matter sector is
coupled to supergravity, and it is the only matter coupling needed for the
universal scalar threshold.

There is no analogous universal low-energy theorem for the coupling of
\(\chi_\mu\) to \(\mathcal N=1\) matter. We parameterize this coupling by
\begin{equation}
{\cal L}_{\chi\text{-matter}}
=
-\,\chi_\mu j^\mu_\chi+\text{h.c.}
\label{eq:sec5_chigeneral_revised}
\end{equation}
and separate derivative and non-derivative structures,
\begin{equation}
j^\mu_\chi
=
j^\mu_{\chi,\partial}
+
j^\mu_{\chi,Y}.
\label{eq:sec5_jchi_split_revised}
\end{equation}
A derivative companion current may contain terms of the form
\begin{equation}
j^\mu_{\chi,\partial}
=
\frac{c_{i\bar j}}{M_\chi}\,
(\partial_\nu\bar\phi^{\bar j})\,\sigma^\nu\bar\sigma^\mu\psi^i
+
\frac{\widetilde C_{\rm gauge}}{M_\chi}\,
\sigma^{\nu\rho}\sigma^\mu\bar\lambda^a F^a_{\nu\rho}
+\cdots ,
\label{eq:sec5_jchi_derivative_revised}
\end{equation}
where \(M_\chi\) denotes the suppression scale of the companion-channel
coupling. A non-derivative companion current may contain
\begin{equation}
j^\mu_{\chi,Y}
=
Y_{ij}\,\sigma^\mu\bar\psi^i\,\phi^j
+
\widetilde Y_{aA}\,\sigma^\mu\bar\lambda^a\,\varphi^A
+
\widehat Y_i\,\sigma^\mu\bar\psi^i\,x
+\cdots .
\label{eq:sec5_jchi_yukawa_revised}
\end{equation}
Here \(\varphi^A\) is a scalar with the gauge quantum numbers needed to make
the operator gauge invariant, and \(x\) denotes a background or spurion field.
Equations \eqref{eq:sec5_jchi_derivative_revised} and
\eqref{eq:sec5_jchi_yukawa_revised} parameterize allowed EFT operators. They
are not generated automatically by the existence of a Dirac spin-\(\tfrac32\)
mass. 

In the explicit one-loop calculations of Sections~\ref{sec:scalar_masses} and
\ref{sec:fermion_masses} we use the derivative, supercurrent-type vertices in
\(j^\mu_{\chi,\partial}\). The non-derivative structures in
\(j^\mu_{\chi,Y}\) are listed to make the EFT classification complete, but
their loop effects are not included in the representative scalar and fermion
mass computations below.

This is analogous in spirit, but not in superspace form, to Dirac-gaugino
model building: an additional fermionic channel is required
\cite{Benakli:2011vb}. The difference is that for spin \(\tfrac32\) the
additional channel cannot be generated by a local \(\mathcal N=1\) spurion
operator of the supersoft type. A local weakly coupled origin normally requires
an enlarged local-supersymmetry structure, hidden super-Higgs dynamics, or a
controlled spin-\(\tfrac32\) EFT below a cutoff.

It is useful to distinguish three possible interpretations of the companion
field \(\chi_\mu\).

\paragraph{Second gravitino from extended or higher-dimensional supergravity.}
If \(\chi_\mu\) is identified with the second gravitino of an underlying
\(\mathcal N=2\) or five-dimensional supergravity, its coupling to matter is
not arbitrary. In a weakly coupled completion the conservative expectation is
that a diagram with two spin-\(\tfrac32\)-matter vertices carries the same
Planck suppression as two gravitational supercurrent vertices,
\begin{equation}
{\cal A}_{\rm grav}\sim \frac{1}{M_P^2},
\label{eq:sec5_conservative_grav_scaling}
\end{equation}
up to wave-function, localization, and normal-derivative factors. This is the
normalization used for the conservative estimates below. What is not fixed by
\eqref{eq:sec5_conservative_grav_scaling} is the identity of the matter
current selected by the second-supersymmetry channel.

\paragraph{Hidden or partially sequestered local supersymmetry.}
More generally, the Scherk--Schwarz interval should be viewed as a calculable
benchmark, not as the only possible origin of the Dirac spin-\(\tfrac32\)
mass. One may instead imagine a compactification or hidden sector in which the
gravitational sector contains two spin-\(\tfrac32\) gauge fields, while the
visible sector preserves only one manifest \(\mathcal N=1\) subalgebra. The
mass mixing of \(\psi_\mu\) and \(\chi_\mu\) may then be generated by aligned
gauging, flux data, or hidden super-Higgs dynamics. From the \(\mathcal N=1\)
EFT viewpoint the companion matter current is not universal, but in a UV
completion it is determined by how the visible fields are embedded into, or
coupled to, the sector carrying the second local supersymmetry.

\paragraph{Massive spin-\(\tfrac32\) EFT below a cutoff.}
A more general EFT possibility is that \(\chi_\mu\) is a massive
Rarita--Schwinger field not coupled as a universal gravitational gauge field,
for example a localized or composite spin-\(\tfrac32\) resonance of a hidden
sector. Then the suppression scale \(M_\chi\) in
\eqref{eq:sec5_jchi_derivative_revised} need not be \(M_P\). In such an EFT
one may write \(M_\chi=\Lambda_\chi\), with \(\Lambda_\chi\) possibly below
\(M_P\), so that the mixed amplitudes can be parametrically larger than in the
conservative supergravity estimate.
This is, however, an additional EFT assumption. It requires a cutoff below
which the massive spin-\(\tfrac32\) description and its Rarita--Schwinger
constraints are under control. In this case the replacement
\(M_P\to\Lambda_\chi\) is not fixed by Scherk--Schwarz geometry or by
\(U(1)_R\) symmetry alone.

We now state the \(U(1)_R\) selection rules used below. We choose the Dirac
basis in which
\begin{equation}
R(\psi_\mu)=+1,
\qquad
R(\chi_\mu)=-1.
\label{eq:sec5_Rgravitini_revised}
\end{equation}
For chiral multiplets,
\begin{equation}
R(Q^i)=r_i,
\qquad
R(\phi^i)=r_i,
\qquad
R(\psi^i)=r_i-1,
\label{eq:sec5_Rchiral_revised}
\end{equation}
and for vector multiplets,
\begin{equation}
R(V^a)=0,
\qquad
R(A_\mu^a)=0,
\qquad
R(\lambda^a)=+1.
\label{eq:sec5_Rvector_revised}
\end{equation}

For the non-derivative chiral coupling
\begin{equation}
\chi_\mu\,\sigma^\mu\bar\psi^i\,\phi^j ,
\end{equation}
\(R\)-neutrality gives
\begin{equation}
-1+(1-r_i)+r_j=0
\qquad\Longleftrightarrow\qquad
r_j=r_i .
\label{eq:sec5_Rrule_chiral_revised}
\end{equation}
For the gauge-sector coupling
\begin{equation}
\chi_\mu\,\sigma^\mu\bar\lambda^a\,\varphi^A ,
\end{equation}
one finds
\begin{equation}
-1-1+R(\varphi^A)=0
\qquad\Longleftrightarrow\qquad
R(\varphi^A)=2 .
\label{eq:sec5_Rrule_gauge_revised}
\end{equation}
For the derivative chiral current in
\eqref{eq:sec5_jchi_derivative_revised}, the operator
\(\chi_\mu(\partial_\nu\bar\phi^{\bar j})\sigma^\nu\bar\sigma^\mu\psi^i\) is
neutral only if
\begin{equation}
-1-r_j+(r_i-1)=0
\qquad\Longleftrightarrow\qquad
r_i-r_j=2 .
\label{eq:sec5_Rrule_derivative_chiral}
\end{equation}
This last condition applies to the companion current written with
\(R(\chi_\mu)=-1\). It is not the selection rule for the ordinary supercurrent
coupled to \(\psi_\mu\). More generally, these conditions test only
\(U(1)_R\)-neutrality of the operator. They do not imply that the
corresponding coupling is generated in a given microscopic realization.

\subsection{Effective assumptions used in the loop analysis}
\label{subsec:minimal_EFT_loops_revised}

The loop analysis in the following sections uses the component EFT
\begin{equation}
{\cal L}_{\rm EFT}
=
{\cal L}_{\rm matter}
+
{\cal L}_{\rm kin}^{(\psi,\chi)}
+
{\cal L}_{m}^{(3/2)}
+
{\cal L}_{\psi J}
+
{\cal L}_{\chi\text{-matter}} .
\label{eq:sec5_totalEFT_revised}
\end{equation}
The role of this EFT is not to provide a full off-shell superspace completion.
Its role is to isolate the component couplings needed for the one-loop scalar
and fermion operators.

The loop analysis uses the following assumptions.

\paragraph{Universal scalar term.}
The universal scalar term uses only the ordinary supercurrent coupling
\eqref{eq:sec5_psicurrent_revised}. It is present for a boundary scalar
coupled to the bulk supergravity multiplet. In the pure anti-periodic
Scherk--Schwarz benchmark, for a canonically normalized boundary scalar and
the standard boundary-coupled five-dimensional supergravity action, the
supersymmetry-breaking threshold is fixed:
\begin{equation}
\delta m_\phi^2\big|_{\omega=1/2}
=
-\frac{31\,\zeta(5)}{128\pi^6}\,
\frac{1}{R^4M_4^2}
=
-\frac{31\,\zeta(5)}{128\pi^7}\,
\frac{1}{R^5M_5^3}.
\label{eq:sec5_univ_scalar_fixed_result}
\end{equation}
Here \(m_\phi^2\) is defined by \(V\supset m_\phi^2|\phi|^2\). Equation
\eqref{eq:sec5_univ_scalar_fixed_result} is the standard Scherk--Schwarz
gravity-mediated scalar threshold
\cite{Gherghetta:2001sa,Rattazzi:2003rj,Antoniadis:2015chx}. In this paper we
use it in a kernel form adapted to the interval computation:
\begin{equation}
\delta m_\phi^2\big|_{\rm univ}
=
\mathcal C_A
\int\frac{d^4p_E}{(2\pi)^4}\,
p_E^2\,\mathcal K_A^{\rm same}(p),
\qquad
\mathcal C_A=\frac{2}{3M_5^3}
=\frac{2\pi R}{3M_4^2}.
\label{eq:sec5_univ_kernel_CA}
\end{equation}
The coefficient \(\mathcal C_A\) is fixed by the complete off-shell
boundary-coupled five-dimensional supergravity action. It includes the contact
terms required by local supersymmetry of the boundary matter action and is not
the coefficient of an isolated cubic supercurrent exchange graph. Extra local soft terms, radion-mediated contributions, or stabilization
effects, if present in a specific ultraviolet completion, must be added
separately. Thus \eqref{eq:scalar_univ_half_result_final} is the universal
Scherk--Schwarz contribution, not a complete scalar-spectrum prediction.

\paragraph{Mixed scalar terms.}
Mixed scalar terms require a nonzero companion current \(j^\mu_\chi\). If
\(j^\mu_\chi=0\), these terms are absent. If \(j^\mu_\chi\neq0\), the resulting
operator need not be a correction to the same diagonal scalar mass. Depending
on the matter topology, it can generate diagonal entries, non-holomorphic
off-diagonal entries, or holomorphic \(B\)-type bilinears. In particular, one
may obtain
\begin{equation}
\phi^\dagger\phi',
\qquad
\phi\phi' .
\end{equation}
The \(U(1)_R\) conditions depend on the type of bilinear. A non-holomorphic
off-diagonal term \(\phi^\dagger\phi'\), with an \(R\)-neutral coefficient, is
allowed only if
\begin{equation}
R(\phi')=R(\phi).
\label{eq:sec5_Rrule_phi_dagger_phi}
\end{equation}
A holomorphic scalar bilinear \(\phi\phi'\), with an \(R\)-neutral coefficient,
is allowed only if
\begin{equation}
R(\phi)+R(\phi')=0 .
\label{eq:sec5_Rrule_phi_phi_soft}
\end{equation}
If instead the same pair appears through a superpotential term
\(W\supset \mu\,\Phi\Phi'\), with \(\mu\) treated as \(R\)-neutral, the
condition is
\begin{equation}
R(\Phi)+R(\Phi')=2 .
\label{eq:sec5_Rrule_superpotential_mass}
\end{equation}
The coefficient of a mixed scalar term therefore depends not only on the
spin-\(\tfrac32\) kernel, but also on the matter channel, possible mass
insertions, localization, and the suppression scale \(M_\chi\).

\paragraph{Dirac fermion mass.}
A Dirac fermion mass for two matter fermions \(\psi_Q\) and \(\psi_\Sigma\)
requires two conditions. First, the fermion bilinear must be \(R\)-neutral:
\begin{equation}
R(\psi_Q)+R(\psi_\Sigma)=0
\qquad\Longleftrightarrow\qquad
r_Q+r_\Sigma=2.
\label{eq:sec5_Rneutral_dirac_fermion}
\end{equation}
Second, the companion current must connect \(\chi_\mu\) to the second matter
channel. If that current is absent, the Dirac fermion mass is zero even when
\eqref{eq:sec5_Rneutral_dirac_fermion} is satisfied.

\paragraph{Normalization of companion-channel amplitudes.}
The tensor structure of the mixed scalar and fermion amplitudes is fixed once
the vertices are specified. The absolute normalization is not fixed by
\(U(1)_R\) alone. In the conservative supergravity interpretation,
\(\chi_\mu\) is the second gravitino of an underlying \(\mathcal N=2\) or
five-dimensional theory, and amplitudes with two spin-\(\tfrac32\)-matter
vertices are suppressed by \(1/M_P^2\), up to wave-function and boundary
kernel factors. A larger normalization is possible only as an additional EFT
assumption, for example through a localized or composite companion
spin-\(\tfrac32\) sector with a lower scale \(\Lambda_\chi\). In that case the
enhancement is not a prediction of the Scherk--Schwarz benchmark.

The content of the loop analysis may therefore be summarized as follows:
\begin{enumerate}
\item the universal scalar term is present for a boundary scalar coupled to the
bulk supergravity multiplet and is fixed in the pure anti-periodic
Scherk--Schwarz benchmark;
\item mixed scalar terms and Dirac fermion masses are absent unless a companion
matter current is present;
\item \(U(1)_R\) determines which operators are allowed, but not whether the
required couplings are generated;
\item the Planck-scale or non-Planckian normalization of companion-channel
amplitudes follows from the ultraviolet interpretation of \(\chi_\mu\), not
from symmetry alone.
\end{enumerate}

\subsection{Anti-periodic Scherk--Schwarz realization on \texorpdfstring{$S^1/\mathbb Z_2$}{S1/Z2}}
\label{subsec:scalar_SS_structure_revised}

We now collect the structures fixed by the pure anti-periodic Scherk--Schwarz
benchmark. The interval is
\begin{equation}
y\in[0,\pi R],
\end{equation}
and we Fourier transform the non-compact coordinates to Euclidean momentum
\(p_E^\mu\), with
\begin{equation}
p\equiv |p_E|=\sqrt{p_E^2}.
\label{eq:sec5_pE_convention}
\end{equation}
We work at
\begin{equation}
\omega=\frac12,
\qquad
M_0=M_\pi=0 .
\label{eq:sec5_SS_geometry_benchmark}
\end{equation}

At the half-twist point it is convenient to work in a basis in which the
profiles associated with the components even and odd at \(y=0\) obey mixed
boundary conditions on the interval:
\begin{equation}
f_n^{(+)}(y)
=
\sqrt{\frac{2}{\pi R}}\,
\cos\!\left(\frac{n+\tfrac12}{R}\,y\right),
\qquad
f_n^{(-)}(y)
=
\sqrt{\frac{2}{\pi R}}\,
\sin\!\left(\frac{n+\tfrac12}{R}\,y\right),
\qquad
m_n=\frac{n+\tfrac12}{R}.
\label{eq:sec5_SS_modefunctions_revised}
\end{equation}
The \(+\) profile is nonzero at \(y=0\) and vanishes at \(y=\pi R\), whereas
the \(-\) profile vanishes at \(y=0\) and is nonzero at \(y=\pi R\). This is
the basis in which the boundary kernels below take their simplest form.

The mixed-representation Green functions are
\begin{equation}
G_+(p;y,y')
=
\frac{\cosh(p\,y_<)\,\sinh\!\big(p(\pi R-y_>)\big)}
{p\,\cosh(\pi pR)},
\qquad
G_-(p;y,y')
=
\frac{\sinh(p\,y_<)\,\cosh\!\big(p(\pi R-y_>)\big)}
{p\,\cosh(\pi pR)},
\label{eq:sec5_Gpm_revised}
\end{equation}
where
\begin{equation}
y_<\equiv \min(y,y'),
\qquad
y_>\equiv \max(y,y') .
\end{equation}

The direct same-boundary kernels are
\begin{equation}
G_+(p;0,0)=\frac{\tanh(\pi pR)}{p},
\qquad
G_-(p;\pi R,\pi R)=\frac{\tanh(\pi pR)}{p}.
\label{eq:sec5_same_boundary_direct_revised}
\end{equation}
The wrong-parity direct kernels vanish:
\begin{equation}
G_-(p;0,0)=0,
\qquad
G_+(p;\pi R,\pi R)=0 .
\label{eq:sec5_wrong_boundary_direct_revised}
\end{equation}
Thus a direct boundary coupling at \(y=0\) sees only the \(+\) tower, while a
direct boundary coupling at \(y=\pi R\) sees only the \(-\) tower.

The supersymmetry-breaking same-boundary kernel entering the universal scalar
threshold is the difference between the anti-periodic and unbroken kernels for
the same direct boundary supercurrent coupling at \(y=0\):
\begin{equation}
\mathcal K_A^{\rm same}(p)
\equiv
G_+^{(\omega=1/2)}(p;0,0)
-
G_+^{(\omega=0)}(p;0,0).
\label{eq:sec5_KA_same_definition}
\end{equation}
With the above conventions this gives
\begin{equation}
\mathcal K_A^{\rm same}(p)
=
\frac{\tanh(\pi pR)}{p}
-
\frac{\coth(\pi pR)}{p}
=
-\frac{2}{p\,\sinh(2\pi pR)} .
\label{eq:sec5_KA_same_explicit}
\end{equation}
This finite kernel is the object that appears in the universal scalar
threshold. Its normalization is fixed by matching to the full
boundary-coupled five-dimensional supergravity result, as explained in
Section~\ref{sec:scalar_masses} and Appendix~\ref{app:scalar_cubic_trace}.

The derivative kernels must be treated separately. At \(y=0\), with the second
endpoint in the open interval or at the opposite boundary, one finds
\begin{equation}
\partial_y G_-(p;y,y')\big|_{y=0}
=
\frac{\cosh\!\big(p(\pi R-y')\big)}{\cosh(\pi pR)},
\qquad
0<y'\leq \pi R .
\label{eq:sec5_derivative_minus_revised}
\end{equation}
Thus
\begin{equation}
\lim_{y'\to0^+}
\partial_y G_-(p;y,y')\big|_{y=0}
=1,
\qquad
\partial_y G_-(p;y,\pi R)\big|_{y=0}
=
\frac{1}{\cosh(\pi pR)} .
\label{eq:sec5_derivative_minus_limits_revised}
\end{equation}
However, if the second endpoint is fixed exactly at the same odd boundary, the
single-derivative kernel is
\begin{equation}
\partial_y G_-(p;y,0)\big|_{y=0}=0 .
\label{eq:sec5_single_derivative_same_boundary_zero}
\end{equation}
There is no contradiction between
\eqref{eq:sec5_derivative_minus_limits_revised} and
\eqref{eq:sec5_single_derivative_same_boundary_zero}. They are different
kernels. The first is a derivative--bulk or derivative--opposite-boundary
kernel followed by a limit. The second is a derivative--direct same-boundary
kernel, and it vanishes because \(f_n^{(-)}(0)=0\) in the spectral
representation.

If both same-boundary vertices couple through normal derivatives of the odd
profile, the relevant object is the double-derivative kernel
\begin{equation}
\mathcal D_{--}^{(0,0)}(p)
\equiv
\sum_{n=0}^{\infty}
\frac{\partial_y f_n^{(-)}(0)\,\partial_{y'}f_n^{(-)}(0)}
{p^2+m_n^2}.
\label{eq:sec5_double_derivative_kernel_definition}
\end{equation}
This kernel contains a local divergent piece. After subtraction of the
\(p\)-independent local term, its finite momentum-dependent part is
\begin{equation}
\mathcal D_{--}^{(0,0)}(p)\Big|_{\rm finite}
=
-\,p\,\tanh(\pi pR),
\label{eq:sec5_double_derivative_kernel_finite}
\end{equation}
up to the sign convention for inward versus outward normal derivatives.

Similarly, at \(y=\pi R\),
\begin{equation}
\partial_y G_+(p;y,y')\big|_{y=\pi R}
=
-\,\frac{\cosh(py')}{\cosh(\pi pR)},
\qquad
0\leq y'<\pi R ,
\label{eq:sec5_derivative_plus_revised}
\end{equation}
and in particular
\begin{equation}
\partial_y G_+(p;y,0)\big|_{y=\pi R}
=
-\frac{1}{\cosh(\pi pR)} .
\label{eq:sec5_derivative_plus_opposite_revised}
\end{equation}

It is useful to name the opposite-boundary derivative--direct kernel, since it
will appear repeatedly in the mixed amplitudes:
\begin{equation}
\mathcal K_{\partial{\rm d}}^{\rm opp}(p)
\equiv
\partial_y G_-(p;y,\pi R)\big|_{y=0}
=
\frac{1}{\cosh(\pi pR)} .
\label{eq:sec5_Kpd_opp}
\end{equation}
More generally, we write mixed scalar amplitudes using a topology-dependent
kernel \(\mathcal K_B^{\mathfrak t}(p)\). For the opposite-boundary
derivative--direct topology, \(\mathfrak t={\rm opp}\),
\begin{equation}
\mathcal K_B^{\mathfrak t={\rm opp}}(p)
=
\frac{1}{\cosh(\pi pR)} .
\label{eq:sec5_KB_opp}
\end{equation}

In the fermion-mass calculation we also encounter a chirality-changing
one-insertion propagator, denoted schematically by
\begin{equation}
G_{+-}^{(1)}(p;y,y') .
\label{eq:sec5_Gpm_one_insertion_definition}
\end{equation}
For the opposite-boundary channel, after stripping off the vertex-dependent
prefactor and possible factors of \(m_{3/2}\), \(R^{-1}\), or normal
derivatives, its finite non-local part is proportional to
\begin{equation}
\frac{1}{\cosh(\pi pR)} .
\label{eq:sec5_Gpm_one_insertion_opp}
\end{equation}

These equations give the non-local kinematical content of the pure
anti-periodic Scherk--Schwarz benchmark relevant for the loop analysis.

\paragraph{Fixed by the geometry.}
The mode functions, the direct boundary kernels, the derivative kernels, the
same-boundary supersymmetry-breaking kernel \(\mathcal K_A^{\rm same}\), the
opposite-boundary kernel \(\mathcal K_{\partial{\rm d}}^{\rm opp}\), and the
non-local part of \(\mathcal K_B^{\mathfrak t={\rm opp}}\) are fixed. In
particular, opposite-boundary communication is exponentially suppressed at
large Euclidean momentum:
\begin{equation}
\frac{1}{\cosh(\pi pR)}
\sim 2e^{-\pi pR},
\qquad
\pi pR\gg1 .
\end{equation}

\paragraph{Not fixed by the geometry alone.}
The geometry does not determine which matter current couples to the
opposite-parity spin-\(\tfrac32\) channel. It also does not prove that an
arbitrary companion field has the same normalization as a supergravity current.
Those statements require a microscopic derivation of the boundary matter
couplings.

\paragraph{Implication for the loop analysis.}
The universal scalar contribution is fixed because it uses the direct
supergravity coupling to the component with nonzero boundary value. Mixed
scalar terms and Dirac fermion masses require an additional companion matter
current. The pure anti-periodic Scherk--Schwarz construction therefore provides
a precise kinematical selection rule: it fixes the allowed kernels, sets the
wrong-parity direct boundary kernels to zero, and leaves the companion matter
current as an explicit dynamical assumption.
















\section{One-loop scalar masses}
\label{sec:scalar_masses}

We now study the scalar sector. Let
\begin{equation}
Q=(\phi,\psi,F)
\label{eq:scalar_visible_chiral}
\end{equation}
be a boundary chiral multiplet localized at \(y=0\) on the interval
\(S^1/\mathbb Z_2\). We define the scalar mass correction by
\begin{equation}
\delta\mathcal L_{\rm scalar}
=
-\,\delta m_\phi^2\,\phi^\dagger\phi ,
\label{eq:scalar_mass_definition}
\end{equation}
so that the scalar potential contains
\(V\supset \delta m_\phi^2|\phi|^2\). Since \(\phi^\dagger\phi\) is neutral
under the residual \(U(1)_R\), such a scalar mass is allowed even when
Majorana masses for matter fermions are forbidden.

There are two classes of one-loop scalar effects:
\begin{equation}
\delta m_\phi^2
=
\delta m_\phi^2\big|_{\rm univ}
+
\delta m_\phi^2\big|_{\rm mix}.
\label{eq:scalar_mass_split_main}
\end{equation}
The universal term is generated by the ordinary gravitino coupling to the
boundary supercurrent. It is present for a boundary scalar coupled to the bulk
supergravity multiplet. The mixed term requires an additional matter current
coupled to the companion spin-\(\tfrac32\) channel. If this companion current
is absent, the mixed contribution vanishes.

We work at one loop and at zero external scalar momentum. The matter fields are
localized on a boundary, and the Scherk--Schwarz background is treated as fixed
in the sense explained in Section~\ref{sec:SS_origin}. Thus
\begin{equation}
\omega=\frac12,
\qquad
M_0=M_\pi=0 ,
\label{eq:scalar_half_twist_benchmark}
\end{equation}
with fixed radius \(R\). Same-boundary thresholds contain local pieces and are
sensitive to local counterterms. Opposite-boundary thresholds are finite
non-local effects.

We use Euclidean momentum space. The Euclidean four-momentum is \(p_E^\mu\),
and
\begin{equation}
p\equiv |p_E|=\sqrt{p_E^2}.
\label{eq:scalar_Euclidean_momentum}
\end{equation}
The extra coordinate is
\begin{equation}
y\in[0,\pi R],
\end{equation}
and the half-twist gravitino mass scale is
\begin{equation}
m_{3/2}=\frac{1}{2R}.
\label{eq:scalar_KK_masses_half}
\end{equation}

The universal scalar mass is the full one-loop Scherk--Schwarz soft threshold
for a boundary scalar. It is not obtained from an isolated local cubic--cubic
diagram. The locally supersymmetric calculation contains both cubic--cubic and
quartic seagull contributions, and the physical soft mass is the difference
between the broken and unbroken backgrounds
\cite{Antoniadis:1997ic,Gherghetta:2001sa,Rattazzi:2003rj,Antoniadis:2015chx}.

The mixed sector is model-dependent in a different sense. The orbifold kernels
are fixed by the Scherk--Schwarz geometry, but the companion matter current and
its normalization are additional dynamical input, as explained in
Section~\ref{sec:coupling_to_matter}. Depending on the matter topology, the
mixed sector can generate diagonal scalar contributions, non-holomorphic
off-diagonal terms, or holomorphic \(B\)-type scalar bilinears.

In the minimal anti-periodic Scherk--Schwarz threshold, the universal scalar
contribution is negative. Taken alone, it gives a negative diagonal
mass-squared. This sign is the sign of one calculable contribution in the
limit where local boundary terms, radion mediation, stabilization effects, and
possible additional sector-dependent thresholds are set aside. It should not be
read as a prediction of a tachyonic physical scalar in an arbitrary completion.
A complete model must make the full real scalar mass matrix positive. We
therefore treat scalar stability as a constraint on completions, not as an
assumption built into the threshold calculation. Other positive contributions
may be present, and mixed diagonal contributions may partly compensate the
universal term. Conversely, off-diagonal mixed entries can destabilize the
scalar mass matrix if they are too large.

\subsection{Universal scalar contribution}
\label{subsec:scalar_universal}

For a canonically normalized boundary scalar coupled to the standard
boundary-coupled five-dimensional supergravity action, the anti-periodic
Scherk--Schwarz threshold is
\begin{equation}
\delta m_\phi^2\big|_{\rm univ,\;\omega=1/2}
=
-\frac{31\,\zeta(5)}{128\pi^6}\,
\frac{1}{R^4M_4^2}
=
-\frac{31\,\zeta(5)}{128\pi^7}\,
\frac{1}{R^5M_5^3},
\label{eq:scalar_univ_half_result_final}
\end{equation}
with
\begin{equation}
M_4^2=\pi R\,M_5^3.
\label{eq:scalar_univ_M4M5_relation}
\end{equation}
Equation~\eqref{eq:scalar_univ_half_result_final} is the universal scalar
threshold used in the phenomenological estimates below
\cite{Antoniadis:1997ic,Gherghetta:2001sa,Rattazzi:2003rj,Antoniadis:2015chx}. Extra local soft
terms, if present in a specific ultraviolet completion, must be added
separately.

It is useful to display how
\eqref{eq:scalar_univ_half_result_final} is encoded in the mixed-representation
kernel. At the half-twist point we use the basis introduced in
Section~\ref{sec:coupling_to_matter}, in which the relevant profiles obey
mixed boundary conditions on the interval:
\begin{equation}
f_n^{(+)}(y)
=
\sqrt{\frac{2}{\pi R}}\,
\cos\!\left(\frac{n+\tfrac12}{R}\,y\right),
\qquad
f_n^{(-)}(y)
=
\sqrt{\frac{2}{\pi R}}\,
\sin\!\left(\frac{n+\tfrac12}{R}\,y\right),
\qquad
n\ge0 .
\label{eq:scalar_univ_mode_functions_half}
\end{equation}
The corresponding Green functions are
\begin{equation}
G_+(p;y,y')
=
\frac{\cosh(py_<)\,\sinh\!\big(p(\pi R-y_>)\big)}
     {p\,\cosh(\pi pR)},
\qquad
G_-(p;y,y')
=
\frac{\sinh(py_<)\,\cosh\!\big(p(\pi R-y_>)\big)}
     {p\,\cosh(\pi pR)},
\label{eq:scalar_univ_Gpm_half}
\end{equation}
where
\begin{equation}
y_<\equiv\min(y,y'),
\qquad
y_>\equiv\max(y,y').
\label{eq:scalar_univ_yless_ygreater}
\end{equation}
For the unbroken theory, \(\omega=0\), the even propagator entering the same
direct boundary supercurrent coupling at \(y=0\) is
\begin{equation}
G_+^{(0)}(p;y,y')
=
\frac{\cosh(py_<)\,\cosh\!\big(p(\pi R-y_>)\big)}
     {p\,\sinh(\pi pR)} .
\label{eq:scalar_univ_Gplus_unbroken}
\end{equation}

The supersymmetry-breaking same-boundary kernel is
\begin{equation}
\mathcal K_A^{\rm same}(p)
\equiv
G_+^{(\omega=1/2)}(p;0,0)-G_+^{(0)}(p;0,0).
\label{eq:scalar_univ_kernel_same_definition}
\end{equation}
Since
\begin{equation}
G_+^{(\omega=1/2)}(p;0,0)=\frac{\tanh(\pi pR)}{p},
\qquad
G_+^{(0)}(p;0,0)=\frac{\coth(\pi pR)}{p},
\label{eq:scalar_univ_same_boundary_values}
\end{equation}
one obtains
\begin{equation}
\mathcal K_A^{\rm same}(p)
=
-\frac{2}{p\,\sinh(2\pi pR)}.
\label{eq:scalar_univ_kernel_same}
\end{equation}

In this kernel convention the universal scalar mass is
\begin{equation}
\delta m_\phi^2\big|_{\rm univ}
=
\mathcal C_A
\int\frac{d^4p_E}{(2\pi)^4}\,
p_E^2\,\mathcal K_A^{\rm same}(p),
\label{eq:scalar_univ_loop_integral_start}
\end{equation}
where the full supergravity normalization is
\begin{equation}
\mathcal C_A
=
\frac{2\pi R}{3M_4^2}
=
\frac{2}{3M_5^3}.
\label{eq:scalar_univ_CA_matching}
\end{equation}
The coefficient \(\mathcal C_A\) is not a free EFT parameter in the minimal
anti-periodic benchmark. It is the matching coefficient that reproduces the
full Scherk--Schwarz scalar threshold, including the cubic--cubic graph, the
quartic seagull graph required by local supersymmetry, and the subtraction of
the unbroken \(\omega=0\) background. It is not the local cubic--cubic tensor
coefficient computed in Appendix~\ref{app:scalar_cubic_trace}.

Using \eqref{eq:scalar_univ_kernel_same}, one finds
\begin{equation}
\delta m_\phi^2\big|_{\rm univ}
=
-2\mathcal C_A
\int\frac{d^4p_E}{(2\pi)^4}\,
\frac{p_E^2}{p\,\sinh(2\pi pR)} .
\label{eq:scalar_univ_loop_integral_half}
\end{equation}
With
\begin{equation}
\int d^4p_E=2\pi^2\int_0^\infty dp\,p^3,
\label{eq:scalar_univ_spherical_measure}
\end{equation}
this becomes
\begin{equation}
\delta m_\phi^2\big|_{\rm univ}
=
-\frac{\mathcal C_A}{4\pi^2}
\int_0^\infty dp\,
\frac{p^4}{\sinh(2\pi pR)} .
\label{eq:scalar_univ_loop_radial}
\end{equation}
Setting
\begin{equation}
x\equiv 2\pi R p,
\qquad
dp=\frac{dx}{2\pi R},
\label{eq:scalar_univ_x_variable}
\end{equation}
gives
\begin{equation}
\delta m_\phi^2\big|_{\rm univ}
=
-\frac{\mathcal C_A}{128\pi^7R^5}
\int_0^\infty dx\,\frac{x^4}{\sinh x}.
\label{eq:scalar_univ_loop_x_integral}
\end{equation}
Using
\begin{equation}
\int_0^\infty dx\,\frac{x^4}{\sinh x}
=
2\,\Gamma(5)\,\Big(1-2^{-5}\Big)\zeta(5)
=
\frac{93}{2}\,\zeta(5),
\label{eq:scalar_univ_x4_over_sinh}
\end{equation}
one obtains
\begin{equation}
\delta m_\phi^2\big|_{\rm univ}
=
-\frac{93\,\mathcal C_A}{256\pi^7}\,
\frac{\zeta(5)}{R^5}.
\label{eq:scalar_univ_half_result_CA}
\end{equation}
Substituting \eqref{eq:scalar_univ_CA_matching} reproduces
\eqref{eq:scalar_univ_half_result_final}.

For reference, the generic-\(\omega\) KK sum may be written as
\begin{equation}
\delta m_\phi^2\big|_{\rm univ}
=
\frac{1}{8\pi^7M_5^3R^5}
\left[
\operatorname{Re}\operatorname{Li}_5\!\left(e^{2\pi i\omega}\right)-\zeta(5)
\right].
\label{eq:scalar_univ_polylog_generic}
\end{equation}
This is the generic-\(\omega\) KK sum, not an analytic continuation of the
closed \(\omega=\tfrac12\) kernel \eqref{eq:scalar_univ_kernel_same}. For small
\(\omega\),
\begin{equation}
\operatorname{Re}\operatorname{Li}_5\!\left(e^{2\pi i\omega}\right)-\zeta(5)
=
-2\pi^2\zeta(3)\,\omega^2+\mathcal O(\omega^4),
\label{eq:scalar_univ_polylog_smallomega}
\end{equation}
and therefore
\begin{equation}
\delta m_\phi^2\big|_{\rm univ}
=
-\frac{\zeta(3)}{4\pi^5M_5^3R^5}\,\omega^2
+\mathcal O(\omega^4).
\label{eq:scalar_univ_smallomega_result}
\end{equation}

Finally, at \(\omega=\tfrac12\),
\begin{equation}
G_+(p;0,\pi R)=0.
\label{eq:scalar_univ_Gplus_opposite_zero}
\end{equation}
Thus there is no opposite-boundary contribution of the same direct
even-channel universal type.

\subsection{Mixed scalar contribution}
\label{subsec:scalar_mixed}

The mixed scalar contribution requires a matter current coupled to the
companion spin-\(\tfrac32\) channel. The Scherk--Schwarz geometry fixes the
boundary kernels, but it does not fix the existence or normalization of this
current. In this section we restrict the explicit mixed-scalar calculations to the
derivative companion currents of Section~\ref{sec:coupling_to_matter}. The
non-derivative currents are allowed EFT structures, but they lead to different
tensor numerators and are not included in the representative loop topologies
computed here.

Consider two chiral multiplets,
\begin{equation}
Q=(q,\psi_Q,F_Q),
\qquad
\widetilde Q=(\tilde q,\psi_{\widetilde Q},F_{\widetilde Q}) .
\label{eq:scalar_mixed_Q_Qtilde_definition}
\end{equation}
We distinguish two kinds of derivative companion-channel vertices. The first
kind is diagonal in the matter multiplet:
\begin{align}
\mathcal L_{\rm cub}^{(Q)}
&=
\frac{\beta_Q}{M_P}\,
(\partial_\nu q^\dagger)\,
\bar\Psi_\mu\gamma^\nu\gamma^\mu P_L\psi_Q
+\text{h.c.},
\label{eq:scalar_mixed_cubic_Q_channel}
\\[1mm]
\mathcal L_{\rm cub}^{(\widetilde Q)}
&=
\frac{\beta_{\widetilde Q}}{M_P}\,
(\partial_\nu \tilde q^\dagger)\,
\bar\Psi_\mu\gamma^\nu\gamma^\mu P_L\psi_{\widetilde Q}
+\text{h.c.}.
\label{eq:scalar_mixed_cubic_Qtilde_channel}
\end{align}
A loop using one \(Q\)-vertex and one \(\widetilde Q\)-vertex naturally
connects \(q\) to \(\tilde q\), not \(q\) to itself. Thus these vertices
contribute to mixed scalar bilinears unless additional matter conversion
vertices are present.

A diagonal mixed correction to \(q^\dagger q\) requires a second vertex which
couples the same scalar \(q\) to the other fermion channel. We therefore also
allow crossed vertices,
\begin{align}
\mathcal L_{\rm cub}^{({\rm mix},Q)}
&=
\frac{\beta_{\rm mix}^{Q}}{M_P}\,
(\partial_\nu q^\dagger)\,
\bar\Psi_\mu\gamma^\nu\gamma^\mu P_L\psi_{\widetilde Q}
+\text{h.c.},
\label{eq:scalar_mixed_cubic_cross_Q}
\\[1mm]
\mathcal L_{\rm cub}^{({\rm mix},\widetilde Q)}
&=
\frac{\beta_{\rm mix}^{\widetilde Q}}{M_P}\,
(\partial_\nu \tilde q^\dagger)\,
\bar\Psi_\mu\gamma^\nu\gamma^\mu P_L\psi_Q
+\text{h.c.}.
\label{eq:scalar_mixed_cubic_cross_Qtilde}
\end{align}
The coefficients
\(\beta_Q\), \(\beta_{\widetilde Q}\), \(\beta_{\rm mix}^{Q}\), and
\(\beta_{\rm mix}^{\widetilde Q}\) are dimensionless EFT parameters. The
crossed vertices are additional data. They are not implied by the diagonal
vertices in
\eqref{eq:scalar_mixed_cubic_Q_channel} and
\eqref{eq:scalar_mixed_cubic_Qtilde_channel}.

Here \(\Psi_\mu\) denotes the spin-\(\tfrac32\) line entering the mixed
diagram. The distinction between the ordinary and companion channels is encoded
in the kernel and in the matter current, not in the notation of the vertex
itself. If \(\chi_\mu\) is the second gravitino of an underlying
\(\mathcal N=2\) or five-dimensional supergravity, the conservative
normalization has two gravitational vertices and therefore a \(1/M_P^2\)
suppression. A larger normalization is an additional EFT assumption.

At \(y=0\), the odd profile vanishes,
\begin{equation}
f_n^{(-)}(0)=0,
\label{eq:scalar_mixed_odd_zero}
\end{equation}
whereas its normal derivative does not,
\begin{equation}
\partial_y f_n^{(-)}(0)\neq0.
\label{eq:scalar_mixed_odd_derivative_nonzero}
\end{equation}
Therefore a direct coupling to the odd tower at \(y=0\) is zero, while a
normal-derivative coupling can be nonzero.

The relevant kernel depends on the microscopic topology. For a
derivative--direct opposite-boundary topology one has
\begin{equation}
\mathcal K_{\partial{\rm d}}^{\rm opp}(p)
\equiv
\partial_y G_-(p;y,\pi R)\big|_{y=0}
=
\frac{1}{\cosh(\pi pR)}.
\label{eq:scalar_mixed_kernel_derivative_opposite}
\end{equation}
This kernel is finite and exponentially suppressed for \(\pi pR\gg1\).

For a derivative--direct same-boundary topology,
\begin{equation}
\partial_yG_-(p;y,0)\big|_{y=0}=0.
\label{eq:scalar_mixed_single_derivative_same_zero}
\end{equation}
Thus this topology gives no same-boundary mixed threshold.

If both same-boundary vertices couple through normal derivatives of the odd
field, the relevant kernel is
\begin{equation}
\mathcal D_{--}^{(0,0)}(p)
\equiv
\sum_{n=0}^{\infty}
\frac{\partial_y f_n^{(-)}(0)\,\partial_{y'}f_n^{(-)}(0)}
{p^2+m_n^2}.
\label{eq:scalar_mixed_double_derivative_kernel}
\end{equation}
This object contains a local divergent piece. After subtracting the
\(p\)-independent local term, its finite momentum-dependent part is
\begin{equation}
\mathcal D_{--}^{(0,0)}(p)\Big|_{\rm finite}
=
-\,p\,\tanh(\pi pR),
\label{eq:scalar_mixed_double_derivative_finite}
\end{equation}
up to the sign convention for inward versus outward normal derivatives.
Same-boundary derivative--derivative topologies are therefore local-threshold
sensitive.

For the finite opposite-boundary amplitudes below we denote the relevant
dimensionless non-local kernel by
\begin{equation}
\mathcal K_B^{\mathfrak t}(p),
\label{eq:scalar_mixed_general_kernel}
\end{equation}
where \(\mathfrak t\) labels the microscopic topology. For the
derivative--direct opposite-boundary topology,
\begin{equation}
\mathcal K_B^{\mathfrak t={\rm opp}}(p)
=
\frac{1}{\cosh(\pi pR)}.
\label{eq:scalar_mixed_general_kernel_opp}
\end{equation}
Same-boundary derivative--derivative topologies are different. Their kernel is
\(\mathcal D_{--}^{(0,0)}(p)\), which has mass dimension one and contains a
local divergent piece. Such terms must be written with the corresponding local
normal-derivative operator normalization and boundary counterterms. In the
estimates below we display them only parametrically.

\subsubsection{Diagonal mixed terms with a perturbative fermion mixing insertion}
\label{subsubsec:scalar_diag_mixed_trace}

A diagonal mixed correction to \(q^\dagger q\) requires two ingredients. First,
one needs a crossed scalar--fermion companion vertex of the form
\eqref{eq:scalar_mixed_cubic_cross_Q}. Second, one needs a conversion between
\(\psi_Q\) and \(\psi_{\widetilde Q}\) along the internal matter line. We
parameterize this conversion by
\begin{equation}
\mathcal L_{\rm mix}
=
-\,\mu_{Q\widetilde Q}\,\bar\psi_Q\psi_{\widetilde Q}
-\,\mu_{\widetilde Q Q}\,\bar\psi_{\widetilde Q}\psi_Q .
\label{eq:scalar_mixed_fermion_mixing_operator}
\end{equation}
For the representative calculation we take
\begin{equation}
\mathcal M_{Q\widetilde Q}=m_{\rm mix}\,\mathbf 1,
\qquad
m_{\rm mix}\in\mathbb R,
\label{eq:scalar_diagmix_vectorlike_mixing}
\end{equation}
and treat \(m_{\rm mix}\) as a perturbative insertion:
\begin{equation}
\langle\psi_Q\bar\psi_{\widetilde Q}\rangle
=
S_Q\,\mathcal M_{Q\widetilde Q}\,S_{\widetilde Q}
+\mathcal O(\mathcal M^3).
\label{eq:scalar_diagmix_mixed_fermion_line}
\end{equation}

This type of insertion is compatible with the residual \(U(1)_R\) only if the
fermion bilinear \(\bar\psi_Q\psi_{\widetilde Q}\) is \(R\)-neutral, namely
\begin{equation}
R(\psi_{\widetilde Q})=R(\psi_Q)
\qquad\Longleftrightarrow\qquad
r_{\widetilde Q}=r_Q .
\label{eq:scalar_diagmix_Rrule_mixing_insertion}
\end{equation}
If the conversion instead comes from a chiral superpotential mass
\(W\supset \mu Q\widetilde Q\), with \(\mu\) treated as \(R\)-neutral, the
condition is
\begin{equation}
r_Q+r_{\widetilde Q}=2.
\label{eq:scalar_diagmix_Rrule_superpotential_conversion}
\end{equation}
These are different realizations of the conversion between the two matter
channels: the first corresponds to a vectorlike fermion bilinear, whereas the
second corresponds to a chiral superpotential mass term.

We use
\begin{equation}
S_{3/2}^{\mu\nu}(p_E)
=
\frac{-\,i\slashed p_E+m_{3/2}}{p_E^2+m_{3/2}^2}\,
\Pi^{\mu\nu}_{3/2},
\qquad
\Pi^{\mu\nu}_{3/2}
=
\delta^{\mu\nu}-\frac13\gamma^\mu\gamma^\nu,
\label{eq:scalar_diagmix_RS_prop}
\end{equation}
and
\begin{equation}
S_Q(p_E)
=
\frac{-\,i\slashed p_E+m_Q}{p_E^2+m_Q^2},
\qquad
S_{\widetilde Q}(p_E)
=
\frac{-\,i\slashed p_E+m_{\widetilde Q}}{p_E^2+m_{\widetilde Q}^2}.
\label{eq:scalar_diagmix_matter_props}
\end{equation}
The vertices entering the diagonal \(q^\dagger q\) correction are
\begin{equation}
\Gamma_\mu^{(Q)}(p_E)
=
p_{E\rho}\gamma^\rho\gamma_\mu P_L,
\qquad
\widetilde\Gamma_\nu^{({\rm mix},Q)}(p_E)
=
p_{E\sigma}\gamma_\nu\gamma^\sigma P_R .
\label{eq:scalar_diagmix_vertices}
\end{equation}
The second vertex is the hermitian-conjugate vertex associated with
\eqref{eq:scalar_mixed_cubic_cross_Q}; this is why the coefficient below is
\(\beta_Q\beta_{\rm mix}^{Q}\), not
\(\beta_Q\beta_{\widetilde Q}\).

For a finite opposite-boundary non-local topology \(\mathfrak t\), the
diagonal mixed contribution may be written as
\begin{equation}
\delta m_q^2\big|_{\rm mix}^{\mathfrak t}
=
\frac{\beta_Q\beta_{\rm mix}^{Q}}{M_P^2}
\int\frac{d^4p_E}{(2\pi)^4}\,
\mathcal K_B^{\mathfrak t}(p)\,
\frac{\mathcal P_B(p_E^2;m_{3/2},m_Q,m_{\widetilde Q},m_{\rm mix})}
{(p_E^2+m_{3/2}^2)(p_E^2+m_Q^2)(p_E^2+m_{\widetilde Q}^2)} .
\label{eq:scalar_diagmix_master}
\end{equation}
For same-boundary derivative--derivative topologies the corresponding local
normal-derivative normalization and counterterms must be kept explicitly; the
parametric estimate is given below.

The numerator is
\begin{align}
\mathcal P_B
&=
m_{\rm mix}\,
p_{E\rho}p_{E\sigma}\,
{\rm Tr}\!\Big[
\gamma^\rho\gamma_\mu P_L
(-\,i\slashed p_E+m_{3/2})
\Pi^{\mu\nu}_{3/2}
\gamma_\nu\gamma^\sigma P_R
(-\,i\slashed p_E+m_Q)
(-\,i\slashed p_E+m_{\widetilde Q})
\Big].
\label{eq:scalar_diagmix_PB_trace_definition}
\end{align}

Using
\begin{equation}
P_L(-\,i\slashed p_E+m_{3/2})
=
-\,i\slashed p_E P_R + m_{3/2}P_L,
\label{eq:scalar_diagmix_PL_reduction}
\end{equation}
the term proportional to \(m_{3/2}P_L\) vanishes because it produces an even
gamma structure between \(P_L\) and \(P_R\). The surviving term is
\begin{align}
\mathcal P_B
&=
-\,i\,m_{\rm mix}\,
p_{E\rho}p_{E\sigma}\,
{\rm Tr}\!\Big[
\gamma^\rho\gamma_\mu\slashed p_E
\Pi^{\mu\nu}_{3/2}
\gamma_\nu\gamma^\sigma P_R
(-\,i\slashed p_E+m_Q)
(-\,i\slashed p_E+m_{\widetilde Q})
\Big].
\label{eq:scalar_diagmix_PB_after_first_step}
\end{align}

The matter part is
\begin{align}
P_R(-\,i\slashed p_E+m_Q)(-\,i\slashed p_E+m_{\widetilde Q})
&=
-\,p_E^2 P_R
- i(m_Q+m_{\widetilde Q})\slashed p_E P_L
+ m_Qm_{\widetilde Q}P_R .
\label{eq:scalar_diagmix_PR_matter_reduction}
\end{align}
The terms proportional to \(P_R\) vanish in the trace. The only surviving term
is proportional to \((m_Q+m_{\widetilde Q})\slashed p_E P_L\). Thus
\begin{align}
\mathcal P_B
&=
-\,m_{\rm mix}(m_Q+m_{\widetilde Q})\,
\frac12\,
p_{E\rho}p_{E\sigma}\,
{\rm Tr}\!\Big[
\gamma^\rho\gamma_\mu\slashed p_E
\Pi^{\mu\nu}_{3/2}
\gamma_\nu\gamma^\sigma\slashed p_E
\Big].
\label{eq:scalar_diagmix_PB_even_trace}
\end{align}

Define
\begin{equation}
T_B
\equiv
p_{E\rho}p_{E\sigma}\,
{\rm Tr}\!\Big[
\gamma^\rho\gamma_\mu\slashed p_E
\Pi^{\mu\nu}_{3/2}
\gamma_\nu\gamma^\sigma\slashed p_E
\Big].
\label{eq:scalar_diagmix_TB_definition}
\end{equation}
Using
\begin{equation}
\Pi^{\mu\nu}_{3/2}
=
\delta^{\mu\nu}-\frac13\gamma^\mu\gamma^\nu,
\label{eq:scalar_diagmix_Pi_split}
\end{equation}
one finds
\begin{equation}
T_B
=
\frac23\,
p_{E\rho}p_{E\sigma}\,
{\rm Tr}\!\Big[
\gamma^\rho\slashed p_E\gamma^\sigma\slashed p_E
\Big]
=
\frac{8}{3}\,p_E^4.
\label{eq:scalar_diagmix_TB_final}
\end{equation}
Therefore
\begin{equation}
\mathcal P_B(p_E^2;m_{3/2},m_Q,m_{\widetilde Q},m_{\rm mix})
=
-\frac43\,
m_{\rm mix}\,(m_Q+m_{\widetilde Q})\,p_E^4,
\label{eq:scalar_diagmix_PB_closed_result}
\end{equation}
up to the overall Euclidean sign convention for the vertices.

For a local same-boundary topology, the magnitude scales as
\begin{equation}
\bigl|\delta m_q^2\bigr|_{\rm mix}^{\rm same}
\sim
\frac{\beta_Q\beta_{\rm mix}^{Q}}{16\pi^2}\,
\frac{|m_{\rm mix}(m_Q+m_{\widetilde Q})|}{M_P^2}\,
\Lambda_B^2 ,
\label{eq:scalar_diagmix_same_scaling}
\end{equation}
where \(\Lambda_B\) is the cutoff of the local threshold. The symbol
\(\sim\) means that the numerical factor \(-4/3\), the sign, the local
normal-derivative normalization, and the counterterm-dependent finite part are
not displayed.

For an opposite-boundary derivative--direct topology,
\begin{equation}
\mathcal K_B^{\mathfrak t={\rm opp}}(p)=\frac{1}{\cosh(\pi pR)},
\label{eq:scalar_diagmix_KB_opp}
\end{equation}
and the contribution is finite:
\begin{equation}
\bigl|\delta m_q^2\bigr|_{\rm mix}^{\rm opp}
\sim
\frac{\beta_Q\beta_{\rm mix}^{Q}}{16\pi^2}\,
\frac{|m_{\rm mix}(m_Q+m_{\widetilde Q})|}{M_P^2R^2}\,
\widehat F(m_{3/2}R,m_QR,m_{\widetilde Q}R),
\label{eq:scalar_diagmix_opp_scaling}
\end{equation}
where \(\widehat F\) is a dimensionless finite function.

The analogous diagonal correction to
\(\tilde q^\dagger\tilde q\) is obtained by
\begin{equation}
\beta_Q\beta_{\rm mix}^{Q}
\longrightarrow
\beta_{\widetilde Q}\beta_{\rm mix}^{\widetilde Q},
\qquad
q\longrightarrow \tilde q .
\label{eq:scalar_diagmix_tilde_replacement}
\end{equation}

\subsubsection{Off-diagonal mixed terms}
\label{subsubsec:scalar_offdiag_mixed_trace}

We now discuss off-diagonal scalar bilinears. There are two representative
possibilities.

First, the diagonal vertices
\eqref{eq:scalar_mixed_cubic_Q_channel} and
\eqref{eq:scalar_mixed_cubic_Qtilde_channel}, together with a perturbative
fermion mixing insertion, generate a non-holomorphic off-diagonal term of the
form \(q^\dagger\tilde q\). This topology uses the coefficients
\(\beta_Q\) and \(\beta_{\widetilde Q}\), but not the crossed vertices
\(\beta_{\rm mix}^{Q,\widetilde Q}\).

Second, a holomorphic off-diagonal bilinear \(q\tilde q\) can be generated if
the two scalar vertices share a common internal fermion. There are then two
versions of the common-fermion topology, depending on whether the common
internal fermion is \(\psi_Q\) or \(\psi_{\widetilde Q}\). We first give the
local numerator for a generic common fermion \(\eta\), and then write the
two-channel result.

We write
\begin{equation}
\delta\mathcal L_{\rm scalar}^{\rm off\text{-}diag}
=
-\,m_{q\tilde q}^2\,q\,\tilde q+\text{h.c.}
\label{eq:scalar_offmix_term_definition}
\end{equation}
and use
\begin{equation}
S_\eta(p_E)
=
\frac{-\,i\slashed p_E+m_\eta}{p_E^2+m_\eta^2}.
\label{eq:scalar_offmix_eta_prop}
\end{equation}
The two vertices are
\begin{equation}
\Gamma_\mu^{(q)}(p_E)
=
p_{E\rho}\gamma^\rho\gamma_\mu P_L,
\qquad
\Gamma_\nu^{(\tilde q)}(p_E)
=
p_{E\sigma}\gamma^\sigma\gamma_\nu P_L .
\label{eq:scalar_offmix_vertices_common_eta}
\end{equation}

The numerator is
\begin{align}
\mathcal P_{q\tilde q}
&=
p_{E\rho}p_{E\sigma}\,
{\rm Tr}\!\Big[
\gamma^\rho\gamma_\mu P_L
(-\,i\slashed p_E+m_{3/2})
\Pi^{\mu\nu}_{3/2}
\gamma^\sigma\gamma_\nu P_L
(-\,i\slashed p_E+m_\eta)
\Big].
\label{eq:scalar_offmix_P_trace_definition}
\end{align}
Using
\begin{equation}
P_L(-\,i\slashed p_E+m_{3/2})
=
-\,i\slashed p_E P_R + m_{3/2}P_L,
\label{eq:scalar_offmix_PL_reduction}
\end{equation}
the term proportional to \(-i\slashed p_E P_R\) vanishes because it contains a
factor \(P_R X_{\rm even}P_L\). The surviving term is proportional to
\(m_{3/2}P_L\). The part of the matter propagator proportional to
\(\slashed p_E\) gives an odd trace, or an epsilon tensor contracted with
symmetric products of \(p_E^\mu\), and vanishes. The mass part gives
\begin{align}
\mathcal P_{q\tilde q}
&=
m_{3/2}m_\eta\,
\frac12\,
p_{E\rho}p_{E\sigma}\,
{\rm Tr}\!\Big[
\gamma^\rho\gamma_\mu
\Pi^{\mu\nu}_{3/2}
\gamma^\sigma\gamma_\nu
\Big].
\label{eq:scalar_offmix_m32_mass_reduced}
\end{align}
Expanding
\begin{equation}
\Pi^{\mu\nu}_{3/2}
=
\delta^{\mu\nu}-\frac13\gamma^\mu\gamma^\nu,
\label{eq:scalar_offmix_Pi_split}
\end{equation}
one obtains
\begin{equation}
p_{E\rho}p_{E\sigma}\,
{\rm Tr}\!\Big[
\gamma^\rho\gamma_\mu
\Pi^{\mu\nu}_{3/2}
\gamma^\sigma\gamma_\nu
\Big]
=
\frac{8}{3}\,p_E^2.
\label{eq:scalar_offmix_T_final}
\end{equation}
Therefore
\begin{equation}
\mathcal P_{q\tilde q}(p_E^2;m_{3/2},m_\eta)
=
\frac43\,m_{3/2}m_\eta\,p_E^2,
\label{eq:scalar_offmix_P_closed_result}
\end{equation}
up to the overall Euclidean sign convention.

For a finite opposite-boundary non-local topology \(\mathfrak t\), the two
common-fermion realizations give
\begin{align}
m_{q\tilde q}^{2\,\mathfrak t}
&=
\frac{1}{M_P^2}
\int\frac{d^4p_E}{(2\pi)^4}\,
\mathcal K_B^{\mathfrak t}(p)
\Bigg[
\beta_Q\beta_{\rm mix}^{\widetilde Q}\,
\frac{\mathcal P_{q\tilde q}(p_E^2;m_{3/2},m_Q)}
{(p_E^2+m_{3/2}^2)(p_E^2+m_Q^2)}
\nonumber\\
&\hspace{3.0cm}
+
\beta_{\widetilde Q}\beta_{\rm mix}^{Q}\,
\frac{\mathcal P_{q\tilde q}(p_E^2;m_{3/2},m_{\widetilde Q})}
{(p_E^2+m_{3/2}^2)(p_E^2+m_{\widetilde Q}^2)}
\Bigg].
\label{eq:scalar_offmix_master_two_channels}
\end{align}
For same-boundary derivative--derivative topologies, the analogous expression
contains local normal-derivative normalization factors and boundary
counterterms and will only be used parametrically.

For a local same-boundary topology,
\begin{align}
|m_{q\tilde q}^{2\,{\rm same}}|
&\sim
\frac{1}{16\pi^2}\,
\frac{|m_{3/2}|}{M_P^2}\,
\Lambda_B^2
\left(
|\beta_Q\beta_{\rm mix}^{\widetilde Q}m_Q|
+
|\beta_{\widetilde Q}\beta_{\rm mix}^{Q}m_{\widetilde Q}|
\right).
\label{eq:scalar_offmix_same_scaling}
\end{align}
For an opposite-boundary derivative--direct topology,
\begin{align}
|m_{q\tilde q}^{2\,{\rm opp}}|
&\sim
\frac{1}{16\pi^2}\,
\frac{|m_{3/2}|}{M_P^2R^2}
\Big[
|\beta_Q\beta_{\rm mix}^{\widetilde Q}m_Q|\,
\widehat G(m_{3/2}R,m_QR)
\nonumber\\
&\hspace{3.2cm}
+
|\beta_{\widetilde Q}\beta_{\rm mix}^{Q}m_{\widetilde Q}|\,
\widehat G(m_{3/2}R,m_{\widetilde Q}R)
\Big],
\label{eq:scalar_offmix_opp_scaling}
\end{align}
with \(\widehat G\) a dimensionless finite function.

The numerator \eqref{eq:scalar_offmix_P_closed_result} is not universal. It
holds only for the common-fermion topology specified above. If the two currents
contain distinct fermions and there is no mixing insertion or common internal
fermion, the off-diagonal loop is absent.

The \(R\)-charge condition depends on the interpretation of the scalar
bilinear. For a non-holomorphic term \(q^\dagger\tilde q\) with an
\(R\)-neutral coefficient one needs
\begin{equation}
R(\tilde q)-R(q)=0.
\label{eq:scalar_offmix_Rrule_nonholomorphic}
\end{equation}
For a holomorphic soft term \(q\tilde q\) with an \(R\)-neutral coefficient one
needs
\begin{equation}
R(q)+R(\tilde q)=0.
\label{eq:scalar_offmix_Rrule_soft}
\end{equation}
If instead the same pair is interpreted as descending from a superpotential
mass term, the corresponding condition is
\begin{equation}
R(Q)+R(\widetilde Q)=2.
\label{eq:scalar_offmix_Rrule_superpotential}
\end{equation}

\subsection{Scalar stability}
\label{subsec:scalar_stability}

The universal contribution is present for any boundary scalar coupled to the
bulk supergravity multiplet. In the minimal anti-periodic Scherk--Schwarz
threshold it is negative in our conventions. The mixed contributions have a
different status. Depending on the matter topology, they can generate diagonal
corrections, non-holomorphic off-diagonal terms, or holomorphic \(B\)-type
terms. Their sign and size depend on the companion matter channel, on possible
fermion mixing insertions, and on localization.

It is useful to separate the different entries explicitly. For two complex
scalars \(q\) and \(\tilde q\), write the quadratic potential as
\begin{equation}
V_2
=
\begin{pmatrix}
q^\dagger & \tilde q^\dagger
\end{pmatrix}
H
\begin{pmatrix}
q\\
\tilde q
\end{pmatrix}
+
\frac12
\left[
\begin{pmatrix}
q & \tilde q
\end{pmatrix}
B
\begin{pmatrix}
q\\
\tilde q
\end{pmatrix}
+\text{h.c.}
\right],
\label{eq:scalar_general_quadratic_potential}
\end{equation}
where
\begin{equation}
H=H^\dagger,
\qquad
B=B^T .
\label{eq:scalar_H_B_properties}
\end{equation}
The hermitian matrix \(H\) contains the diagonal scalar masses and the
non-holomorphic mixing \(q^\dagger\tilde q\). The symmetric matrix \(B\)
contains holomorphic scalar bilinears such as \(q\tilde q\). This is the most 
general quadratic form for two complex scalars. In a given
\(U(1)_R\)-symmetric model some of the entries displayed in \(H\) and \(B\)
are absent. For example, a non-holomorphic entry \(q^\dagger\tilde q\)
requires \(R(q)=R(\tilde q)\), whereas a holomorphic entry \(q\tilde q\)
requires the total \(R\)-charge to be compensated by the coefficient or by the
supersymmetry-breaking insertion. The stability conditions below should
therefore be applied after imposing these selection rules. Here
\begin{equation}
H=
\begin{pmatrix}
m_q^2 & C_{q\tilde q}\\[1mm]
C_{q\tilde q}^\ast & m_{\tilde q}^2
\end{pmatrix},
\qquad
B=
\begin{pmatrix}
B_{qq} & B_{q\tilde q}\\[1mm]
B_{q\tilde q} & B_{\tilde q\tilde q}
\end{pmatrix}.
\label{eq:scalar_H_B_two_field}
\end{equation}
The diagonal entries are sums of all diagonal contributions,
\begin{equation}
m_q^2
=
m_{q,{\rm univ}}^2
+
m_{q,{\rm mix,diag}}^2
+
m_{q,{\rm local}}^2
+
m_{q,{\rm stab}}^2
+\cdots ,
\label{eq:scalar_diagonal_sum_q}
\end{equation}
and similarly for \(m_{\tilde q}^2\). The universal threshold is only one term
in this sum.

The exact stability condition is positivity of the real scalar mass matrix. In
terms of real fields,
\begin{equation}
q=\frac{1}{\sqrt2}(q_1+iq_2),
\qquad
\tilde q=\frac{1}{\sqrt2}(\tilde q_1+i\tilde q_2),
\label{eq:scalar_real_decomposition}
\end{equation}
the quadratic potential can be written as
\begin{equation}
V_2
=
\frac12
\Phi_R^T\,\mathcal M_R^2\,\Phi_R,
\qquad
\Phi_R=(q_1,\tilde q_1,q_2,\tilde q_2)^T .
\label{eq:scalar_real_mass_matrix_definition}
\end{equation}
With \(\Phi=(q,\tilde q)^T\), the real mass matrix is
\begin{equation}
\mathcal M_R^2
=
\begin{pmatrix}
{\rm Re}(H+B) & -\,{\rm Im}(H-B)\\[1mm]
{\rm Im}(H+B) & {\rm Re}(H-B)
\end{pmatrix}.
\label{eq:scalar_real_mass_matrix}
\end{equation}
A stable quadratic vacuum requires
\begin{equation}
\mathcal M_R^2>0
\label{eq:scalar_general_stability_condition}
\end{equation}
or, allowing exactly flat directions, \(\mathcal M_R^2\ge0\) with higher-order
terms stabilizing the flat directions.

Several useful limits follow immediately.

First, if holomorphic bilinears are absent, \(B=0\), the condition reduces to
positivity of the hermitian matrix \(H\). For
\begin{equation}
H=
\begin{pmatrix}
m_q^2 & C_{q\tilde q}\\[1mm]
C_{q\tilde q}^\ast & m_{\tilde q}^2
\end{pmatrix},
\label{eq:scalar_nonholomorphic_matrix}
\end{equation}
the eigenvalues are
\begin{equation}
m_\pm^2
=
\frac12\left[
m_q^2+m_{\tilde q}^2
\pm
\sqrt{
(m_q^2-m_{\tilde q}^2)^2
+
4|C_{q\tilde q}|^2
}
\right].
\label{eq:scalar_mass_eigenvalues_nonholomorphic}
\end{equation}
Stability is equivalent to
\begin{equation}
m_q^2>0,
\qquad
m_{\tilde q}^2>0,
\qquad
m_q^2m_{\tilde q}^2>|C_{q\tilde q}|^2 .
\label{eq:scalar_stability_nonholomorphic}
\end{equation}
Equivalently,
\begin{equation}
m_q^2+m_{\tilde q}^2>0,
\qquad
m_q^2m_{\tilde q}^2-|C_{q\tilde q}|^2>0 .
\label{eq:scalar_stability_trace_det}
\end{equation}
Thus a non-holomorphic off-diagonal entry cannot make a negative diagonal
direction stable by itself. If one diagonal entry is negative and the other is
positive, the determinant is negative; if both are negative, the trace is
negative. Once the diagonal entries are fixed, the off-diagonal entry can only
lower the smaller eigenvalue.

Second, if the only off-diagonal term is a holomorphic bilinear
\begin{equation}
V_2
=
m_q^2|q|^2
+
m_{\tilde q}^2|\tilde q|^2
+
\left(
B_{q\tilde q}\,q\tilde q+\text{h.c.}
\right),
\label{eq:scalar_holomorphic_only_potential}
\end{equation}
then, after a phase rotation of one field, the two real blocks have eigenvalues
controlled by
\begin{equation}
\begin{pmatrix}
m_q^2 & |B_{q\tilde q}|\\
|B_{q\tilde q}| & m_{\tilde q}^2
\end{pmatrix},
\qquad
\begin{pmatrix}
m_q^2 & -|B_{q\tilde q}|\\
-|B_{q\tilde q}| & m_{\tilde q}^2
\end{pmatrix}.
\label{eq:scalar_holomorphic_blocks}
\end{equation}
The stability condition is again
\begin{equation}
m_q^2>0,
\qquad
m_{\tilde q}^2>0,
\qquad
m_q^2m_{\tilde q}^2>|B_{q\tilde q}|^2 .
\label{eq:scalar_stability_holomorphic}
\end{equation}
Thus a holomorphic \(B\)-term also cannot cure a negative diagonal mass by
itself. If too large, it produces a tachyonic real scalar.

Third, the mixed sector can nevertheless help through its diagonal entries. The
representative diagonal contribution computed in
Section~\ref{subsubsec:scalar_diag_mixed_trace} corrects \(m_q^2\) or
\(m_{\tilde q}^2\), not only the off-diagonal entries. If this diagonal
contribution is positive and sufficiently large, it can compensate the negative
minimal Scherk--Schwarz universal threshold. Schematically,
\begin{equation}
m_q^2
=
m_{q,{\rm univ}}^2
+
m_{q,{\rm mix,diag}}^2
+
m_{q,{\rm other}}^2
\label{eq:scalar_diag_balance}
\end{equation}
can be positive even when
\(m_{q,{\rm univ}}^2<0\). The same applies to \(\tilde q\). Once the diagonal
entries are positive, the off-diagonal non-holomorphic and holomorphic entries
must satisfy the corresponding determinant bounds.

For example, if \(B=0\), stability requires
\begin{equation}
\left(
m_{q,{\rm univ}}^2
+
m_{q,{\rm mix,diag}}^2
+
m_{q,{\rm other}}^2
\right)
\left(
m_{\tilde q,{\rm univ}}^2
+
m_{\tilde q,{\rm mix,diag}}^2
+
m_{\tilde q,{\rm other}}^2
\right)
>
|C_{q\tilde q}|^2,
\label{eq:scalar_stability_with_sources}
\end{equation}
with both diagonal factors positive. This is the precise sense in which the
mixed sector can either help or hurt: its diagonal part can raise the diagonal
masses, while its off-diagonal part can lower the lightest eigenvalue.

The result is therefore a stability criterion, not a general sign statement
about the mixed sector. In the minimal anti-periodic Scherk--Schwarz threshold
the universal diagonal contribution is negative. A consistent vacuum requires
the remaining diagonal contributions, from the mixed sector, local
counterterms, or the stabilization potential, to make the diagonal entries of
the quadratic form positive. Once this is achieved, the non-holomorphic and
holomorphic off-diagonal entries must still satisfy the determinant bounds of
the full real scalar mass matrix. The constraint is model-dependent: it
restricts the allowed matter couplings and stabilization data, but does not by
itself exclude the Dirac-gravitino construction.

\section{One-loop fermion masses from the Dirac gravitino sector}
\label{sec:fermion_masses}

We now turn to the fermion sector. Unlike the scalar sector, it does not contain
a universal Scherk--Schwarz threshold generated by the ordinary gravitino
coupling alone. In the anti-periodic Scherk--Schwarz setup, the gravitational
Majorana gaugino mass vanishes at \(\omega=\tfrac12\), while scalar masses
remain nonzero
\cite{Gherghetta:2001sa,Rattazzi:2003rj,Antoniadis:1997ic}. Fermion masses
therefore require additional structure.

Let
\begin{equation}
Q=(\phi,\psi,F),
\qquad
\Sigma=(s,\xi,F_\Sigma)
\label{eq:fermion_visible_multiplets_section}
\end{equation}
be two visible chiral multiplets. We ask whether the Dirac-gravitino sector can
induce the Dirac bilinear
\begin{equation}
{\cal L}_{m_D}
=
-\,m_D\,\psi\xi+\text{h.c.}
\label{eq:fermion_target_mass_term}
\end{equation}
at one loop.

Three conditions are necessary.

\paragraph{First condition: \(R\)-neutrality.}
The bilinear \(\psi\xi\) is allowed only if it is neutral under the residual
\(U(1)_R\). If
\begin{equation}
R(\psi)=r_Q-1,
\qquad
R(\xi)=r_\Sigma-1,
\end{equation}
then
\begin{equation}
R(\psi\xi)=0
\qquad\Longleftrightarrow\qquad
(r_Q-1)+(r_\Sigma-1)=0
\qquad\Longleftrightarrow\qquad
r_Q+r_\Sigma=2 .
\label{eq:fermion_R_neutrality_condition}
\end{equation}
We impose this condition in what follows.

\paragraph{Second condition: a companion matter current.}
The ordinary gravitino coupling alone does not generate the Dirac mass
\eqref{eq:fermion_target_mass_term}. A second matter current coupled to the
companion spin-\(\tfrac32\) field is also required. If this current is absent,
the one-loop Dirac mass vanishes.

\paragraph{Third condition: closure of the internal bosonic line.}
The one-loop diagram contains two derivative matter vertices and one internal
bosonic propagator. The bosonic line must therefore close. This can happen in
one of two ways:
\begin{enumerate}
\item[(i)] the two vertices involve the same scalar field;
\item[(ii)] the two vertices involve distinct scalars, but there is a nonzero
mixed scalar propagator between them.
\end{enumerate}
If neither condition is satisfied, the one-loop Dirac mass vanishes.

The calculation separates three ingredients:
\begin{enumerate}
\item the tensor coefficient fixed by four-dimensional spin-\(\tfrac32\)
algebra;
\item the momentum kernels fixed by the anti-periodic Scherk--Schwarz geometry;
\item the normalization of the companion-channel matter coupling and of the
bosonic-line closure, which are not fixed by the geometry alone.
\end{enumerate}
This separation fixes the scaling with \(M_P\).

\subsection{Four-dimensional EFT setup: explicit assumptions and tensor reduction}
\label{subsec:fermion_4d_explicit}

We begin with the four-dimensional EFT. Since both \(\psi_\mu\) and
\(\chi_\mu\) are spin-\(\tfrac32\) fields, we take the relevant matter couplings
to be of supercurrent type \cite{Freedman:2012zz}. We do not use a
non-derivative Yukawa-like coupling of the form
\(\chi_\mu\sigma^\mu\bar\xi\,\phi\).

We write
\begin{equation}
{\cal L}_{\psi J}
=
-\frac{\kappa_Q}{\sqrt2}\,
(\partial_\nu\phi^\dagger)\,
\bar\psi_\mu\,\gamma^\nu\gamma^\mu P_L\,\psi
+\text{h.c.},
\label{eq:fermion_vertex_Q_general}
\end{equation}
and
\begin{equation}
{\cal L}_{\chi J_2}
=
-\frac{\kappa_\Sigma}{\sqrt2}\,
(\partial_\nu s^\dagger)\,
\bar\xi\,\gamma^\nu\gamma^\mu P_L\,\chi_\mu
+\text{h.c.}.
\label{eq:fermion_vertex_Sigma_general}
\end{equation}
The coefficients \(\kappa_Q\) and \(\kappa_\Sigma\) have mass dimension
\(-1\).

Two normalizations must be distinguished.

\paragraph{Pure EFT normalization.}
If \(\chi_\mu\) is treated as an independent low-energy spin-\(\tfrac32\)
field, then \(\kappa_\Sigma\) is a free EFT parameter. No definite Planck
scaling follows.

\paragraph{Supergravity normalization.}
If \(\chi_\mu\) is identified with the second gravitino of an underlying
\(\mathcal N=2\) or five-dimensional supergravity, then the conservative
normalization is
\begin{equation}
\kappa_Q=\frac{1}{M_P},
\qquad
\kappa_\Sigma=\frac{g_\Sigma}{M_P},
\label{eq:fermion_supergravity_normalization}
\end{equation}
with \(g_\Sigma\) dimensionless. Any diagram with two explicit boundary
spin-\(\tfrac32\)-matter vertices then carries
\begin{equation}
\kappa_Q\kappa_\Sigma
=
\frac{g_\Sigma}{M_P^2}.
\label{eq:fermion_MP2_expectation}
\end{equation}
This statement is independent of the value of the Scherk--Schwarz twist. The
special point \(\omega=\tfrac12\) changes the spectrum and the \(R\)-symmetry
selection rules; it does not change the Planck suppression of a local boundary
vertex.

We work in Euclidean momentum space. The loop momentum is \(k_E^\mu\), and
\begin{equation}
k\equiv |k_E|=\sqrt{k_E^2}.
\label{eq:fermion_4d_k_definition}
\end{equation}

The internal bosonic line is denoted by
\begin{equation}
\Delta_{\phi s}(k_E).
\label{eq:fermion_general_scalar_line}
\end{equation}
If the same scalar propagates between the two vertices, then
\begin{equation}
\Delta_{\phi s}(k_E)
=
\frac{1}{k_E^2+m_B^2}.
\label{eq:fermion_same_scalar_case}
\end{equation}
If the two vertices involve distinct scalars, then \(\Delta_{\phi s}\) is a
mixed scalar propagator. For example, with a scalar mixing insertion
\(\mu_{\phi s}^2\,\phi^\dagger s+\text{h.c.}\), one has perturbatively
\begin{equation}
\Delta_{\phi s}(k_E)
=
\frac{\mu_{\phi s}^2}
{(k_E^2+m_\phi^2)(k_E^2+m_s^2)}
+\mathcal O\big((\mu_{\phi s}^2)^3\big).
\label{eq:fermion_mixed_scalar_propagator_example}
\end{equation}
If no scalar mixing is present, then
\begin{equation}
\Delta_{\phi s}(k_E)=0
\qquad\Longrightarrow\qquad
\Sigma_{\psi\xi}^{4D}(0)=0.
\label{eq:fermion_no_scalar_closure}
\end{equation}

The reduced mixed spin-\(\tfrac32\) propagator is taken to be
\begin{equation}
S^{\psi\chi}_{\mu\nu}(k_E)
=
\frac{m_{3/2}}{k_E^2+m_{3/2}^2}
\left(
\delta_{\mu\nu}-\frac13\gamma_\mu\gamma_\nu
\right).
\label{eq:fermion_reduced_mixed_propagator}
\end{equation}
This is not the full off-shell Rarita--Schwinger propagator. It is the
mass-projecting part of the mixed spin-\(\tfrac32\) propagator relevant for
the zero-external-momentum Dirac self-energy. The full numerator also contains
a term proportional to \(-i\slashed k_E\). With the chiral derivative vertices
\eqref{eq:fermion_vertex_Q_general} and
\eqref{eq:fermion_vertex_Sigma_general}, that term does not contribute to the
local chirality-flipping operator \(\psi\,\xi\). It gives a derivative or
kinetic structure, equivalently a contribution proportional to the external
momentum, rather than a mass insertion. The Dirac mass is therefore obtained
from the \(m_{3/2}\) part of the numerator. Terms that vanish in this mass
projection are not displayed.

At zero external momentum, the one-loop self-energy is
\begin{equation}
\Sigma_{\psi\xi}^{4D}(0)
=
\frac{\kappa_Q\kappa_\Sigma}{2}
\int\frac{d^4k_E}{(2\pi)^4}\;
\Delta_{\phi s}(k_E)\,
k_{E\rho}k_{E\sigma}\,
\gamma^\rho\gamma^\mu\,
S^{\psi\chi}_{\mu\nu}(k_E)\,
\gamma^\sigma\gamma^\nu .
\label{eq:fermion_4d_selfenergy_start}
\end{equation}
Substituting \eqref{eq:fermion_reduced_mixed_propagator} gives
\begin{equation}
\Sigma_{\psi\xi}^{4D}(0)
=
\frac{\kappa_Q\kappa_\Sigma\,m_{3/2}}{2}
\int\frac{d^4k_E}{(2\pi)^4}\;
\frac{\Delta_{\phi s}(k_E)}{k_E^2+m_{3/2}^2}\;
{\cal N}_F(k_E),
\label{eq:fermion_4d_selfenergy_after_prop}
\end{equation}
where
\begin{equation}
{\cal N}_F(k_E)
=
k_{E\rho}k_{E\sigma}\,
\gamma^\rho\gamma^\mu
\left(
\delta_{\mu\nu}-\frac13\gamma_\mu\gamma_\nu
\right)
\gamma^\sigma\gamma^\nu .
\label{eq:fermion_4d_numerator_definition}
\end{equation}

The numerator reduces as follows:
\begin{equation}
{\cal N}_F
=
k_{E\rho}k_{E\sigma}
\left[
\gamma^\rho\gamma^\nu\gamma^\sigma\gamma_\nu
-\frac13\,
\gamma^\rho\gamma_\mu\gamma^\mu\gamma_\nu\gamma^\sigma\gamma^\nu
\right].
\label{eq:fermion_4d_numerator_step1}
\end{equation}
Using
\begin{equation}
\gamma^\nu\gamma^\sigma\gamma_\nu=-2\gamma^\sigma,
\qquad
\gamma_\mu\gamma^\mu=4,
\label{eq:fermion_4d_gamma_identities}
\end{equation}
one obtains
\begin{equation}
{\cal N}_F
=
k_{E\rho}k_{E\sigma}
\left[
-2\,\gamma^\rho\gamma^\sigma
+\frac{8}{3}\,\gamma^\rho\gamma^\sigma
\right]
=
\frac23\,
k_{E\rho}k_{E\sigma}\gamma^\rho\gamma^\sigma .
\label{eq:fermion_4d_numerator_step2}
\end{equation}
Since
\begin{equation}
k_{E\rho}k_{E\sigma}\gamma^\rho\gamma^\sigma
=
\slashed{k}_E\slashed{k}_E
=
k_E^2,
\label{eq:fermion_4d_kslash_square}
\end{equation}
we find
\begin{equation}
{\cal N}_F=\frac23\,k_E^2.
\label{eq:fermion_4d_numerator_final}
\end{equation}
Thus the tensor coefficient is
\begin{equation}
c_F\equiv \frac23.
\label{eq:fermion_cF_definition}
\end{equation}

The self-energy becomes
\begin{equation}
\Sigma_{\psi\xi}^{4D}(0)
=
\frac{c_F\,\kappa_Q\kappa_\Sigma\,m_{3/2}}{2}
\int\frac{d^4k_E}{(2\pi)^4}\;
\frac{k_E^2\,\Delta_{\phi s}(k_E)}{k_E^2+m_{3/2}^2}.
\label{eq:fermion_4d_selfenergy_reduced}
\end{equation}

For the simple scalar propagator
\(\Delta_{\phi s}(k_E)=1/(k_E^2+m_B^2)\), this gives
\begin{equation}
\Sigma_{\psi\xi}^{4D}(0)
=
\frac{c_F\,\kappa_Q\kappa_\Sigma\,m_{3/2}}{2}
\int\frac{d^4k_E}{(2\pi)^4}\;
\frac{k_E^2}{(k_E^2+m_B^2)(k_E^2+m_{3/2}^2)} .
\label{eq:fermion_4d_selfenergy_massive_scalar}
\end{equation}
Using
\begin{equation}
\frac{k_E^2}{(k_E^2+m_B^2)(k_E^2+m_{3/2}^2)}
=
\frac{m_B^2}{m_B^2-m_{3/2}^2}\,
\frac{1}{k_E^2+m_B^2}
-
\frac{m_{3/2}^2}{m_B^2-m_{3/2}^2}\,
\frac{1}{k_E^2+m_{3/2}^2},
\label{eq:fermion_4d_partial_fraction}
\end{equation}
and a hard Euclidean cutoff \(\Lambda\), one finds
\begin{align}
\Sigma_{\psi\xi}^{4D}(0)
&=
\frac{c_F\,\kappa_Q\kappa_\Sigma\,m_{3/2}}{32\pi^2}
\Bigg[
\Lambda^2
-
\frac{
m_B^4\ln\!\left(1+\frac{\Lambda^2}{m_B^2}\right)
-
m_{3/2}^4\ln\!\left(1+\frac{\Lambda^2}{m_{3/2}^2}\right)
}{
m_B^2-m_{3/2}^2
}
\Bigg].
\label{eq:fermion_4d_cutoff_general}
\end{align}
For \(m_B=0\),
\begin{equation}
\Sigma_{\psi\xi}^{4D}(0)
=
\frac{c_F\,\kappa_Q\kappa_\Sigma\,m_{3/2}}{32\pi^2}
\left[
\Lambda^2
-
m_{3/2}^2\ln\!\left(1+\frac{\Lambda^2}{m_{3/2}^2}\right)
\right].
\label{eq:fermion_4d_cutoff_massless_scalar}
\end{equation}
With the supergravity normalization
\eqref{eq:fermion_supergravity_normalization}, this becomes
\begin{equation}
\Sigma_{\psi\xi}^{4D}(0)
=
\frac{g_\Sigma\,m_{3/2}}{48\pi^2M_P^2}
\left[
\Lambda^2
-
m_{3/2}^2\ln\!\left(1+\frac{\Lambda^2}{m_{3/2}^2}\right)
\right]
\qquad (m_B=0).
\label{eq:fermion_4d_cutoff_supergravity}
\end{equation}

The four-dimensional EFT result implies the local scaling
\begin{equation}
m_D^{4D}
\sim
\kappa_Q\kappa_\Sigma\,m_{3/2}\Lambda_{\rm eff}^2.
\label{eq:fermion_4d_planck_scaling_general}
\end{equation}
With supergravity-normalized vertices,
\begin{equation}
m_D^{4D}
\sim
\frac{m_{3/2}\Lambda_{\rm eff}^2}{M_P^2}.
\label{eq:fermion_4d_planck_scaling}
\end{equation}
Thus two explicit Planck-suppressed matter vertices give \(1/M_P^2\), not
\(1/M_P\). The half-twist point is relevant for the Dirac-gravitino degeneracy
and the \(U(1)_R\) selection rule, not for the power of \(M_P\) carried by local
boundary vertices.

\subsection{Anti-periodic Scherk--Schwarz realization at \texorpdfstring{$\omega=\tfrac12$}{w=1/2}}
\label{subsec:fermion_SS_rewrite}

We now specialize to the anti-periodic Scherk--Schwarz benchmark, using the
mode functions and boundary kernels reviewed in
Sections~\ref{sec:SS_origin} and \ref{sec:coupling_to_matter}
\cite{Gherghetta:2001sa,Rattazzi:2003rj,Antoniadis:1997ic}. The interval
coordinate is \(y\in[0,\pi R]\), and the Euclidean momentum modulus is
\begin{equation}
p\equiv |p_E|=\sqrt{p_E^2}.
\label{eq:fermion_SS_p_definition}
\end{equation}

At \(\omega=\tfrac12\), the relevant towers are
\begin{equation}
f_n^{(+)}(y)
=
\sqrt{\frac{2}{\pi R}}\cos\!\left(\frac{n+\tfrac12}{R}\,y\right),
\qquad
f_n^{(-)}(y)
=
\sqrt{\frac{2}{\pi R}}\sin\!\left(\frac{n+\tfrac12}{R}\,y\right),
\qquad
m_n=\frac{n+\tfrac12}{R}.
\label{eq:fermion_SS_mode_functions}
\end{equation}
The \(+\) tower couples directly on the boundary \(y=0\). The \(-\) tower
couples directly on the boundary \(y=\pi R\). At a given boundary, the
wrong-parity tower can enter only through normal derivatives or through
bulk-mediated kernels.

The Green functions are
\begin{equation}
G_+(p;y,y')
=
\frac{\cosh(p\,y_<)\,\sinh\!\big(p(\pi R-y_>)\big)}
{p\,\cosh(\pi pR)},
\qquad
G_-(p;y,y')
=
\frac{\sinh(p\,y_<)\,\cosh\!\big(p(\pi R-y_>)\big)}
{p\,\cosh(\pi pR)}.
\label{eq:fermion_SS_green_functions}
\end{equation}
The useful derivative kernels are
\begin{equation}
\partial_y G_-(p;y,y')\Big|_{y=0}
=
\frac{\cosh\!\big(p(\pi R-y')\big)}{\cosh(\pi pR)},
\qquad
-\partial_y G_+(p;y,y')\Big|_{y=\pi R}
=
\frac{\cosh(py')}{\cosh(\pi pR)}.
\label{eq:fermion_SS_derivative_kernels}
\end{equation}
Opposite-boundary communication is exponentially suppressed as
\(\sech(\pi pR)\) for \(\pi pR\gg1\).

The mixed even/odd propagation with one bulk Dirac mass insertion is
\begin{equation}
G_{+-}^{(1)}(p;y,y')
=
m_{3/2}
\int_0^{\pi R} dz\;
G_+(p;y,z)\,G_-(p;z,y').
\label{eq:fermion_SS_mixed_kernel_definition}
\end{equation}
This definition fixes the momentum dependence of the threshold kernels. It does
not fix the normalization of the boundary matter currents.

The two kernels needed below are
\begin{equation}
\partial_{y'}G_{+-}^{(1)}(p;0,y')\Big|_{y'=0}
=
\frac{m_{3/2}}{2p^2}\,\tanh^2(\pi pR),
\label{eq:fermion_SS_same_kernel}
\end{equation}
and
\begin{equation}
G_{+-}^{(1)}(p;0,\pi R)
=
\frac{m_{3/2}}{2p^3\cosh(\pi pR)}
\Big(\pi pR-\tanh(\pi pR)\Big).
\label{eq:fermion_SS_opp_kernel}
\end{equation}

We parameterize the remaining normalization freedom by
\begin{equation}
{\cal A}_{\rm same},
\qquad
{\cal A}_{\rm opp}.
\label{eq:fermion_SS_effective_normalizations}
\end{equation}
These coefficients include the two boundary-current normalizations, the tensor
factor \(c_F=\tfrac23\), and the scalar-line closure factor. Their mass
dimensions are fixed by the formulas below:
\begin{equation}
[{\cal A}_{\rm same}]=-2,
\qquad
[{\cal A}_{\rm opp}]=-1.
\label{eq:fermion_SS_A_dimensions}
\end{equation}
For example, in a conservative supergravity normalization,
\({\cal A}_{\rm same}\) is of order \(M_P^{-2}\), whereas
\({\cal A}_{\rm opp}\) must contain one additional mass scale in order to have
dimension \(-1\).

For two fermions localized on the same boundary, we write
\begin{equation}
\Sigma_{\psi\xi}^{\rm same}(0)
=
{\cal A}_{\rm same}
\int\frac{d^4p_E}{(2\pi)^4}\;
\partial_{y'}G_{+-}^{(1)}(p;0,y')\Big|_{y'=0}.
\label{eq:fermion_SS_same_selfenergy_start}
\end{equation}
Using \eqref{eq:fermion_SS_same_kernel},
\begin{equation}
\Sigma_{\psi\xi}^{\rm same}(0)
=
{\cal A}_{\rm same}\,
\frac{m_{3/2}}{2}
\int\frac{d^4p_E}{(2\pi)^4}\;
\frac{\tanh^2(\pi pR)}{p^2}.
\label{eq:fermion_SS_same_selfenergy_explicit}
\end{equation}
After angular integration,
\begin{equation}
\Sigma_{\psi\xi}^{\rm same}(0)
=
\frac{{\cal A}_{\rm same}\,m_{3/2}}{16\pi^2}
\int_0^\infty dp\;p\,\tanh^2(\pi pR).
\label{eq:fermion_SS_same_radial}
\end{equation}
This has the ultraviolet behaviour of a local threshold, since
\(\tanh^2(\pi pR)\to1\) at large \(p\). We split
\begin{equation}
\int_0^\infty dp\;p\,\tanh^2(\pi pR)
=
\int_0^\infty dp\;p
+
\int_0^\infty dp\;p\,\big(\tanh^2(\pi pR)-1\big).
\label{eq:fermion_SS_same_split}
\end{equation}
The first term is absorbed into a local boundary counterterm. The finite
remainder is
\begin{equation}
\int_0^\infty dp\;p\,\big(\tanh^2(\pi pR)-1\big)
=
-\frac{\ln2}{\pi^2R^2}.
\label{eq:fermion_SS_same_finite_integral}
\end{equation}
Therefore
\begin{equation}
\Sigma_{\psi\xi}^{\rm same,fin}(0)
=
-\frac{{\cal A}_{\rm same}\,\ln2}{16\pi^4}\,
\frac{m_{3/2}}{R^2}.
\label{eq:fermion_SS_same_finite_result}
\end{equation}
Using \(m_{3/2}=1/(2R)\), this becomes
\begin{equation}
m_{D,\,{\rm same}}^{\rm fin}
=
-\frac{{\cal A}_{\rm same}\,\ln2}{4\pi^4}\,
m_{3/2}^3 .
\label{eq:fermion_SS_same_final_mass}
\end{equation}

For fermions localized on opposite boundaries, we write
\begin{equation}
\Sigma_{\psi\xi}^{\rm opp}(0)
=
{\cal A}_{\rm opp}
\int\frac{d^4p_E}{(2\pi)^4}\;
G_{+-}^{(1)}(p;0,\pi R).
\label{eq:fermion_SS_opp_selfenergy_start}
\end{equation}
Using \eqref{eq:fermion_SS_opp_kernel},
\begin{equation}
\Sigma_{\psi\xi}^{\rm opp}(0)
=
{\cal A}_{\rm opp}\,
\frac{m_{3/2}}{2}
\int\frac{d^4p_E}{(2\pi)^4}\;
\frac{\pi pR-\tanh(\pi pR)}{p^3\cosh(\pi pR)}.
\label{eq:fermion_SS_opp_selfenergy_explicit}
\end{equation}
After angular integration,
\begin{equation}
\Sigma_{\psi\xi}^{\rm opp}(0)
=
\frac{{\cal A}_{\rm opp}\,m_{3/2}}{16\pi^2}
\int_0^\infty dp\;
\frac{\pi pR-\tanh(\pi pR)}{\cosh(\pi pR)}.
\label{eq:fermion_SS_opp_radial}
\end{equation}
The integral is finite. With \(x=\pi pR\),
\begin{equation}
\Sigma_{\psi\xi}^{\rm opp}(0)
=
\frac{{\cal A}_{\rm opp}\,m_{3/2}}{16\pi^3R}
\int_0^\infty dx\;
\frac{x-\tanh x}{\cosh x}.
\label{eq:fermion_SS_opp_x_integral}
\end{equation}
The integral is
\begin{equation}
{\cal I}_{\rm opp}
\equiv
\int_0^\infty dx\;
\frac{x-\tanh x}{\cosh x}
=
2G_{\rm Cat}-1
\simeq 0.83193,
\label{eq:fermion_SS_Iopp_exact}
\end{equation}
where \(G_{\rm Cat}\) is Catalan's constant. Hence
\begin{equation}
\Sigma_{\psi\xi}^{\rm opp}(0)
=
\frac{{\cal A}_{\rm opp}(2G_{\rm Cat}-1)}{16\pi^3}\,
\frac{m_{3/2}}{R},
\label{eq:fermion_SS_opp_finite_result}
\end{equation}
or, using \(m_{3/2}=1/(2R)\),
\begin{equation}
m_{D,\,{\rm opp}}
=
\frac{{\cal A}_{\rm opp}(2G_{\rm Cat}-1)}{8\pi^3}\,
m_{3/2}^2.
\label{eq:fermion_SS_opp_final_mass}
\end{equation}

Equations \eqref{eq:fermion_SS_same_final_mass} and
\eqref{eq:fermion_SS_opp_final_mass} are exact for the stated kernels and
effective coefficients \({\cal A}_{\rm same}\) and \({\cal A}_{\rm opp}\). The
finite coefficients \(\ln2\) and \(2G_{\rm Cat}-1\) are fixed. The remaining
input is the normalization of the companion-channel matter coupling and of the
scalar-line closure.

If the companion channel has a conservative supergravity origin, the natural
dimensional estimates are
\begin{equation}
{\cal A}_{\rm same}
=
\frac{a_{\rm same}}{M_P^2},
\qquad
{\cal A}_{\rm opp}
=
\frac{a_{\rm opp}\,m_{3/2}}{M_P^2},
\label{eq:fermion_SS_supergravity_A_scaling}
\end{equation}
with \(a_{\rm same}\) and \(a_{\rm opp}\) dimensionless coefficients determined
by the microscopic boundary construction. The factor \(m_{3/2}\) in
\({\cal A}_{\rm opp}\) is required by dimensional analysis if the only mass
scale in the non-local threshold is the Scherk--Schwarz scale \(1/R=2m_{3/2}\).

With \eqref{eq:fermion_SS_supergravity_A_scaling}, one obtains
\begin{equation}
m_{D,\,{\rm same}}^{\rm fin}
=
-\frac{a_{\rm same}\ln2}{4\pi^4}\,
\frac{m_{3/2}^3}{M_P^2},
\label{eq:fermion_SS_same_supergravity_scaling}
\end{equation}
and
\begin{equation}
m_{D,\,{\rm opp}}
=
\frac{a_{\rm opp}(2G_{\rm Cat}-1)}{8\pi^3}\,
\frac{m_{3/2}^3}{M_P^2}.
\label{eq:fermion_SS_opp_supergravity_scaling}
\end{equation}
Thus, with supergravity-normalized companion couplings, the Dirac fermion mass
scales as
\begin{equation}
m_D
\sim
\frac{1}{16\pi^2}\,
\frac{m_{3/2}^3}{M_P^2},
\label{eq:fermion_SS_conservative_scaling}
\end{equation}
up to dimensionless coefficients and topology-dependent numerical factors.

A larger scaling,
\begin{equation}
m_D\sim \frac{1}{16\pi^2}\frac{m_{3/2}^2}{M_P},
\label{eq:fermion_SS_enhanced_scaling}
\end{equation}
requires an effective companion-channel normalization with one less power of
\(M_P\). In the present EFT this is an additional normalization assumption, not
a consequence of the Scherk--Schwarz kernels.

\subsection{Alternative topologies and the role of mass insertions}
\label{subsec:fermion_alternative_topologies_rewrite}

It remains to discuss the alternative topology in which the derivative
wrong-parity coupling is replaced by additional insertions of the Dirac
gravitino mass along the internal spin-\(\tfrac32\) line.

In the anti-periodic Scherk--Schwarz realization, a direct coupling on the
boundary \(y=0\) is to the even tower. The odd tower vanishes at that boundary.
The mixed even/odd channel at the same boundary is therefore realized by one
direct even-channel coupling and one derivative coupling to the wrong-parity
channel. This is the leading geometric realization of the mixed
first/second-supersymmetry diagram.

If one keeps only direct same-boundary couplings and inserts the Dirac
gravitino mass along the internal line, the topology is different. In the
standard Scherk--Schwarz construction it reduces to an even-even boundary
kernel. Projected onto a Majorana channel, that contribution is of the type
forbidden at \(\omega=\tfrac12\) by the residual \(U(1)_R\). Projected onto the
Dirac channel, it does not reproduce the derivative wrong-parity kernel
computed above. It can contribute only to local, counterterm-sensitive
same-boundary terms unless an additional microscopic coupling is specified.

Thus the derivative wrong-parity coupling is the leading way in which the odd
tower enters a boundary amplitude in the explicit Scherk--Schwarz geometry.

The conclusions of this section are:
\begin{enumerate}
\item The four-dimensional tensor algebra fixes
\begin{equation}
c_F=\frac23.
\end{equation}

\item A one-loop Dirac fermion mass requires both a companion matter current and
a closed internal scalar line.

\item If both spin-\(\tfrac32\)-matter vertices are supergravity-normalized,
the natural Planck suppression is \(1/M_P^2\). This statement is independent of
whether \(\omega=\tfrac12\).

\item The anti-periodic Scherk--Schwarz geometry fixes the finite kernels and
the numerical coefficients \(\ln2\) and \(2G_{\rm Cat}-1\).

\item Same-boundary terms have the ultraviolet behaviour of local thresholds
and require boundary counterterms. Opposite-boundary terms are finite and
non-local.

\item With conservative supergravity normalization, the radiative Dirac fermion
mass scales as \(m_{3/2}^3/M_P^2\). A larger scaling
\(m_{3/2}^2/M_P\) is an additional EFT assumption, not a result of the minimal
anti-periodic Scherk--Schwarz construction.
\end{enumerate}
















\section{Phenomenological implications: hierarchies, neutrinos, and modulini}
\label{sec:pheno}

The results of the previous sections can be organized by separating statements
which are fixed within the Scherk--Schwarz supergravity benchmark from
statements which depend on additional matter couplings.

The first fixed effect is the scalar threshold. The operator \(|\phi|^2\) is
neutral under the residual \(U(1)_R\), and the ordinary gravitino coupling to
the visible-sector supercurrent is present for any boundary scalar coupled to
the bulk supergravity multiplet. In the anti-periodic Scherk--Schwarz
benchmark, the corresponding one-loop threshold has the parametric form
\begin{equation}
m_{\phi,\,{\rm univ}}^2
\sim
\frac{1}{16\pi^2}\,
C_{\rm univ}\,
\frac{m_{3/2}^4}{M_P^2},
\label{eq:pheno_scalar_univ}
\end{equation}
where \(C_{\rm univ}\) is a dimensionless coefficient including the
Scherk--Schwarz threshold normalization. In the minimal contribution computed
in Section~\ref{sec:scalar_masses}, \(C_{\rm univ}<0\) with our conventions
\cite{Antoniadis:1997ic,Gherghetta:2001sa,Rattazzi:2003rj,Antoniadis:2015chx}.
Thus the universal threshold is tachyonic in the minimal setup. The physical
scalar spectrum is obtained only after adding all diagonal and off-diagonal
contributions, including possible local, anomaly-mediated, radion-mediated,
gauge-mediated, mixed-sector, and stabilization effects. Positive scalar
masses can be obtained in deformed brane-to-brane setups with additional
localized gravitational terms, modified radion stabilization, or extra bulk
multiplets \cite{Rattazzi:2003rj}. Such ingredients are model-dependent and
are not part of the minimal universal threshold.

The fermion sector has a different status. In the exact \(U(1)_R\)-symmetric
limit, Majorana masses are forbidden. A Dirac fermion mass can be generated
only if the low-energy matter sector contains the required partner channel and
if the companion spin-\(\tfrac32\) field couples to the corresponding second
matter current. If this extra channel is absent, the one-loop Dirac fermion
mass is absent.

The same additional matter structure can also generate mixed scalar terms.
These terms are not fixed by the ordinary supergravity coupling. Depending on
the topology, they may contribute to diagonal scalar entries,
non-holomorphic off-diagonal entries, or holomorphic \(B\)-type bilinears.
Diagonal entries can help compensate the negative universal threshold, while
off-diagonal entries are constrained by the determinant bounds of the full real
scalar mass matrix.

The normalization of the fermion threshold is not fixed by the geometry alone.
The anti-periodic Scherk--Schwarz background fixes the even/odd kernels and the
finite non-local threshold functions. It does not fix the absolute
normalization of the companion-channel matter current. If the companion field
is identified with the second gravitino of an underlying \(\mathcal N=2\) or
five-dimensional supergravity, then two explicit spin-\(\tfrac32\)-matter
vertices carry the conservative suppression \(1/M_P^2\). In that case,
\begin{equation}
m_f^{\rm Dirac}
\sim
\frac{1}{16\pi^2}\,
\alpha_f\,
\frac{m_{3/2}^3}{M_P^2},
\label{eq:pheno_fermion_conservative}
\end{equation}
up to dimensionless coefficients, localization factors, and details of the
scalar-line closure. Equivalently, in a local four-dimensional cutoff estimate,
\begin{equation}
m_f^{\rm Dirac}
\sim
\frac{1}{16\pi^2}\,
\alpha_f\,
\frac{m_{3/2}\Lambda_{\rm eff}^2}{M_P^2},
\label{eq:pheno_fermion_conservative_cutoff}
\end{equation}
where \(\Lambda_{\rm eff}\) is the scale cutting off the local threshold, or
the corresponding non-local scale in the Scherk--Schwarz realization. In the
minimal anti-periodic Scherk--Schwarz threshold, \(\Lambda_{\rm eff}\) is of
order \(m_{3/2}\), giving \eqref{eq:pheno_fermion_conservative}.

A larger estimate,
\begin{equation}
m_f^{\rm Dirac}
\sim
\frac{1}{16\pi^2}\,
\alpha_f\,
\frac{m_{3/2}^{2}}{M_P},
\label{eq:pheno_fermion_enhanced}
\end{equation}
can be used as an EFT benchmark. It requires a companion-channel normalization
stronger than the conservative supergravity one. It is therefore an additional
EFT assumption, not a consequence of the minimal anti-periodic Scherk--Schwarz
construction.

Mixed scalar entries can be summarized by the template
\begin{equation}
m_{\phi\phi'}^{2\,{\rm mix}}
\sim
\frac{1}{16\pi^2}\,
C_{\rm mix}\,
\frac{{\cal M}_{\rm mix}}{M_P^2}\,
\Lambda_{\rm mix}^2,
\label{eq:pheno_scalar_mix_general}
\end{equation}
where \({\cal M}_{\rm mix}\) is the mass product appearing in the numerator of
the relevant loop topology. For the representative diagonal mixed term of
Section~\ref{sec:scalar_masses},
\begin{equation}
{\cal M}_{\rm mix}
\sim
m_{\rm mix}(m_Q+m_{\widetilde Q}),
\label{eq:pheno_scalar_mix_diagonal_mass_product}
\end{equation}
whereas for the representative holomorphic off-diagonal topology with a common
internal fermion \(\eta\),
\begin{equation}
{\cal M}_{\rm mix}
\sim
m_{3/2}m_\eta .
\label{eq:pheno_scalar_mix_offdiag_mass_product}
\end{equation}
Equation~\eqref{eq:pheno_scalar_mix_general} is not a second universal scalar
mass formula. It is a template for model-dependent entries in the scalar mass
matrix. Its phenomenological effect depends on whether the entry is diagonal,
non-holomorphic off-diagonal, or holomorphic.

Finally, in the exact \(U(1)_R\)-symmetric limit,
\begin{equation}
m_f^{\rm Maj}=0.
\label{eq:pheno_majorana_zero}
\end{equation}
After small \(R\)-breaking, Majorana masses reappear:
\begin{equation}
m_f^{\rm Maj}
\sim
\epsilon_R^n\,m_f^{\rm ref},
\label{eq:pheno_majorana_breaking_general}
\end{equation}
where \(m_f^{\rm ref}\) is the scale of the corresponding Majorana operator.
For gauginos, for example, an anomaly-mediated contribution may be written
schematically as
\begin{equation}
M_{1/2}^{\rm Maj}
\sim
\epsilon_R^n\,
\frac{1}{16\pi^2}\,
\widetilde\alpha_f\,
F_C^{\rm eff},
\label{eq:pheno_majorana_breaking}
\end{equation}
with \(F_C^{\rm eff}\) the effective compensator \(F\)-term in the low-energy
description. If \(F_C^{\rm eff}\sim m_{3/2}\), the \(R\)-breaking parameter
must be small enough that the induced Majorana splitting does not dominate the
Dirac threshold.

\subsection*{Hierarchy in the conservative supergravity scaling}
\label{subsec:pheno_hierarchy_conservative}

We first use the conservative scaling
\eqref{eq:pheno_fermion_conservative}. Eliminating \(m_{3/2}\) in favour of
the Dirac fermion mass gives
\begin{equation}
m_{3/2}
\sim
\left[
16\pi^2\,
\frac{M_P^2\,m_f^{\rm Dirac}}{\alpha_f}
\right]^{1/3}.
\label{eq:pheno_m32_from_mf_conservative}
\end{equation}
The characteristic magnitude of the universal scalar threshold is
\begin{equation}
|m_{\phi,\,{\rm univ}}|
\sim
\frac{\sqrt{|C_{\rm univ}|}}{4\pi}\,
\frac{m_{3/2}^2}{M_P}.
\label{eq:pheno_scalar_univ_from_m32_conservative}
\end{equation}
Substituting \eqref{eq:pheno_m32_from_mf_conservative} gives
\begin{equation}
|m_{\phi,\,{\rm univ}}|
\sim
(16\pi^2)^{1/6}\,
\frac{\sqrt{|C_{\rm univ}|}}{\alpha_f^{2/3}}\,
M_P^{1/3}\,
\big(m_f^{\rm Dirac}\big)^{2/3}.
\label{eq:pheno_scalar_univ_vs_mf_conservative}
\end{equation}
Thus, in the conservative supergravity scaling, the hierarchy of magnitudes is
\begin{equation}
m_f^{\rm Dirac}
\ll
|m_{\phi,\,{\rm univ}}|
\ll
m_{3/2}
\label{eq:pheno_conservative_hierarchy}
\end{equation}
for the light fermion masses considered below.

For orientation, taking \(M_P=2.4\times10^{18}\,{\rm GeV}\) and
\(\alpha_f=|C_{\rm univ}|=1\), one finds
\begin{align}
m_{3/2}
&\simeq
9.7\times10^{12}\,{\rm GeV}\,
\left(\frac{m_f^{\rm Dirac}}{{\rm GeV}}\right)^{1/3},
\label{eq:pheno_m32_numeric_conservative}
\\[1mm]
|m_{\phi,\,{\rm univ}}|
&\simeq
3.1\times10^{6}\,{\rm GeV}\,
\left(\frac{m_f^{\rm Dirac}}{{\rm GeV}}\right)^{2/3}.
\label{eq:pheno_scalar_numeric_conservative}
\end{align}
Equivalently,
\begin{align}
m_f^{\rm Dirac}=0.05~{\rm eV}
&\quad\Longrightarrow\quad
m_{3/2}\sim 3.6\times10^9~{\rm GeV},
\qquad
|m_{\phi,\,{\rm univ}}|\sim 0.4~{\rm GeV},
\label{eq:pheno_neutrino_conservative_numbers}
\\[1mm]
m_f^{\rm Dirac}=1~{\rm keV}
&\quad\Longrightarrow\quad
m_{3/2}\sim 1.0\times10^{11}~{\rm GeV},
\qquad
|m_{\phi,\,{\rm univ}}|\sim 3\times10^2~{\rm GeV},
\label{eq:pheno_keV_conservative_numbers}
\\[1mm]
m_f^{\rm Dirac}=1~{\rm GeV}
&\quad\Longrightarrow\quad
m_{3/2}\sim 1.0\times10^{13}~{\rm GeV},
\qquad
|m_{\phi,\,{\rm univ}}|\sim 3\times10^6~{\rm GeV},
\label{eq:pheno_GeV_conservative_numbers}
\\[1mm]
m_f^{\rm Dirac}=1~{\rm TeV}
&\quad\Longrightarrow\quad
m_{3/2}\sim 1.0\times10^{14}~{\rm GeV},
\qquad
|m_{\phi,\,{\rm univ}}|\sim 3\times10^8~{\rm GeV}.
\label{eq:pheno_TeV_conservative_numbers}
\end{align}
These values are not predictions for the physical scalar spectrum. They are
benchmarks for the magnitude of the universal threshold only. They assume
conservative supergravity normalization, order-one threshold coefficients, and
no additional scalar-sector contributions.

The conservative scaling has a direct implication. A small radiative Dirac
fermion mass requires a heavy gravitino, while the associated universal scalar
threshold remains parametrically larger than the fermion mass. The scalar
sector is therefore not protected by the smallness of \(m_f^{\rm Dirac}\).
Positive diagonal contributions from stabilization, local counterterms, radion
mediation, or mixed-sector thresholds may be needed to obtain a stable scalar
spectrum.

The mixed scalar sector is the main model-dependent constraint. The same
low-energy structure that opens the Dirac fermion channel can also generate
scalar entries. These entries are not fixed by \(m_f^{\rm Dirac}\) alone.
Diagonal mixed entries can help compensate a negative universal threshold,
whereas non-holomorphic or holomorphic off-diagonal entries can lower the
lightest scalar eigenvalue. The relevant condition is positivity of the full
real scalar mass matrix.

\subsection*{Enhanced benchmark}
\label{subsec:pheno_hierarchy_enhanced}

For comparison, one may also consider the enhanced benchmark
\eqref{eq:pheno_fermion_enhanced}. Then
\begin{equation}
m_{3/2}
\sim
4\pi\,
\sqrt{\frac{M_P\,m_f^{\rm Dirac}}{\alpha_f}},
\label{eq:pheno_m32_from_mf_enhanced}
\end{equation}
and the magnitude of the universal scalar threshold becomes
\begin{equation}
|m_{\phi,\,{\rm univ}}|
\sim
4\pi\,
\frac{\sqrt{|C_{\rm univ}|}}{\alpha_f}\,
m_f^{\rm Dirac}.
\label{eq:pheno_scalar_univ_vs_mf_enhanced}
\end{equation}
For order-one coefficients this gives
\begin{equation}
|m_{\phi,\,{\rm univ}}|
\sim
10\,m_f^{\rm Dirac}.
\label{eq:pheno_loop_hierarchy_enhanced}
\end{equation}
This relation is useful as a comparison point. It is not the conservative
Scherk--Schwarz result. It follows only after assuming the stronger
companion-channel normalization in \eqref{eq:pheno_fermion_enhanced}.

\subsection*{Scalar mixing and stability}
\label{subsec:pheno_scalar_mixing_stability}

The mixed scalar sector contains several possible entries. In the notation of
Section~\ref{subsec:scalar_stability}, the quadratic potential for two complex
scalars is
\begin{equation}
V_2
=
\begin{pmatrix}
\phi^\dagger & \phi'^\dagger
\end{pmatrix}
H
\begin{pmatrix}
\phi\\
\phi'
\end{pmatrix}
+
\frac12
\left[
\begin{pmatrix}
\phi & \phi'
\end{pmatrix}
B
\begin{pmatrix}
\phi\\
\phi'
\end{pmatrix}
+\text{h.c.}
\right],
\label{eq:pheno_scalar_HB_potential}
\end{equation}
with \(H=H^\dagger\) and \(B=B^T\). Diagonal mixed terms contribute to the
diagonal entries of \(H\). Non-holomorphic scalar mixing contributes to the
off-diagonal entries of \(H\). Holomorphic \(B\)-type terms contribute to
\(B\).

A useful parametric template for a mixed scalar entry is
\begin{equation}
m_{\phi\phi'}^{2\,{\rm mix}}
\sim
\frac{1}{16\pi^2}\,
C_{\rm mix}\,
\frac{{\cal M}_{\rm mix}}{M_P^2}\,
\Lambda_{\rm mix}^2 .
\label{eq:pheno_scalar_mix_entry}
\end{equation}
Its interpretation depends on the topology. For diagonal entries,
\(m_{\phi\phi'}^{2\,{\rm mix}}\) should be read as a correction to
\(m_\phi^2\) or \(m_{\phi'}^2\). For non-holomorphic or holomorphic
off-diagonal entries, it should be read as an entry in \(H\) or \(B\),
respectively. Its size is not determined by the Dirac fermion mass alone. It
depends on which scalar propagates in the loop, whether fermion or scalar
mixing insertions are present, whether the diagram is same-boundary or
opposite-boundary, and which mass insertions appear in the numerator.

In the simpler limit in which holomorphic bilinears are absent, \(B=0\), the
relevant hermitian matrix is
\begin{equation}
H
=
\begin{pmatrix}
m_{\phi}^2 & C_{\phi\phi'} \\
C_{\phi\phi'}^\ast & m_{\phi'}^2
\end{pmatrix}.
\label{eq:pheno_scalar_matrix}
\end{equation}
The eigenvalues are
\begin{equation}
m_\pm^2
=
\frac12\Big[
m_{\phi}^2+m_{\phi'}^2
\pm
\sqrt{
\big(m_{\phi'}^2-m_{\phi}^2\big)^2
+
4|C_{\phi\phi'}|^2
}
\Big].
\label{eq:pheno_scalar_eigenvalues}
\end{equation}
Stability in this limit requires
\begin{equation}
m_{\phi}^2>0,
\qquad
m_{\phi'}^2>0,
\qquad
m_{\phi}^2m_{\phi'}^2>|C_{\phi\phi'}|^2 .
\label{eq:pheno_tachyon_condition}
\end{equation}
Thus a non-holomorphic off-diagonal entry cannot cure a negative diagonal
mass-squared by itself. It can only lower the smaller eigenvalue once the
diagonal entries are fixed.

If a holomorphic bilinear is present,
\begin{equation}
V_2
\supset
B_{\phi\phi'}\,\phi\phi'
+\text{h.c.},
\label{eq:pheno_holomorphic_B_term}
\end{equation}
the stability condition must be imposed on the full real scalar mass matrix.
For the simple case with only diagonal hermitian masses and one holomorphic
off-diagonal entry, the condition is
\begin{equation}
m_{\phi}^2>0,
\qquad
m_{\phi'}^2>0,
\qquad
m_{\phi}^2m_{\phi'}^2>|B_{\phi\phi'}|^2 .
\label{eq:pheno_holomorphic_stability_condition}
\end{equation}
Again, the holomorphic entry can destabilize the spectrum if it is too large.

The mixed sector therefore has two roles. Its diagonal entries can be necessary
to offset the negative universal Scherk--Schwarz threshold, while its
off-diagonal entries are constrained by the determinant bounds of the scalar
mass matrix. A viable model must satisfy both requirements:
\begin{equation}
m_{\phi,{\rm diag}}^2>0
\quad\text{for the relevant diagonal entries},
\qquad
\mathcal M_R^2>0
\quad\text{for the full real scalar mass matrix}.
\label{eq:pheno_scalar_stability_summary}
\end{equation}
For moduli or modulus-like fields, this quadratic condition is only the first
test; the final conclusion depends on the full stabilization potential.

\subsection*{Suppressing scalar mixing}
\label{subsec:suppressing_scalar_mixing}

The extra fermion channel can be accompanied by scalar bilinears. It is useful
to state when scalar mixing can be suppressed without eliminating the Dirac
fermion mass.

Let \(Q\) and \(\Sigma\) be chiral multiplets with scalar components
\((\phi,\phi')\) and fermions \((\psi,\xi)\). The Dirac fermion mass is
associated with
\begin{equation}
m_D\,\psi\,\xi+\text{h.c.}
\end{equation}
The scalar terms may be non-holomorphic,
\begin{equation}
C_{\rm mix}\,\phi^\dagger\phi' + \text{h.c.},
\label{eq:Cmix_nonholomorphic}
\end{equation}
or holomorphic,
\begin{equation}
B_{\rm mix}\,\phi\,\phi' + \text{h.c.}
\label{eq:Bmix_holomorphic}
\end{equation}

For \(R(Q)=r_Q\) and \(R(\Sigma)=r_\Sigma\), the scalars carry charges
\(r_Q\) and \(r_\Sigma\), while the fermions carry charges \(r_Q-1\) and
\(r_\Sigma-1\). The Dirac fermion bilinear is neutral if
\begin{equation}
(r_Q-1)+(r_\Sigma-1)=0
\qquad\Longleftrightarrow\qquad
r_Q+r_\Sigma=2.
\label{eq:Rneutrality_dirac_mass}
\end{equation}
The non-holomorphic scalar bilinear \(\phi^\dagger\phi'\) carries charge
\begin{equation}
R(\phi^\dagger\phi')=-r_Q+r_\Sigma.
\label{eq:Rcharge_nonholomorphic_scalar}
\end{equation}
Therefore
\begin{equation}
r_Q+r_\Sigma=2,
\qquad
r_Q\neq r_\Sigma,
\label{eq:R_condition_Cmix_forbidden}
\end{equation}
allows the Dirac fermion mass while forbidding the non-holomorphic scalar
mixing \eqref{eq:Cmix_nonholomorphic}.

The holomorphic bilinear \(\phi\phi'\) carries charge
\begin{equation}
R(\phi\phi')=r_Q+r_\Sigma .
\end{equation}
Thus the same condition that allows the Dirac fermion mass gives
\(R(\phi\phi')=2\). Whether a holomorphic \(B\)-type term is allowed depends on
the \(R\)-charge carried by the supersymmetry-breaking insertion or by the
coefficient multiplying the operator. The residual \(U(1)_R\) symmetry alone
does not remove all holomorphic scalar mixing.

A second control mechanism is sequestering. Same-boundary amplitudes contain
local pieces and require local counterterms. Opposite-boundary amplitudes are
finite and non-local. Localizing \(Q\) and \(\Sigma\) on opposite boundaries
therefore removes the local same-boundary contribution. It does not make the
mixing vanish, but it replaces a local threshold by a finite
boundary-to-boundary threshold.

Further suppression can come from orbifold parities, internal symmetries, or
selection rules in the higher-dimensional matter sector. These are
model-dependent and must be checked in any explicit construction.

\subsection*{Moduli, modulini, and dark matter}
\label{subsec:pheno_moduli_modulini_DM}

The neutral sector provides a natural application: moduli, modulini, and other
singlet fermions. Gauge constraints are weaker there, and the extra partner
channel required for a Dirac mass is easier to realize. At the same time,
moduli are sensitive to the full stabilization potential, so the scalar-matrix
constraint is especially important.

In the conservative scaling \eqref{eq:pheno_fermion_conservative}, a target
Dirac modulino mass
\begin{equation}
m_{\tilde m}^{\rm Dirac}\sim 1~{\rm keV}
\label{eq:pheno_modulino_keV_input}
\end{equation}
corresponds parametrically to
\begin{equation}
m_{3/2}\sim 1.0\times10^{11}~{\rm GeV},
\qquad
|m_{\phi,\,{\rm univ}}|\sim 3\times10^2~{\rm GeV},
\label{eq:pheno_modulino_keV_conservative}
\end{equation}
for order-one coefficients. In the enhanced benchmark
\eqref{eq:pheno_fermion_enhanced}, the same target mass gives instead
\begin{equation}
m_{3/2}\sim 2\times10^7~{\rm GeV},
\qquad
|m_{\phi,\,{\rm univ}}|\sim 10~{\rm keV}.
\label{eq:pheno_modulino_keV_enhanced}
\end{equation}
Here the second number is an order-of-magnitude value; for
\(\alpha_f=|C_{\rm univ}|=1\) it is \(4\pi\,{\rm keV}\). Thus the conservative
scaling has a heavier gravitino and a larger universal scalar threshold than
the enhanced benchmark for the same Dirac modulino mass.

A keV-scale Dirac modulino could be relevant for warm-dark-matter model
building. The present EFT does not determine the relic abundance, production
history, lifetime, or structure-formation constraints. It only gives a
radiative mass scale once the required neutral Dirac channel is present.

For a heavier neutral fermion,
\begin{equation}
m_{\tilde m}^{\rm Dirac}\sim 1~{\rm GeV},
\end{equation}
the conservative scaling gives
\begin{equation}
m_{3/2}\sim 1.0\times10^{13}~{\rm GeV},
\qquad
|m_{\phi,\,{\rm univ}}|\sim 3\times10^6~{\rm GeV},
\label{eq:pheno_modulino_GeV_conservative}
\end{equation}
again for order-one coefficients. This regime is more naturally associated
with hidden-sector or decaying-dark-matter phenomenology than with warm dark
matter.

In these benchmarks the gravitino is much heavier than the light neutral
fermion generated by the mechanism. The framework therefore points more
naturally to modulino- or singlet-like dark matter than to light gravitino dark
matter.

The scalar sector remains the main constraint. If the neutral Dirac channel is
present, the same structure can induce diagonal and off-diagonal scalar
entries. In a modulus sector this does not automatically exclude the
construction, because the full stabilization potential can dominate the
quadratic approximation. However, any viable model must show that the full
scalar mass matrix is non-tachyonic around the stabilized vacuum.

\subsection*{Radiative Dirac neutrinos}
\label{subsec:pheno_neutrinos}

The neutrino case illustrates the parametric consequences of a very small
Dirac mass. It also provides a natural setting in which the residual
\(U(1)_R\) can be tied to a leptonic symmetry. This is the logic used in
\(R\)-symmetric constructions where the \(R\)-symmetry is identified with
lepton number, or with a closely related leptonic symmetry
\cite{Frugiuele:2012pe}. Suppose
\begin{equation}
m_\nu^{\rm Dirac}\sim 0.05~{\rm eV}.
\label{eq:pheno_neutrino_input}
\end{equation}
In the conservative scaling \eqref{eq:pheno_fermion_conservative}, this gives
\begin{equation}
m_{3/2}\sim 3.6\times10^9~{\rm GeV},
\qquad
|m_{\phi,\,{\rm univ}}|\sim 0.4~{\rm GeV},
\label{eq:pheno_neutrino_conservative}
\end{equation}
for order-one coefficients. In the enhanced benchmark
\eqref{eq:pheno_fermion_enhanced}, one instead obtains
\begin{equation}
m_{3/2}\sim 1.4\times10^5~{\rm GeV},
\qquad
|m_{\phi,\,{\rm univ}}|\sim 0.6~{\rm eV}.
\label{eq:pheno_neutrino_enhanced}
\end{equation}
Thus the conservative scaling gives a heavier gravitino and a larger universal
scalar threshold than the enhanced benchmark. The enhanced benchmark has the
simple loop-factor relation
\(|m_{\phi,\,{\rm univ}}|\sim 4\pi m_\nu^{\rm Dirac}\), but it assumes the
stronger companion-channel normalization.

Radiative Dirac neutrino masses are possible in either scaling, but the
associated supergravity scale is different. The conservative scaling is a
high-scale scenario, while the enhanced benchmark gives a lower gravitino scale
and a smaller universal scalar threshold for the same neutrino mass.

The scalar sector is again a separate constraint. The neutrino mass does not
determine the mixed scalar entries. If the second channel couples to light
scalar directions, the induced scalar matrix must be controlled by
\(R\)-charges, localization, orbifold selection rules, additional symmetries,
or stabilization effects. A neutrino model based on this mechanism therefore
requires both
\begin{equation}
m_\nu^{\rm Dirac}\sim 0.05~{\rm eV}
\end{equation}
and a stable scalar sector.

\subsection*{What is fixed, what is assumed}
\label{subsec:pheno_generic_assumed_predictive}

The phenomenological content can be summarized as follows.

First, the universal scalar threshold is fixed within the minimal
anti-periodic Scherk--Schwarz benchmark. It follows from the ordinary gravitino
coupling and from the \(R\)-neutrality of \(|\phi|^2\). In the minimal
benchmark it is negative, and the physical spectrum requires additional
diagonal contributions or stabilization effects. In the exact \(U(1)_R\)
limit, Majorana fermion masses are absent. These statements do not require the
extra companion matter channel.

Second, the Dirac fermion mass is conditional. It requires an additional
low-energy matter channel coupled to the companion spin-\(\tfrac32\) field and
a closed scalar line in the loop. This is not a universal consequence of the
Dirac gravitino itself.

Third, the Planck scaling of the Dirac fermion mass depends on the
normalization of the companion channel. With conservative supergravity
normalization,
\begin{equation}
m_f^{\rm Dirac}
\sim
\frac{1}{16\pi^2}\,
\alpha_f\,
\frac{m_{3/2}^3}{M_P^2}.
\end{equation}
For a fixed light Dirac fermion mass, this scaling implies a heavier gravitino
and a larger universal scalar threshold than the enhanced benchmark. The
enhanced scaling \(m_{3/2}^2/M_P\) is an additional EFT assumption. 
It should therefore be used as a benchmark for stronger companion-channel
normalization, not as a prediction of the minimal Scherk--Schwarz construction.

Fourth, the mixed scalar sector is the main model-building constraint. It is
not universal, but it is tied to the same extra structure that allows the
fermion mass. Its diagonal entries can help compensate the universal scalar
threshold, while its off-diagonal entries are constrained by the positivity of
the full scalar mass matrix.

The anti-periodic Scherk--Schwarz benchmark fixes the even/odd kernels,
distinguishes same-boundary local thresholds from opposite-boundary finite
thresholds, and explains why the odd gravitino enters through the
second-supersymmetry structure. It does not fix the normalization of the
companion matter current.















\section{Small \texorpdfstring{$U(1)_R$}{U(1)R} breaking}
\label{sec:small_R_breaking}

The previous sections were organized in the exact \(U(1)_R\)-symmetric limit.
This limit is useful because it separates Dirac and Majorana structures. In a
complete quantum-gravity or string-theory embedding, however, this continuous
\(U(1)_R\) should not be interpreted as an exact global symmetry. It should be
viewed as an organizing symmetry of the supergravity limit. In an explicit
compactification it may be gauged, Higgsed, Stueckelberg-broken, reduced to a
discrete \(R\)-symmetry, or broken by non-perturbative effects.

A favorable possibility is that a discrete \(R\)-symmetry remains,
\begin{equation}
U(1)_R
\longrightarrow
\mathbb Z_N^R .
\label{eq:Rbreaking_discrete_remnant}
\end{equation}
If the discrete symmetry is gauged, it can remain exact in a quantum-gravity
embedding and can protect an approximate Dirac structure
\cite{Krauss:1988zc,Banks:1991xj,Ibanez:1991pr,Ibanez:1991hv}. This protection
is weaker than that of the continuous symmetry. Operators forbidden by the
continuous \(U(1)_R\) can reappear at higher order if enough insertions of
\(R\)-breaking fields make the operator neutral under \(\mathbb Z_N^R\). Thus
one expects operators of the schematic form
\begin{equation}
\Delta\mathcal L
\supset
\left(\frac{X}{M}\right)^n
\mathcal O_{\Delta R}
+\text{h.c.},
\label{eq:Rbreaking_higher_dim_operator}
\end{equation}
with
\begin{equation}
\epsilon_R^n
\equiv
\left(\frac{\langle X\rangle}{M}\right)^n .
\label{eq:Rbreaking_epsilon_definition}
\end{equation}
The integer \(n\) is fixed by the charge assignments under the residual
discrete symmetry.

The Dirac gravitino then becomes pseudo-Dirac. In the light
spin-\(\tfrac32\) sector the mass matrix can be written schematically as
\begin{equation}
\mathcal M_{3/2}
=
\begin{pmatrix}
\delta m_1 & m_D \\
m_D & \delta m_2
\end{pmatrix},
\label{eq:pseudo_dirac_gravitino_matrix}
\end{equation}
where \(m_D\) is the \(U(1)_R\)-preserving Dirac entry and
\(\delta m_{1,2}\) are \(R\)-breaking Majorana entries. The pseudo-Dirac
description is valid when
\begin{equation}
|\delta m_{1,2}|\ll m_D .
\label{eq:pseudo_dirac_gravitino_condition}
\end{equation}
The same statement applies to visible or hidden fermions. Once \(R\)-breaking
is present, Majorana masses can be generated with the parametric form
\begin{equation}
m_f^{\rm Maj}
\sim
\epsilon_R^n\,m_f^{\rm ref},
\label{eq:Rbreaking_majorana_generic}
\end{equation}
where \(m_f^{\rm ref}\) is the scale of the operator that becomes allowed after
\(R\)-breaking.

Anomaly mediation provides a useful reference point because it generates
Majorana gaugino masses in an ordinary \(\mathcal N=1\) description
\cite{Randall:1998uk,Giudice:1998xp}. For gauginos we use the conventional
notation \(M_{1/2}\), whereas for generic matter fermions and modulini we write
\(m_f^{\rm Maj}\). Denoting the effective compensator \(F\)-term by
\(F_C^{\rm eff}\), the standard gaugino contribution is
\begin{equation}
M_a^{\rm AMSB}
=
\frac{\beta_{g_a}}{g_a}\,
F_C^{\rm eff}.
\label{eq:AMSB_gaugino_standard}
\end{equation}
Since \(\beta_{g_a}/g_a\sim {\cal O}(1/16\pi^2)\), this contribution is
parametrically
\begin{equation}
M_a^{\rm AMSB}
\sim
\frac{1}{16\pi^2}\,
\widetilde\alpha_a\,
F_C^{\rm eff},
\label{eq:AMSB_gaugino_parametric}
\end{equation}
with \(\widetilde\alpha_a\) an order-one loop coefficient. This is a Majorana
mass. It is absent in the exact \(U(1)_R\) limit, and with controlled
\(R\)-breaking it is multiplied by the appropriate \(R\)-breaking insertion:
\begin{equation}
M_a^{\rm AMSB}
\sim
\epsilon_R^n\,
\frac{1}{16\pi^2}\,
\widetilde\alpha_a\,
F_C^{\rm eff}.
\label{eq:AMSB_gaugino_Rbreaking}
\end{equation}
Thus anomaly mediation is not a source of Dirac masses in this setup. It is a
source of pseudo-Dirac splitting. We do not attempt to compute \(F_C^{\rm eff}\) in a stabilized radion sector.
The formulae involving anomaly mediation should therefore be read as parametric
bounds for a given low-energy value of \(F_C^{\rm eff}\), not as predictions of
the Scherk--Schwarz benchmark before stabilization.

The physical Dirac mass should be written schematically as
\begin{equation}
m_f^{\rm Dirac,phys}
=
m_f^{\rm DG}
+
m_f^{\rm Dirac,match}
+
\delta_{\rm RG}m_f^{\rm Dirac}
+\cdots ,
\label{eq:physical_dirac_mass_with_matching}
\end{equation}
where \(m_f^{\rm DG}\) denotes the Dirac-gravitino threshold computed in the
previous sections. The term \(m_f^{\rm Dirac,match}\) denotes possible
threshold contributions generated at the matching scale by additional
\(R\)-symmetric microscopic interactions, while
\(\delta_{\rm RG}m_f^{\rm Dirac}\) denotes ordinary running and finite
low-energy threshold corrections. The dangerous Majorana splitting is
\begin{equation}
m_f^{\rm Maj,AMSB}
\sim
\epsilon_R^n\,
\frac{1}{16\pi^2}\,
\widetilde\alpha_f\,
F_C^{\rm eff}.
\label{eq:AMSB_majorana_splitting}
\end{equation}
Approximate Dirac behavior requires
\begin{equation}
m_f^{\rm Maj,AMSB}\ll m_f^{\rm Dirac,phys}.
\label{eq:AMSB_majorana_bound}
\end{equation}
If the Dirac-gravitino threshold is the dominant contribution to the Dirac
mass, this reduces to
\begin{equation}
m_f^{\rm Maj,AMSB}\ll m_f^{\rm DG}.
\label{eq:AMSB_majorana_bound_DG_dominated}
\end{equation}

For conservative supergravity normalization,
\begin{equation}
m_f^{\rm DG}
\sim
\frac{1}{16\pi^2}\,
\alpha_f\,
\frac{m_{3/2}^3}{M_P^2}.
\label{eq:DG_threshold_conservative_reminder}
\end{equation}
If \(F_C^{\rm eff}\) is of order \(m_{3/2}\), the condition
\eqref{eq:AMSB_majorana_bound} becomes
\begin{equation}
\epsilon_R^n
\ll
\frac{\alpha_f}{\widetilde\alpha_f}\,
\frac{m_{3/2}^2}{M_P^2}.
\label{eq:AMSB_Rbreaking_bound_conservative}
\end{equation}
This is a strong bound. In a stronger companion-channel benchmark with
\begin{equation}
m_f^{\rm DG}
\sim
\frac{1}{16\pi^2}\,
\alpha_f\,
\frac{m_{3/2}^2}{M_P},
\label{eq:DG_threshold_enhanced_local}
\end{equation}
the corresponding bound is
\begin{equation}
\epsilon_R^n
\ll
\frac{\alpha_f}{\widetilde\alpha_f}\,
\frac{m_{3/2}}{M_P}.
\label{eq:AMSB_Rbreaking_bound_enhanced}
\end{equation}
The allowed amount of \(R\)-breaking therefore depends directly on the
normalization of the Dirac threshold. In sequestered or no-scale
Scherk--Schwarz realizations, \(F_C^{\rm eff}\) can be suppressed or replaced
by model-dependent radion and threshold contributions. In that case the bound
must be evaluated in the explicit stabilization model.

The scalar sector is less protected. Even in the exact \(U(1)_R\) limit,
diagonal scalar masses are allowed. Anomaly-mediated scalar masses are also
allowed:
\begin{equation}
m_i^{2\,{\rm AMSB}}
=
-\frac14 |F_C^{\rm eff}|^2
\frac{d\gamma_i}{d\ln\mu},
\label{eq:AMSB_scalar_mass}
\end{equation}
up to convention-dependent normalizations. The physical scalar masses therefore
have the schematic form
\begin{equation}
m_i^2
=
m_{i,{\rm DG}}^2
+
m_i^{2\,{\rm AMSB}}
+
m_i^{2\,{\rm radion}}
+
m_i^{2\,{\rm local}}
+\cdots .
\label{eq:physical_scalar_mass_with_AMSB}
\end{equation}
Their sign and size are determined by the complete scalar potential and by
local threshold terms, not by the \(R\)-symmetry alone.

The Higgs sector provides an independent reason to allow controlled
\(R\)-breaking, or to enlarge the Higgs sector. In Dirac-gaugino and
\(R\)-symmetric models, the minimal MSSM Higgs sector is too restrictive if the
\(R\)-symmetry is kept exact. One possibility is to work with an extended Higgs
sector, as in the \(R\)-symmetric MSSM and related Dirac-gaugino constructions
\cite{Kribs:2007ac,Benakli:2011vb}. Another possibility is to use the extended
supersymmetric structure itself to obtain Higgs alignment, or alignment without
decoupling, in models closely related to Dirac-gaugino setups
\cite{Benakli:2018vqz,Benakli:2018vjk,Benakli:2018ldd,Benakli:2019yaq}.
A further possibility is to introduce controlled \(R\)-breaking in the Higgs
sector. A standard singlet example contains
\begin{equation}
W\supset \frac{\kappa}{3}S^3 ,
\label{eq:Higgs_Rbreaking_kappaS3}
\end{equation}
which gives schematically
\begin{equation}
\mu_{\rm eff}\sim \lambda_S\langle S\rangle,
\qquad
B\mu_{\rm eff}\sim \kappa\,\lambda_S\langle S\rangle^2 .
\label{eq:Higgs_mu_Bmu_schematic}
\end{equation}
The same \(R\)-breaking source can generate Majorana masses. Controlled
\(R\)-breaking may therefore be useful for the Higgs sector, but it must remain
small enough not to erase the pseudo-Dirac hierarchy.

If no useful discrete remnant survives, Majorana masses and scalar mixings are
allowed subject only to locality, sequestering, gauge symmetries, or accidental
small coefficients. In that case the Dirac limit is not symmetry-protected.

Thus the relevant question in a string embedding is not whether the continuous
\(U(1)_R\) is exact. It is whether the compactification leaves a sufficiently
strong discrete \(R\)-symmetry, or sufficiently small \(R\)-breaking effects,
to preserve the pseudo-Dirac hierarchy while keeping the Higgs and scalar
sectors under control.

\subsection*{A small departure from \texorpdfstring{$\omega=\tfrac12$}{omega=1/2}}
\label{subsec:pheno_near_half_twist}

The point \(\omega=\tfrac12\) is special because the two lightest
spin-\(\tfrac32\) states are degenerate and the low-energy gravitino sector is
naturally organized as a Dirac gravitino. In the pure boundary-condition
problem, the effective \(U(1)_R\) forbids diagonal Majorana masses. We now
describe a nearby regime in which this selection rule is weakly broken.
 
We parameterize the departure from the Dirac point by
\begin{equation}
\omega=\frac12-\epsilon_\omega,
\qquad
|\epsilon_\omega|\ll1 .
\label{eq:omega_near_half_definition}
\end{equation}
The parameter \(\epsilon_\omega\) captures only the part of the localized
\(R\)-breaking that is encoded in the effective gravitino boundary condition.
Additional localized fermionic degrees of freedom, such as pseudo-Goldstinos,
are not included in this parametrization.

A schematic localized \(R\)-breaking gravitino mass term has the form
\begin{equation}
{\cal L}_{\rm brane}
\supset
-\frac12\,\delta(y-y_i)\,
M_i^{(R)}\,\psi_\mu\sigma^{\mu\nu}\psi_\nu
+\text{h.c.},
\label{eq:brane_Rbreaking_gravitino_mass}
\end{equation}
with notation adapted to four-dimensional Weyl gravitini. Such localized terms
modify the gravitino boundary conditions and may be described as boundary jump
conditions or, in the restricted Scherk--Schwarz-type case, as a generalized
twist
\cite{Bagger:2001qi,Bagger:2001ep,Gherghetta:2001sa,Meissner:2002dg,
Biggio:2002rb,Rattazzi:2003rj,Benakli:2007zza}.

At tree level, this deformation turns the light spin-\(\tfrac32\) sector into
a pseudo-Dirac system. One may write
\begin{equation}
{\cal M}_{3/2}
=
\begin{pmatrix}
\delta m_L & m_D \\
m_D & \delta m_R
\end{pmatrix},
\qquad
m_D=\frac{1}{2R},
\label{eq:near_half_gravitino_mass_matrix}
\end{equation}
where \(\delta m_L\) and \(\delta m_R\) are induced by localized
\(R\)-breaking. Equivalently, in the effective-twist description the two light
masses are
\begin{equation}
m_1=\frac{\omega}{R}
=
\frac{1}{2R}-\frac{\epsilon_\omega}{R},
\qquad
m_2=\frac{1-\omega}{R}
=
\frac{1}{2R}+\frac{\epsilon_\omega}{R}.
\label{eq:near_half_gravitino_masses}
\end{equation}
Hence
\begin{equation}
m_D=\frac{m_1+m_2}{2}=\frac{1}{2R},
\qquad
\Delta m_{3/2}=m_2-m_1=\frac{2\epsilon_\omega}{R}.
\label{eq:near_half_pseudo_dirac_split}
\end{equation}
Thus \(\epsilon_\omega\) measures the Majorana splitting of the would-be Dirac
gravitino within this effective-twist description.

This description has a limited scope. A localized supersymmetry-breaking
sector is not, in general, equivalent to a shifted Scherk--Schwarz twist. In
particular, a boundary \(F\)-term sector can contain localized would-be
Goldstinos and leave pseudo-Goldstino combinations in the low-energy spectrum.
These additional fermions and their mixings are not encoded in
\(\epsilon_\omega\).

The localized \(R\)-breaking source can also couple directly to matter on the
same boundary. For fields localized on \(y=y_i\), one may have
\begin{equation}
{\cal L}_{\rm brane}
\supset
\delta(y-y_i)
\left[
\frac12\,M_{ab}^{(R)}\,\lambda^a\lambda^b
+
\frac12\,M_{IJ}^{(R)}\,\chi^I\chi^J
+
B_{ij}^{(R)}\,\phi^i\phi^j
+
A_{ijk}^{(R)}\,\phi^i\phi^j\phi^k
+\text{h.c.}
\right],
\label{eq:brane_direct_matter_Rbreaking}
\end{equation}
whenever the corresponding operators are gauge invariant and compatible with
the remaining symmetries. Here \(\lambda^a\) are gauginos and \(\chi^I\) denote
gauge-singlet or vector-like fermions for which a localized Majorana mass is
allowed. If such operators are present with unsuppressed coefficients, they are
the leading \(R\)-breaking effects in the matter sector.

We now isolate the sequestered limit in which the departure from
\(\omega=\tfrac12\) is transmitted to matter only through the gravitational
sector. In this limit we set to zero the non-universal sources which could
otherwise dominate the matter spectrum: companion-channel matter couplings,
unsuppressed direct boundary \(R\)-breaking matter operators, and mixings
between the matter fermion under study and light pseudo-Goldstino modes. Under
these assumptions, the leading induced fermion masses for localized matter are
Majorana masses transmitted by gravitational loops.

The first condition removes the non-universal radiative Dirac masses discussed
in Section~\ref{sec:fermion_masses}. The second excludes direct localized
matter masses of the form \eqref{eq:brane_direct_matter_Rbreaking}. The third
excludes additional neutral-fermion mixing effects from the localized
supersymmetry-breaking sector. Under these assumptions, the leading induced
fermion masses for localized matter are Majorana masses transmitted by
gravitational loops.

For brane fields, the gravitationally transmitted Majorana gaugino mass has the
form
\begin{equation}
M_{1/2}^{\rm grav}(\omega)
=
\frac{1}{R}\,
\frac{1}{(M_P R)^2}\,
\frac{1}{(4\pi)^2}\,
f_{1/2}(\omega),
\label{eq:near_half_gaugino_mass_general}
\end{equation}
with
\begin{equation}
f_{1/2}(\omega)
=
\frac{3i}{8\pi^3}
\left[
{\rm Li}_4(e^{-2\pi i\omega})
-
{\rm Li}_4(e^{2\pi i\omega})
\right].
\label{eq:near_half_fhalf_function}
\end{equation}
At the Dirac point,
\begin{equation}
f_{1/2}\!\left(\frac12\right)=0,
\label{eq:fhalf_half_zero}
\end{equation}
as required by the residual \(U(1)_R\). Expanding around
\(\omega=\tfrac12-\epsilon_\omega\), one obtains
\begin{equation}
f_{1/2}\!\left(\frac12-\epsilon_\omega\right)
=
\frac{9\,\zeta(3)}{8\pi^2}\,
\epsilon_\omega
+
{\cal O}(\epsilon_\omega^3).
\label{eq:fhalf_near_half_expansion}
\end{equation}
Therefore
\begin{equation}
M_{1/2}^{\rm grav}
\simeq
\frac{9\,\zeta(3)}{8\pi^2}\,
\frac{\epsilon_\omega}{(4\pi)^2}\,
\frac{1}{R(M_P R)^2}.
\label{eq:near_half_gaugino_mass_expanded}
\end{equation}
Equivalently, with \(m_{3/2}\simeq 1/(2R)\),
\begin{equation}
M_{1/2}^{\rm grav}
\sim
\epsilon_\omega\,
\frac{1}{16\pi^2}\,
\frac{m_{3/2}^3}{M_P^2},
\label{eq:near_half_gaugino_mass_scaling}
\end{equation}
up to a numerical coefficient.

The scalar sector has a different expansion. The universal scalar mass does not
vanish at \(\omega=\tfrac12\). It is controlled by
\begin{equation}
f_0(\omega)
=
\frac{3}{2\pi^4}
\left[
2\zeta(5)
-
{\rm Li}_5(e^{2\pi i\omega})
-
{\rm Li}_5(e^{-2\pi i\omega})
\right].
\label{eq:near_half_f0_function}
\end{equation}
At the anti-periodic point,
\begin{equation}
f_0\!\left(\frac12\right)
=
\frac{93}{16\pi^4}\,\zeta(5),
\label{eq:f0_half_value}
\end{equation}
and the first correction is quadratic:
\begin{equation}
f_0\!\left(\frac12-\epsilon_\omega\right)
=
f_0\!\left(\frac12\right)
+
{\cal O}(\epsilon_\omega^2).
\label{eq:f0_near_half_expansion}
\end{equation}
Thus a small localized \(R\)-breaking source generates a tree-level
pseudo-Dirac splitting in the gravitino sector. In the sequestered
gravitational limit described above, the same source induces Majorana masses
for localized fermions only through gravitational loops, and therefore only at
order \(\epsilon_\omega\). The universal scalar masses are unchanged at first
order. Majorana masses for localized fermions at order \(\epsilon_\omega\). The
universal scalar masses are unchanged at first order.

The resulting parametric pattern is
\begin{equation}
m_{\rm scalar}^2
\sim
\frac{1}{16\pi^2}\,
\frac{m_{3/2}^4}{M_P^2},
\qquad
M_{1/2}^{\rm grav}
\sim
\epsilon_\omega\,
\frac{1}{16\pi^2}\,
\frac{m_{3/2}^3}{M_P^2}.
\label{eq:near_half_split_soft_pattern}
\end{equation}
Equivalently,
\begin{equation}
|m_{\rm scalar}|
\sim
\frac{1}{4\pi}\,
\frac{m_{3/2}^2}{M_P},
\qquad
\frac{M_{1/2}^{\rm grav}}{|m_{\rm scalar}|}
\sim
\frac{\epsilon_\omega}{4\pi}\,
\frac{m_{3/2}}{M_P},
\label{eq:near_half_gaugino_scalar_hierarchy}
\end{equation}
up to numerical threshold coefficients and the sign of the scalar mass-squared.
The Majorana gaugino mass is suppressed both by the small \(R\)-breaking
displacement and by the gravitational ratio \(m_{3/2}/M_P\).

In an MSSM-like sector, and in the absence of direct localized \(R\)-breaking
matter operators, this gives
\begin{equation}
|m_{\rm scalar}|
\sim
\frac{1}{4\pi}\frac{m_{3/2}^2}{M_P},
\qquad
M_{1/2}
\sim
\epsilon_\omega\,
\frac{1}{16\pi^2}
\frac{m_{3/2}^3}{M_P^2},
\qquad
\frac{M_{1/2}}{|m_{\rm scalar}|}
\sim
\frac{\epsilon_\omega}{4\pi}\frac{m_{3/2}}{M_P}.
\label{eq:near_half_MSSM_like_pattern}
\end{equation}
If direct localized gaugino masses are allowed, the leading estimate is instead
\begin{equation}
M_{1/2}\sim M_{ab}^{(R)},
\label{eq:near_half_direct_gaugino_mass}
\end{equation}
with no loop or volume suppression beyond that already present in the
microscopic coefficient \(M_{ab}^{(R)}\).

The same logic applies to gauge-singlet fermions such as modulini. If a
modulino Majorana mass is forbidden at \(\omega=\tfrac12\), then a small
localized \(R\)-breaking deformation can generate, through indirect
gravitational transmission,
\begin{equation}
m_{\tilde m}^{\rm Maj}
\sim
\epsilon_\omega\,
\frac{1}{16\pi^2}
\frac{m_{3/2}^3}{M_P^2},
\label{eq:near_half_modulino_majorana}
\end{equation}
up to model-dependent coefficients. If the localized \(R\)-breaking source
couples directly to the modulino, the leading contribution is instead
\begin{equation}
m_{\tilde m}^{\rm Maj}\sim M_{\tilde m}^{(R)}.
\label{eq:near_half_direct_modulino_mass}
\end{equation}
If the boundary supersymmetry-breaking sector contains a light
pseudo-Goldstino, there can also be neutral-fermion mixing,
\begin{equation}
{\cal L}_{\rm mix}
\supset
-\,\mu_{\tilde m G}\,\tilde m\,G_{\rm pseudo}
+\text{h.c.},
\label{eq:near_half_modulino_pseudogoldstino_mixing}
\end{equation}
which must be diagonalized together with the modulino mass matrix. In that
case \eqref{eq:near_half_modulino_majorana} is only the indirect gravitational
contribution.

The near-\(\omega=\tfrac12\) regime is therefore distinct from the
non-universal companion-channel mechanism for radiative Dirac masses. It uses a
small localized \(R\)-breaking deformation of the gravitino sector to generate
a pseudo-Dirac gravitino. In the sequestered limit, the same deformation is
communicated to localized matter only through gravitational loops, producing
small Majorana masses suppressed by \(\epsilon_\omega\) and by
\(m_{3/2}/M_P\).

\subsection*{Sequestered moduli, light modulini, and the role of \(R\)-breaking}
\label{subsec:sequestered_moduli_modulini_Rbreaking}

We now consider a restricted application which is distinct from the standard
bulk-gaugino case. If gauge multiplets propagate in the bulk, the
Scherk--Schwarz twist gives tree-level gaugino masses of order \(1/R\). This
case has been studied extensively and is not the one considered here. Instead
we consider a neutral multiplet
\begin{equation}
T=(t,\tilde t,F_T)
\end{equation}
localized on one boundary, while the visible sector may receive its dominant
soft terms from a supersymmetry-breaking source localized on the other
boundary. This is in the same spirit as extra-dimensional sequestering or shining
mechanisms, where a distant brane source communicates a hierarchically small
supersymmetry-breaking effect through bulk propagation
\cite{ArkaniHamed:1999dc}.

Let
\begin{equation}
M_c\equiv \frac1R .
\end{equation}
A localized supersymmetry-breaking source on the visible boundary shifts the
effective Scherk--Schwarz angle. Near the anti-periodic point we write
\begin{equation}
\omega_{\rm eff}
=
\frac12-\epsilon_\omega .
\end{equation}
If the localized source gives an ordinary four-dimensional gravitino mass
contribution of order
\begin{equation}
\delta m_{3/2}^{\rm bdry}
\sim
\frac{F_\pi}{M_P},
\end{equation}
then the induced displacement is parametrically
\begin{equation}
\epsilon_\omega
\sim
\frac{\delta m_{3/2}^{\rm bdry}}{M_c}.
\label{eq:moduli_epsilon_from_boundary}
\end{equation}
If the same boundary source gives visible soft masses
\begin{equation}
m_{\rm soft}^{(\pi)}\sim \frac{F_\pi}{M_P},
\end{equation}
then
\begin{equation}
\epsilon_\omega
\sim
\frac{m_{\rm soft}^{(\pi)}}{M_c}.
\label{eq:moduli_epsilon_soft}
\end{equation}
Thus a TeV-scale boundary breaking gives a small departure from
\(\omega=\tfrac12\) whenever the compactification scale is high.

The scalar \(t\) localized on the distant boundary receives the usual
gravitational brane-to-brane threshold. Parametrically,
\begin{equation}
m_t^2
\sim
\frac{C_T}{16\pi^2}\,
\frac{M_c^4}{M_P^2},
\qquad
|m_t|
\sim
\frac{\sqrt{|C_T|}}{4\pi}\,
\frac{M_c^2}{M_P}.
\label{eq:modulus_scalar_mass}
\end{equation}
Near \(\omega=\tfrac12\) this result is unchanged at first order in
\(\epsilon_\omega\), because the scalar threshold is even around the
anti-periodic point. The sign of \(C_T\) is model-dependent; in the minimal
universal contribution discussed above it is negative.

The fermion \(\tilde t\) is more protected. At the exact anti-periodic point,
the residual \(U(1)_R\) forbids a localized Majorana mass,
\begin{equation}
m_{\tilde t}^{\rm Maj}=0
\qquad
(\omega=\tfrac12).
\label{eq:modulino_majorana_zero}
\end{equation}
A small localized \(R\)-breaking displacement allows such a mass. If the
localized modulus communicates with the breaking only through gravitational
loops, the estimate is
\begin{equation}
m_{\tilde t}^{\rm Maj}
\sim
\epsilon_\omega\,
\frac{c_T}{16\pi^2}\,
\frac{M_c^3}{M_P^2}.
\label{eq:modulino_majorana_epsilon}
\end{equation}
Using \eqref{eq:moduli_epsilon_soft}, this becomes
\begin{equation}
m_{\tilde t}^{\rm Maj}
\sim
\frac{c_T}{16\pi^2}\,
m_{\rm soft}^{(\pi)}
\frac{M_c^2}{M_P^2}.
\label{eq:modulino_majorana_soft}
\end{equation}
Numerically,
\begin{equation}
m_{\tilde t}^{\rm Maj}
\sim
10^{-5}\,{\rm eV}\,
c_T
\left(\frac{m_{\rm soft}^{(\pi)}}{1\,{\rm TeV}}\right)
\left(\frac{M_c}{10^{11}\,{\rm GeV}}\right)^2 .
\label{eq:modulino_majorana_numerical}
\end{equation}
Thus the \(R\)-breaking Majorana mass of a sequestered modulino can be very
small.

The hierarchy with respect to the scalar modulus is
\begin{equation}
\frac{m_{\tilde t}^{\rm Maj}}{|m_t|}
\sim
\frac{c_T}{\sqrt{|C_T|}}\,
\frac{\epsilon_\omega}{4\pi}\,
\frac{M_c}{M_P}
\sim
\frac{c_T}{\sqrt{|C_T|}}\,
\frac{m_{\rm soft}^{(\pi)}}{4\pi M_P}.
\label{eq:modulino_scalar_ratio}
\end{equation}
After using \(\epsilon_\omega\sim m_{\rm soft}^{(\pi)}/M_c\), the ratio is
independent of \(M_c\). For TeV-scale visible soft terms it is of order
\(10^{-16}\)--\(10^{-17}\), up to the coefficients \(c_T\) and \(C_T\).

This estimate has two consequences. First, a localized modulino can remain much
lighter than its scalar partner. This is consistent with the assumptions above:
supersymmetry is already broken, and the residual \(U(1)_R\) protects the
fermion more strongly than the scalar. Second, whether such a light fermion is
phenomenologically acceptable is a separate question. If the modulino is stable
or very long lived, its cosmological relevance depends on its abundance, its
temperature relative to the visible sector, and its couplings to ordinary
matter. A very light neutral fermion can contribute to dark radiation or to the
hot/warm component of the cosmic energy density if it is thermally populated.
If it is never efficiently produced, or if it belongs to a sufficiently
sequestered hidden sector, the bound can be much weaker.

If the \(R\)-breaking Majorana mass
\eqref{eq:modulino_majorana_epsilon} is too small for the desired
phenomenology, one must add a different source of modulino mass. The natural
option within the present framework is the non-universal Dirac channel studied
in Section~\ref{sec:fermion_masses}. This requires additional structure: a
second neutral fermion \(\tilde t'\), an \(R\)-neutral bilinear
\begin{equation}
R(\tilde t)+R(\tilde t')=0,
\end{equation}
and a companion matter current coupled to the second spin-\(\tfrac32\) channel.
If these ingredients are present, a Dirac modulino mass can be generated. With
pure supergravity normalization one expects
\begin{equation}
m_{\tilde t}^{\rm Dirac}
\sim
\frac{\alpha_T}{16\pi^2}\,
\frac{M_c^3}{M_P^2},
\label{eq:modulino_dirac_conservative}
\end{equation}
whereas a larger value,
\begin{equation}
m_{\tilde t}^{\rm Dirac}
\sim
\frac{\alpha_T}{16\pi^2}\,
\frac{M_c^2}{M_P},
\label{eq:modulino_dirac_optimistic}
\end{equation}
requires a stronger effective companion-channel coupling and is not the minimal
supergravity prediction.

Thus the minimal sequestered setup gives an \(R\)-breaking Majorana mass
controlled by \(\epsilon_\omega\). The Dirac mass is a further model-building
option, not a universal effect. It becomes relevant if the modulino generated
by \eqref{eq:modulino_majorana_epsilon} is too light, or if one wants a
neutral fermion with a mass in a phenomenologically preferred range.















\section{Conclusion}
\label{sec:conclusion}

We have studied the conditions under which a four-dimensional low-energy theory
contains a Dirac gravitino, and the consequences for the matter sector. The
central point is that degeneracy of two massive spin-\(\tfrac32\) fields is not
sufficient. A Dirac interpretation requires a symmetry which distinguishes the
two fields and forbids diagonal Majorana entries. In the constructions
considered in this paper this role is played by a residual \(U(1)_R\).

In four-dimensional \(\mathcal N=2\) gauged supergravity, this structure is
obtained when the gravitino shift
\begin{equation}
P_\Lambda^a L^\Lambda
\end{equation}
is aligned along a single direction in \(SU(2)_R\) space. The alignment leaves a
\(U(1)_R\) subgroup unbroken in the spin-\(\tfrac32\) sector. In a basis
adapted to this subgroup, the two gravitini have opposite \(U(1)_R\) charges;
the off-diagonal Dirac mass is allowed, while diagonal Majorana masses are
forbidden. We described this structure in two simple settings: a vector-sector
Fayet--Iliopoulos gauging and a hypermultiplet gauging. These examples were not
used as complete vacuum constructions, but as local supergravity realizations
of the aligned shift and of the residual \(U(1)_R\).

The same organization appears in Scherk--Schwarz compactification on
\(S^1/\mathbb Z_2\). At the anti-periodic point
\begin{equation}
\omega=\frac12 ,
\end{equation}
the two lightest spin-\(\tfrac32\) levels are degenerate, with
\begin{equation}
m_{3/2}=\frac{1}{2R}.
\end{equation}
The low-energy spin-\(\tfrac32\) sector must then be kept as a pair and is
organized as a Dirac gravitino
\cite{Antoniadis:2004dt,Benakli:2011vb,Benakli:2014daa}. The orbifold fixes
the boundary selection rules. On a given boundary the even gravitino has a
direct boundary coupling, while the odd gravitino vanishes there and can enter
only through normal-derivative couplings, bulk propagation, or the
second-supersymmetry channel.

We also formulated the low-energy description in \(\mathcal N=1\) language.
This language is appropriate for the matter sector and for the scalar geometry.
The spin-\(\tfrac32\) sector is more constrained. There is no local
gauge-invariant \(\mathcal N=1\) superspace bilinear, analogous to the
Dirac-gaugino operator, which generates the required Dirac gravitino mixing.
The superspace description requires projection onto the irreducible transverse
superspin-\(\tfrac32\) sector. The resulting expression is non-local in
superspace because the projector contains inverse powers of \(\Box\). This is a
non-locality of the superspace representation of the projected sector, not a
non-locality of the component interaction restricted to physical
spin-\(\tfrac32\) modes.

The matter couplings separate into universal and non-universal parts. The
ordinary gravitino couples universally to the matter supercurrent. In the
Scherk--Schwarz benchmark this is the direct boundary coupling to the even
wavefunction. The companion spin-\(\tfrac32\) channel is different. Its
coupling to matter is not fixed by the existence of the Dirac gravitino alone.
In a complete \(\mathcal N=2\) or five-dimensional model it must be derived
from the underlying matter sector. In a more general EFT it is an additional
low-energy datum, constrained by Lorentz invariance, gauge invariance,
dimensional analysis, and the residual \(U(1)_R\).

This distinction controls the loop effects. The scalar sector contains a
universal contribution because \(\phi^\dagger\phi\) is neutral under the
residual \(U(1)_R\), and because the ordinary supercurrent coupling is always
present. In the anti-periodic Scherk--Schwarz benchmark the universal threshold
has the parametric form
\begin{equation}
m_{\phi,{\rm univ}}^2
\sim
\frac{1}{16\pi^2}\,
\frac{m_{3/2}^4}{M_P^2},
\end{equation}
with a negative sign for the minimal contribution in our conventions. The
coefficient is the standard Scherk--Schwarz boundary-scalar threshold
\cite{Gherghetta:2001sa,Rattazzi:2003rj,Antoniadis:2015chx}. It is fixed by
the full locally supersymmetric boundary coupling, including the contact terms
required by local supersymmetry, and not by an isolated cubic exchange graph.

Mixed scalar terms have a different status. They require the companion-channel
matter current and are therefore not fixed by the ordinary supergravity
coupling. Depending on the matter topology and on possible mass insertions,
they can contribute to diagonal scalar entries, non-holomorphic off-diagonal
entries, or holomorphic \(B\)-type bilinears. The scalar constraint is
therefore the positivity of the full real scalar mass matrix, not the sign or
size of the universal threshold alone.

The fermion sector is more restrictive. In the exact \(U(1)_R\) limit,
Majorana masses are absent. A Dirac fermion mass requires an \(R\)-neutral
bilinear, a second matter current coupled to the companion spin-\(\tfrac32\)
field, and a closed internal matter topology. If these ingredients are absent,
the one-loop Dirac mass is absent. With conservative supergravity
normalization, two explicit spin-\(\tfrac32\)-matter vertices give
\begin{equation}
m_f^{\rm Dirac}
\sim
\frac{1}{16\pi^2}\,
\frac{m_{3/2}^3}{M_P^2},
\end{equation}
up to dimensionless coefficients and localization factors. A larger scaling,
such as \(m_{3/2}^2/M_P\), can be used as an EFT benchmark only if the
companion-channel normalization is stronger than the conservative
supergravity one. It is not fixed by the Scherk--Schwarz kernels alone.

The companion-channel normalization is therefore not a prediction of the
Dirac-gravitino structure alone. In the minimal supergravity interpretation we
use the conservative Planck-suppressed normalization. Stronger normalizations
should be regarded as EFT benchmarks and require an explicit microscopic
origin.

The calculation separates the data fixed by symmetry and geometry from the
data which are model-dependent. The four-dimensional spin-\(\tfrac32\) tensor
algebra fixes the numerator structures. In particular, the representative
fermion self-energy gives
\begin{equation}
c_F=\frac23 .
\end{equation}
The Scherk--Schwarz geometry fixes the even and odd Green functions, the
same-boundary and opposite-boundary kernels, and the finite non-local threshold
functions. It does not fix the existence or the normalization of the companion
matter current. This current is the main model-dependent input in the matter
sector.

We also considered a small displacement from the anti-periodic point,
\begin{equation}
\omega=\frac12-\epsilon_\omega .
\end{equation}
This parameterizes the part of localized \(R\)-breaking which can be encoded in
an effective gravitino boundary condition. In this effective-twist description,
the would-be Dirac gravitino becomes pseudo-Dirac, with splitting
\begin{equation}
\Delta m_{3/2}=\frac{2\epsilon_\omega}{R}.
\end{equation}
This description does not cover a general boundary supersymmetry-breaking
sector, which may contain localized Goldstino or pseudo-Goldstino degrees of
freedom. In the sequestered limit in which direct localized \(R\)-breaking
matter operators and pseudo-Goldstino mixings are absent, the induced Majorana
gaugino masses are transmitted only through gravitational loops and scale as
\begin{equation}
M_{1/2}^{\rm grav}
\sim
\epsilon_\omega\,
\frac{1}{16\pi^2}\,
\frac{m_{3/2}^3}{M_P^2}.
\end{equation}
The universal scalar mass is unchanged at first order in \(\epsilon_\omega\).
The near-\(\omega=\tfrac12\) regime is therefore a pseudo-Dirac limit distinct
from the companion-channel mechanism for radiative Dirac masses.

The phenomenological implications follow from these separations. The universal
scalar threshold is present in the anti-periodic benchmark, but by itself it
does not determine the physical scalar spectrum. Dirac fermion masses are
conditional on additional \(R\)-compatible matter structure. Majorana masses
are forbidden in the exact \(U(1)_R\) limit, but reappear when \(R\)-breaking
moves the theory away from the Dirac point. The mixed scalar sector is the main
model-dependent constraint, because the same companion-channel structure
required for Dirac fermion masses can also generate scalar entries. These
entries can help if they raise diagonal scalar masses, but they can destabilize
the spectrum if they generate large off-diagonal terms.

Several questions remain open. The analysis used a low-energy EFT and the
anti-periodic Scherk--Schwarz benchmark, not a complete stabilized
compactification. The physical scalar eigenvalues depend on the full
potential, including local counterterms, higher-order terms, anomaly- or
radion-mediated contributions, and moduli stabilization. It remains to build
explicit matter sectors which realize the companion-channel couplings, to
derive their normalization in complete \(\mathcal N=2\) or five-dimensional
models, and to check whether the mixed scalar terms can be forbidden,
sequestered, or otherwise controlled while keeping the desired Dirac or
pseudo-Dirac fermion spectrum.

\appendix

\section[Superspin projectors]{Superspin projectors on a real vector superfield}
\label{app:projectors}

In this appendix we collect the superspin-projector conventions used in
Section~\ref{sec:N1_language_and_obstruction}. We do not review the full
projector formalism. We only record the decomposition of a real vector
superfield into irreducible superspin sectors and the explicit projectors that
enter the discussion of the Dirac spin-\(\tfrac32\) mixing. Our conventions
follow the standard linearized \(4D,\mathcal N=1\) superspace projector
formalism
\cite{Gates:1983nr,Buchbinder:1998twe}.

Let \(V_{\alpha\dot\alpha}\) be a real vector superfield,
\begin{equation}
V_{\alpha\dot\alpha}
=
\bar V_{\alpha\dot\alpha}.
\label{eq:app_real_vector_superfield}
\end{equation}
Its decomposition into irreducible superspin sectors is
\begin{equation}
V_{\alpha\dot\alpha}
=
\Big(
\Pi^L_0+\Pi^L_{1/2}+\Pi^T_{1/2}+\Pi^T_1+\Pi^T_{3/2}
\Big)V_{\alpha\dot\alpha},
\label{eq:app_projector_decomposition}
\end{equation}
where the superscripts \(L\) and \(T\) denote longitudinal and transverse
sectors. Equation~\eqref{eq:app_projector_decomposition} is the resolution of
the identity on the space of real vector superfields.

The projector needed explicitly in the body of the paper is the transverse
superspin-\(\tfrac32\) projector. With
\begin{equation}
\Box\equiv \partial^a\partial_a ,
\label{eq:app_box_definition}
\end{equation}
it may be written as
\begin{equation}
\big(\Pi^T_{3/2}V\big)_{\alpha\dot\alpha}
=
-\frac{1}{8}\,
\Box^{-2}\,
\partial_{\dot\alpha}{}^{\beta}\,
D^\gamma \bar D^2 D_{(\gamma}\,
\partial_{\alpha}{}^{\dot\beta}
V_{\beta)\dot\beta}.
\label{eq:app_Pi32_explicit}
\end{equation}
Parentheses denote symmetrization with weight one,
\begin{equation}
X_{(\alpha\beta)}
\equiv
\frac12\big(X_{\alpha\beta}+X_{\beta\alpha}\big).
\label{eq:app_symmetrization_convention}
\end{equation}
With this convention no additional factorial appears in
\eqref{eq:app_Pi32_explicit}.

For completeness we also record the longitudinal superspin-0 projector which
appears in the linearized old-minimal action
\eqref{eq:sec4_OMaction}:
\begin{equation}
\big(\Pi^L_0V\big)_{\alpha\dot\alpha}
=
-\frac{1}{32}\,
\partial_{\alpha\dot\alpha}\,
\Box^{-2}\,
\{D^2,\bar D^2\}\,
\partial^{\beta\dot\beta}V_{\beta\dot\beta}.
\label{eq:app_PiL0_explicit}
\end{equation}

The projectors obey the standard algebra
\begin{equation}
\Pi_i\Pi_j=\delta_{ij}\Pi_i,
\qquad
\sum_i\Pi_i = 1,
\label{eq:app_projector_algebra}
\end{equation}
on real vector superfields. In particular,
\begin{equation}
(\Pi^T_{3/2})^2=\Pi^T_{3/2}.
\label{eq:app_projector_idempotent}
\end{equation}
The transverse superspin-\(\tfrac32\) projector also satisfies
\begin{equation}
\partial^{\alpha\dot\alpha}
\big(\Pi^T_{3/2}V\big)_{\alpha\dot\alpha}=0.
\label{eq:app_projector_transverse}
\end{equation}
It removes the lower-superspin and pure-gauge components of
\(V_{\alpha\dot\alpha}\). This is the property used in
Section~\ref{subsec:projected_nonlocal}: the mixing is formed only after the
ordinary supergravity prepotential and the companion current have been
restricted to the same irreducible transverse superspin-\(\tfrac32\) sector.

We will also use the self-adjointness of the projectors with respect to the
full superspace measure. For real vector superfields \(A_{\alpha\dot\alpha}\)
and \(B_{\alpha\dot\alpha}\),
\begin{equation}
\int d^8z\,
A^{\alpha\dot\alpha}
\big(\Pi^T_{3/2}B\big)_{\alpha\dot\alpha}
=
\int d^8z\,
\big(\Pi^T_{3/2}A\big)^{\alpha\dot\alpha}
B_{\alpha\dot\alpha},
\label{eq:app_projector_self_adjoint}
\end{equation}
up to the usual boundary terms. This identity justifies the two equivalent
forms of the projected coupling used in
Section~\ref{subsec:projected_nonlocal}.

Two comments are useful.

First, \(\Pi^T_{3/2}\) acts on a real vector superfield. It is therefore the
appropriate object for projecting the linearized supergravity prepotential
\(H_{\alpha\dot\alpha}\), or any current with the same index structure, onto
its physical superspin-\(\tfrac32\) component. The obstruction discussed in the
main text is that a local bilinear constructed from unprojected prepotentials
contains lower-superspin and gauge-variant components. The projector removes
these components.

Second, the projection is non-local in superspace. This follows directly from
the inverse powers of \(\Box\) in \eqref{eq:app_Pi32_explicit}. Any superspace
operator in which \(\Pi^T_{3/2}\) appears explicitly is therefore non-local as
a superspace expression. This is the sense in which the Dirac-gravitino mixing
constructed in Section~\ref{sec:N1_language_and_obstruction} is non-local in
\(\mathcal N=1\) superspace. After restriction to the physical
spin-\(\tfrac32\) component, however, the corresponding component mixing is a
local Dirac mass term.

The use of the projector in the main text can be summarized as
\begin{equation}
H_{\alpha\dot\alpha}
\;\longrightarrow\;
\big(\Pi^T_{3/2}H\big)_{\alpha\dot\alpha},
\qquad
J_{\alpha\dot\alpha}
\;\longrightarrow\;
\big(\Pi^T_{3/2}J\big)_{\alpha\dot\alpha}.
\label{eq:app_projector_summary}
\end{equation}
The Dirac-gravitino mixing is then formed only between the common irreducible
transverse superspin-\(\tfrac32\) components.

\section{Mixed-representation propagators on
\texorpdfstring{$S^1/\mathbb Z_2$}{S1/Z2}}
\label{app:SS_kernels}

In this appendix we derive the mixed-representation Green functions used in the
Scherk--Schwarz threshold calculations of
Sections~\ref{sec:scalar_masses} and~\ref{sec:fermion_masses}. We work at the
anti-periodic point \(\omega=\tfrac12\). The two towers relevant for the
boundary amplitudes obey Neumann--Dirichlet and Dirichlet--Neumann boundary
conditions on the interval
\begin{equation}
0\leq y\leq \pi R .
\label{eq:app_interval_definition}
\end{equation}
Only the non-compact four-dimensional coordinates are Fourier transformed. We
use Euclidean momentum and write
\begin{equation}
p\equiv |p_E|=\sqrt{p_E^2}.
\label{eq:app_pE_convention}
\end{equation}
The Green functions below are therefore Green functions of the one-dimensional
operator
\begin{equation}
-\partial_y^2+p^2 .
\end{equation}

At \(\omega=\tfrac12\), the normalized mode functions are
\begin{equation}
f_n^{(+)}(y)
=
\sqrt{\frac{2}{\pi R}}\,
\cos\!\left(\frac{n+\tfrac12}{R}y\right),
\qquad
f_n^{(-)}(y)
=
\sqrt{\frac{2}{\pi R}}\,
\sin\!\left(\frac{n+\tfrac12}{R}y\right),
\qquad
m_n=\frac{n+\tfrac12}{R},
\label{eq:app_mode_functions}
\end{equation}
with \(n\geq0\). They obey
\begin{equation}
\partial_y f_n^{(+)}(0)=0,
\qquad
f_n^{(+)}(\pi R)=0,
\label{eq:app_bc_plus_modes}
\end{equation}
and
\begin{equation}
f_n^{(-)}(0)=0,
\qquad
\partial_y f_n^{(-)}(\pi R)=0 .
\label{eq:app_bc_minus_modes}
\end{equation}
Thus the \(+\) tower has Neumann--Dirichlet boundary conditions, while the
\(-\) tower has Dirichlet--Neumann boundary conditions.

We define
\begin{equation}
\left(-\partial_y^2+p^2\right)G_\pm(p;y,y')
=
\delta(y-y')
\label{eq:app_green_equation}
\end{equation}
with boundary conditions
\begin{align}
&\partial_y G_+(p;y,y')\big|_{y=0}=0,
\qquad
G_+(p;\pi R,y')=0,
\label{eq:app_bc_plus_green}
\\[2mm]
&G_-(p;0,y')=0,
\qquad
\partial_y G_-(p;y,y')\big|_{y=\pi R}=0 .
\label{eq:app_bc_minus_green}
\end{align}

\subsection*{The \texorpdfstring{$+$}{+} tower}
\label{app:derivation_Gplus}

For \(y\neq y'\), \(G_+\) satisfies
\begin{equation}
\left(-\partial_y^2+p^2\right)G_+(p;y,y')=0 .
\label{eq:app_homogeneous_eq_plus}
\end{equation}
For \(y<y'\), the Neumann condition at \(y=0\) selects \(\cosh(py)\). For
\(y>y'\), the Dirichlet condition at \(y=\pi R\) selects
\(\sinh[p(\pi R-y)]\). We therefore write
\begin{equation}
G_+(p;y,y')
=
\begin{cases}
A(y')\,\cosh(py), & 0\leq y<y', \\[2mm]
B(y')\,\sinh\!\big(p(\pi R-y)\big), & y'<y\leq \pi R .
\end{cases}
\label{eq:app_Gplus_piecewise}
\end{equation}
Continuity at \(y=y'\) gives
\begin{equation}
A(y')\cosh(py')
=
B(y')\sinh\!\big(p(\pi R-y')\big).
\label{eq:app_Gplus_continuity}
\end{equation}
Integrating \eqref{eq:app_green_equation} across \(y=y'\) gives
\begin{equation}
\partial_yG_+(p;y,y')\big|_{y=y'_+}
-
\partial_yG_+(p;y,y')\big|_{y=y'_-}
=
-1 .
\label{eq:app_Gplus_jump}
\end{equation}
Using \eqref{eq:app_Gplus_piecewise}, this becomes
\begin{equation}
-\,pB(y')\cosh\!\big(p(\pi R-y')\big)
-
pA(y')\sinh(py')
=
-1 .
\label{eq:app_Gplus_jump_explicit}
\end{equation}

Introduce a common numerator \(N(y')\) by
\begin{equation}
A(y')=\frac{N(y')}{\cosh(py')},
\qquad
B(y')=\frac{N(y')}{\sinh\!\big(p(\pi R-y')\big)} .
\label{eq:app_Gplus_commonN}
\end{equation}
Then
\begin{equation}
pN(y')\left[
\coth\!\big(p(\pi R-y')\big)+\tanh(py')
\right]
=
1 .
\label{eq:app_Gplus_N_equation}
\end{equation}
Using
\begin{equation}
\coth a+\tanh b
=
\frac{\cosh(a+b)}{\sinh a\,\cosh b},
\qquad
a=p(\pi R-y'),\qquad b=py',
\label{eq:app_coth_tanh_identity}
\end{equation}
one obtains
\begin{equation}
N(y')
=
\frac{\sinh\!\big(p(\pi R-y')\big)\cosh(py')}
     {p\,\cosh(\pi pR)} .
\label{eq:app_Gplus_N_solution}
\end{equation}
Thus
\begin{equation}
A(y')
=
\frac{\sinh\!\big(p(\pi R-y')\big)}
     {p\,\cosh(\pi pR)},
\qquad
B(y')
=
\frac{\cosh(py')}
     {p\,\cosh(\pi pR)} .
\label{eq:app_Gplus_AB_solution}
\end{equation}
Equivalently,
\begin{equation}
G_+(p;y,y')
=
\frac{\cosh(p\,y_<)\,
\sinh\!\big(p(\pi R-y_>)\big)}
{p\,\cosh(\pi pR)},
\label{eq:app_Gplus_final}
\end{equation}
where
\begin{equation}
y_<\equiv \min(y,y'),
\qquad
y_>\equiv \max(y,y') .
\label{eq:app_yless_ygreater_plus}
\end{equation}

\subsection*{The \texorpdfstring{$-$}{-} tower}
\label{app:derivation_Gminus}

The derivation of \(G_-\) is obtained by interchanging the boundary
conditions. For \(y<y'\), the Dirichlet condition at \(y=0\) selects
\(\sinh(py)\). For \(y>y'\), the Neumann condition at \(y=\pi R\) selects
\(\cosh[p(\pi R-y)]\). Thus
\begin{equation}
G_-(p;y,y')
=
\begin{cases}
\widetilde A(y')\,\sinh(py), & 0\leq y<y', \\[2mm]
\widetilde B(y')\,\cosh\!\big(p(\pi R-y)\big), & y'<y\leq \pi R .
\end{cases}
\label{eq:app_Gminus_piecewise}
\end{equation}
Continuity gives
\begin{equation}
\widetilde A(y')\sinh(py')
=
\widetilde B(y')\cosh\!\big(p(\pi R-y')\big),
\label{eq:app_Gminus_continuity}
\end{equation}
and the jump condition gives
\begin{equation}
-\,p\widetilde B(y')\sinh\!\big(p(\pi R-y')\big)
-
p\widetilde A(y')\cosh(py')
=
-1 .
\label{eq:app_Gminus_jump_explicit}
\end{equation}
Introduce
\begin{equation}
\widetilde A(y')=\frac{\widetilde N(y')}{\sinh(py')},
\qquad
\widetilde B(y')=\frac{\widetilde N(y')}{\cosh\!\big(p(\pi R-y')\big)} .
\label{eq:app_Gminus_commonN}
\end{equation}
Then
\begin{equation}
p\widetilde N(y')\left[
\tanh\!\big(p(\pi R-y')\big)+\coth(py')
\right]
=
1 .
\label{eq:app_Gminus_N_equation}
\end{equation}
Using
\begin{equation}
\tanh a+\coth b
=
\frac{\cosh(a+b)}{\cosh a\,\sinh b},
\qquad
a=p(\pi R-y'),\qquad b=py',
\label{eq:app_tanh_coth_identity}
\end{equation}
one finds
\begin{equation}
\widetilde N(y')
=
\frac{\cosh\!\big(p(\pi R-y')\big)\sinh(py')}
     {p\,\cosh(\pi pR)} .
\label{eq:app_Gminus_N_solution}
\end{equation}
Hence
\begin{equation}
\widetilde A(y')
=
\frac{\cosh\!\big(p(\pi R-y')\big)}
     {p\,\cosh(\pi pR)},
\qquad
\widetilde B(y')
=
\frac{\sinh(py')}
     {p\,\cosh(\pi pR)} .
\label{eq:app_Gminus_AB_solution}
\end{equation}
Thus
\begin{equation}
G_-(p;y,y')
=
\frac{\sinh(p\,y_<)\,
\cosh\!\big(p(\pi R-y_>)\big)}
{p\,\cosh(\pi pR)} .
\label{eq:app_Gminus_final}
\end{equation}

Equations \eqref{eq:app_Gplus_final} and \eqref{eq:app_Gminus_final} obey the
required boundary conditions and the common jump condition
\begin{equation}
\partial_yG_\pm(p;y,y')\big|_{y=y'_+}
-
\partial_yG_\pm(p;y,y')\big|_{y=y'_-}
=
-1 .
\label{eq:app_Gpm_jump_final}
\end{equation}

\subsection*{Boundary values and derivative kernels}
\label{app:boundary_derivative_kernels}

The direct boundary values are
\begin{equation}
G_+(p;0,0)=\frac{\tanh(\pi pR)}{p},
\qquad
G_-(p;0,y')=0,
\label{eq:app_boundary_values_zero}
\end{equation}
and
\begin{equation}
G_-(p;\pi R,\pi R)=\frac{\tanh(\pi pR)}{p},
\qquad
G_+(p;\pi R,y')=0 .
\label{eq:app_boundary_values_piR}
\end{equation}
The vanishing of \(G_-\) at \(y=0\), and of \(G_+\) at \(y=\pi R\), is the
orbifold parity selection rule in this basis.

The single normal derivative of the odd Green function at \(y=0\), with the
second endpoint away from the same odd boundary, follows from the branch
\(0<y<y'\):
\begin{equation}
G_-(p;y,y')
=
\frac{\sinh(py)\,\cosh\!\big(p(\pi R-y')\big)}
     {p\,\cosh(\pi pR)} ,
\qquad 0<y<y' .
\label{eq:app_Gminus_branch_near_zero}
\end{equation}
Therefore
\begin{equation}
\partial_yG_-(p;y,y')\big|_{y=0}
=
\frac{\cosh\!\big(p(\pi R-y')\big)}
     {\cosh(\pi pR)},
\qquad
0<y'\leq \pi R .
\label{eq:app_dGminus_boundary_general}
\end{equation}
In particular,
\begin{equation}
\partial_yG_-(p;y,\pi R)\big|_{y=0}
=
\frac{1}{\cosh(\pi pR)} .
\label{eq:app_dGminus_opposite_boundary}
\end{equation}

If the second endpoint is fixed at \(y'=0\), the single-derivative kernel
vanishes. In the spectral representation,
\begin{equation}
\partial_yG_-(p;y,0)\big|_{y=0}
=
\sum_{n=0}^{\infty}
\frac{\partial_y f_n^{(-)}(0)\,f_n^{(-)}(0)}
     {p^2+m_n^2}
=0,
\label{eq:app_single_derivative_same_boundary_zero}
\end{equation}
because \(f_n^{(-)}(0)=0\). Thus a single derivative on one endpoint does not
produce a same-boundary kernel for an odd field.

If both same-boundary vertices couple to the odd field through a normal
derivative, the relevant object is instead
\begin{equation}
\mathcal D_{--}^{(0,0)}(p)
\equiv
\lim_{y,y'\to0^+}
\partial_y\partial_{y'}G_-(p;y,y') .
\label{eq:app_double_derivative_limit_definition}
\end{equation}
Equivalently,
\begin{equation}
\mathcal D_{--}^{(0,0)}(p)
=
\sum_{n=0}^{\infty}
\frac{\partial_y f_n^{(-)}(0)\,
      \partial_{y'}f_n^{(-)}(0)}
     {p^2+m_n^2}.
\label{eq:app_double_derivative_spectral_definition}
\end{equation}
Using \eqref{eq:app_mode_functions}, this becomes
\begin{equation}
\mathcal D_{--}^{(0,0)}(p)
=
\frac{2}{\pi R}
\sum_{n=0}^{\infty}
\frac{\big((n+\tfrac12)/R\big)^2}
     {p^2+\big((n+\tfrac12)/R\big)^2}.
\label{eq:app_double_derivative_raw_sum}
\end{equation}
This sum contains a \(p\)-independent local divergence. Writing
\begin{equation}
\frac{m_n^2}{p^2+m_n^2}
=
1-\frac{p^2}{p^2+m_n^2},
\label{eq:app_double_derivative_split}
\end{equation}
the first term is local and counterterm-sensitive. The finite
\(p\)-dependent part is
\begin{equation}
\mathcal D_{--}^{(0,0)}(p)\Big|_{\rm finite}
=
-\,p\,\tanh(\pi pR),
\label{eq:app_double_derivative_finite_part}
\end{equation}
when both derivatives are taken with respect to the interval coordinate near
\(y=0\). A convention in which one or both derivatives are outward-normal
derivatives changes the corresponding sign. The important point is that the
same-boundary derivative--derivative kernel is local-threshold sensitive and is
not the same object as the finite derivative--direct opposite-boundary kernel
\eqref{eq:app_dGminus_opposite_boundary}.

At the other boundary one similarly obtains
\begin{equation}
\partial_yG_+(p;y,y')\big|_{y=\pi R}
=
-\,\frac{\cosh(py')}{\cosh(\pi pR)},
\qquad
0\leq y'<\pi R .
\label{eq:app_dGplus_boundary_general}
\end{equation}
The sign depends on whether the derivative is taken with respect to the
interval coordinate \(y\) or to the outward normal at \(y=\pi R\).

The boundary kernels therefore depend on the microscopic boundary coupling.
Direct, derivative--direct, and derivative--derivative couplings give different
kernels. Same-boundary derivative--derivative kernels contain local pieces;
opposite-boundary derivative--direct kernels are finite and non-local.

\subsection*{Spectral representation}
\label{app:final_result_spectral_form}

The closed forms derived above are equivalent to the spectral sums
\begin{equation}
G_\pm(p;y,y')
=
\sum_{n=0}^{\infty}
\frac{f_n^{(\pm)}(y)\,f_n^{(\pm)}(y')}
     {p^2+m_n^2},
\label{eq:app_spectral_decomposition}
\end{equation}
with \(f_n^{(\pm)}\) and \(m_n\) given in
\eqref{eq:app_mode_functions}. For example,
\begin{align}
G_+(p;0,0)
&=
\frac{2}{\pi R}
\sum_{n=0}^{\infty}
\frac{1}
     {p^2+\big((n+\tfrac12)/R\big)^2}
\nonumber\\
&=
\frac{\tanh(\pi pR)}{p},
\label{eq:app_spectral_check_Gplus_00}
\end{align}
where
\begin{equation}
\sum_{n=0}^{\infty}
\frac{1}{(n+\tfrac12)^2+a^2}
=
\frac{\pi}{2a}\tanh(\pi a)
\label{eq:app_half_integer_sum_identity}
\end{equation}
has been used. This confirms the normalization of the mode expansion and the
closed-form Green functions.

\section{Off-shell boundary origin of the universal Scherk--Schwarz scalar threshold}
\label{app:scalar_cubic_trace}

In this appendix we explain the normalization used for the universal scalar
threshold in Section~\ref{sec:scalar_masses}. In the conventions of the present
paper,
\begin{equation}
\mathcal C_A
=
\frac{2}{3M_5^3}
=
\frac{2\pi R}{3M_4^2},
\qquad
M_4^2=\pi R\,M_5^3 .
\label{eq:app_CA_target}
\end{equation}
This coefficient multiplies the same-boundary Scherk--Schwarz kernel as
\begin{equation}
\delta m_\phi^2\big|_{\rm univ}
=
\mathcal C_A
\int\frac{d^4p_E}{(2\pi)^4}\,
p_E^2\,\mathcal K_A^{\rm same}(p).
\label{eq:app_scalar_kernel_intro}
\end{equation}
It is not the coefficient of an isolated cubic--cubic
spin-\(\tfrac32\) exchange graph. It is the normalization of the complete
locally supersymmetric boundary coupling of a chiral multiplet to the
projected off-shell five-dimensional supergravity multiplet.

The scalar threshold is the standard Scherk--Schwarz gravity-mediated scalar
mass obtained in the off-shell bulk--brane framework of
Ref.~\cite{Rattazzi:2003rj}. The underlying off-shell construction is based on
Zucker's five-dimensional off-shell supergravity
\cite{Zucker:1999ej}, together with the induced four-dimensional
\(\mathcal N=1\) intermediate multiplet on the fixed planes. The corresponding
boundary tensor-calculus rules are those of the Sohnius--West alternative
minimal formulation of \(4D,\mathcal N=1\) supergravity
\cite{Sohnius:1981tp,Sohnius:1981ac}. We use only the part of this formalism
needed to identify the boundary kinetic multiplet and its normalization.

The purpose of the appendix is not to rederive the full off-shell
bulk--brane calculation. Rather, we translate the standard result into the
same-boundary kernel normalization used in the main text and make explicit why
the local cubic--cubic tensor coefficient is not the physical universal
normalization.

Schematically, the physical universal scalar mass has the form
\begin{equation}
\delta m_\phi^2\big|_{\rm univ}
=
\Big[
\delta m_\phi^2\big|_{\rm cubic-cubic}
+
\delta m_\phi^2\big|_{\rm contact}
+
\delta m_\phi^2\big|_{\rm auxiliaries}
\Big]_{\omega}
-
\Big[
\delta m_\phi^2
\Big]_{\omega=0}.
\label{eq:app_full_local_susy_sum}
\end{equation}
The separation into cubic--cubic, contact, and auxiliary pieces is not
separately invariant. It depends on the off-shell completion, the auxiliary
field basis, and the organization of the boundary multiplet. The sum is fixed
by local supersymmetry. Therefore \(\mathcal C_A\) must be identified from the
complete off-shell boundary structure, or equivalently by matching to the full
Scherk--Schwarz scalar threshold. It cannot be extracted from the local cubic
exchange graph alone.

We nevertheless record a representative cubic--cubic numerator at the end of
the appendix. This gives a local check of the gamma-matrix algebra of
derivative supercurrent vertices. The same reduced spin-\(\tfrac32\) trace
appears in the mixed scalar topologies of Section~\ref{sec:scalar_masses}. The
contact term is not used as an independent input anywhere in the paper; only
the complete locally supersymmetric sum enters the universal
Scherk--Schwarz threshold.

\subsection*{Boundary intermediate multiplet and boundary matter}
\label{app:boundary_intermediate_multiplet}

We start from off-shell five-dimensional supergravity on
\(S^1/\mathbb Z_2\). At a fixed plane, the orbifold projection keeps the even
components of the bulk fields. These fields form an off-shell
four-dimensional \(\mathcal N=1\) supergravity multiplet with \(16+16\)
components, the intermediate multiplet
\cite{Zucker:1999ej,Rattazzi:2003rj}.

In a notation adapted to the fixed plane, this multiplet may be written as
\begin{equation}
I
=
\big(
e_m{}^a,\,
\psi_m;\,
b_a,\,
a_m,\,
\lambda,\,
S,\,
t^1,\,
t^2
\big).
\label{eq:app_intermediate_multiplet}
\end{equation}
The fields \(e_m{}^a\) and \(\psi_m\) are the induced vierbein and the even
boundary gravitino. The remaining fields are boundary combinations of the
five-dimensional auxiliary fields. In particular,
\begin{align}
b_a
&=
v_{a\dot 5},
\label{eq:app_ba_definition}
\\[1mm]
a_m
&=
-\frac12
\left(
V_m^3
-
\frac{2}{\sqrt3}\,\widehat F_{m5}\,e^5{}_{\dot5}
+
4e_m{}^a v_{a\dot5}
\right),
\label{eq:app_am_definition}
\\[1mm]
S
&=
C
-\frac12 e^5{}_{\dot5}
\left(
\partial_5 t^3
-
\bar\lambda\,\tau^3\psi_5
+
V_5^1 t^2
-
V_5^2 t^1
\right).
\label{eq:app_S_definition}
\end{align}
The orbifold projection breaks the bulk \(SU(2)_R\) at the fixed plane to the
local chiral \(U(1)\) gauged by \(a_m\). This is the \(U(1)_R\) of the induced
intermediate multiplet.

Boundary matter couples to the complete off-shell boundary multiplet \(I\),
not to an on-shell truncation of the bulk fields. This is the origin of the
contact terms that accompany the cubic supercurrent coupling. These terms are
required by local supersymmetry and belong to the same boundary D-density as
the kinetic term of the brane matter multiplet.

The relevant component rule is the D-density of a general multiplet
\begin{equation}
C_{\rm gen}
=
\big(
C;\,\chi;\,H,K,U_a;\,\chi';\,D
\big).
\label{eq:app_general_multiplet}
\end{equation}
In the Sohnius--West intermediate-multiplet tensor calculus
\cite{Sohnius:1981tp,Sohnius:1981ac}, it takes the form
\begin{align}
[C_{\rm gen}]_D
&=
eD
-4e\Big(
CS+\bar\lambda\chi
-HM-KN
+2C(M^2+N^2)
\Big)
+4e\,b^aU_a
\nonumber\\
&\quad
+\text{fermionic completion}
+\text{total derivative}.
\label{eq:app_SW_Ddensity}
\end{align}
This is the off-shell density which gives the locally supersymmetric completion
of the boundary matter kinetic term.

Let \(Q=(\phi,\psi,F)\) be a chiral multiplet localized at \(y=0\). The
boundary action is obtained from a D-density of the form
\begin{equation}
\mathcal L_{\rm bdry}
=
\delta(y)\,
\big[
\Omega_0(Q,Q^\dagger)\,S_0S_0^\dagger
\big]_D .
\label{eq:app_boundary_D_density}
\end{equation}
Here \(S_0\) is the boundary chiral compensator. Near the origin in field
space,
\begin{equation}
\Omega_0(Q,Q^\dagger)
=
\phi^\dagger\phi+\cdots .
\label{eq:app_Omega0_expansion}
\end{equation}
The boundary D-density therefore contains, in one locally supersymmetric
package, the scalar and fermion kinetic terms of \(Q\), the cubic coupling of
the matter supercurrent to the induced gravitino, the scalar--scalar--gravity
contact terms, and the auxiliary-field terms required by the off-shell
multiplet.

This is the reason why the universal scalar threshold is not determined by the
cubic--cubic graph alone. A separate cubic graph has a well-defined local
gamma-matrix numerator once a projector convention is chosen, but the physical
threshold is obtained only after the complete off-shell boundary coupling and
the unbroken-background subtraction are included.

\subsection*{Quadratic boundary kinetic multiplet and normalization}
\label{app:quartic_boundary_coupling}

We now identify the normalization of the part of the boundary action which is
quadratic in the projected supergravity fields and proportional to
\(|\phi|^2\). The lowest component of the general multiplet entering the
density is
\begin{equation}
C_\Omega\big|_{\rm lowest}
=
|\phi|^2+\cdots .
\label{eq:app_COmega_lowest}
\end{equation}
Thus the scalar background multiplies the same quadratic operator that appears
in the linearized kinetic multiplet of the induced intermediate supergravity.

Let \(\Omega\) denote the real gravitational function appearing in the
four-dimensional superconformal D-density. In the boundary-coupled theory it
contains localized pieces,
\begin{equation}
\Omega
=
\Omega_{\rm bulk}
+
\delta(y)\,\Omega_0
+
\delta(y-\pi R)\,\Omega_\pi ,
\qquad
\Omega_0(Q,Q^\dagger)=\phi^\dagger\phi+\cdots .
\label{eq:app_Omega_definition}
\end{equation}
The component D-density contains the Einstein term
\begin{equation}
\mathcal L_{\rm grav}
\supset
\frac{\Omega}{6}\,R
\label{eq:app_Omega_R_normalization}
\end{equation}
in the conventions used here. Together with its gravitino, auxiliary-vector
and auxiliary-scalar partners, this term defines the quadratic boundary kinetic
multiplet.

Restoring the five-dimensional normalization, and diagonalizing the quadratic
terms into transverse spin sectors, the scalar-dependent quadratic boundary
coupling can be written as
\begin{align}
\mathcal L_{\rm bdry}^{(2)}
&=
\delta(y)\,
\rho
\Big[
p_E^2\,h_{\mu\nu}
\left(
\mathcal P_{(2)}^{\mu\nu,\alpha\beta}
-\frac23\mathcal P_{(1)}^{\mu\nu}\mathcal P_{(1)}^{\alpha\beta}
\right)
h_{\alpha\beta}
\nonumber\\
&\hspace{2.8cm}
+
2\,\bar\Psi_\mu\slashed p_E
\left(
\mathcal P_{(3/2)}^{\mu\nu}
-
2\mathcal P_{(1/2)}^{\mu\nu}
\right)P_L\Psi_\nu
+
\frac13\,V_\mu V^\mu
\Big],
\label{eq:app_offshell_quartic_structure_full}
\end{align}
with
\begin{equation}
\rho
\equiv
\frac{|\phi|^2}{3M_5^3}.
\label{eq:app_rho_definition}
\end{equation}
The projectors \(\mathcal P_{(2)}\), \(\mathcal P_{(1)}\),
\(\mathcal P_{(3/2)}\), and \(\mathcal P_{(1/2)}\) are transverse spin
projectors. The lower-spin pieces are displayed because they are part of the
off-shell kinetic multiplet, although they do not contribute to the projected
universal threshold.

The explicit projectors are
\begin{align}
\mathcal P_{(1)}^{\mu\nu}
&=
\delta^{\mu\nu}
-
\frac{p_E^\mu p_E^\nu}{p_E^2},
\label{eq:app_P1}
\\[1mm]
\mathcal P_{(1/2)}^{\mu\nu}
&=
\frac13
\left(
\gamma^\mu-\frac{p_E^\mu\slashed p_E}{p_E^2}
\right)
\left(
\gamma^\nu-\frac{p_E^\nu\slashed p_E}{p_E^2}
\right),
\label{eq:app_Phalf}
\\[1mm]
\mathcal P_{(3/2)}^{\mu\nu}
&=
\mathcal P_{(1)}^{\mu\nu}
-
\mathcal P_{(1/2)}^{\mu\nu},
\label{eq:app_Pthreehalf}
\\[1mm]
\mathcal P_{(2)}^{\mu\nu,\alpha\beta}
&=
\frac12
\mathcal P_{(1)}^{\mu\alpha}
\mathcal P_{(1)}^{\nu\beta}
+
\frac12
\mathcal P_{(1)}^{\mu\beta}
\mathcal P_{(1)}^{\nu\alpha}
-
\frac13
\mathcal P_{(1)}^{\mu\nu}
\mathcal P_{(1)}^{\alpha\beta}.
\label{eq:app_Ptwo}
\end{align}
They obey the usual idempotence and transversality relations. In particular,
\begin{equation}
p_{E\mu}\mathcal P_{(3/2)}^{\mu\nu}=0,
\qquad
\gamma_\mu \mathcal P_{(3/2)}^{\mu\nu}=0 .
\label{eq:app_projector_properties}
\end{equation}
The combination
\begin{equation}
\mathcal P_{(2)}^{\mu\nu,\alpha\beta}
-\frac23\mathcal P_{(1)}^{\mu\nu}\mathcal P_{(1)}^{\alpha\beta}
\label{eq:app_graviton_projector_combination}
\end{equation}
is the spin-2 plus scalar-trace structure appearing in the intermediate
multiplet kinetic density. The projector \(\mathcal P_{(2)}\) itself denotes
only the transverse traceless spin-2 part.

For the universal threshold, the Scherk--Schwarz supersymmetry-breaking part
of the same-boundary propagator is inserted into the quadratic boundary
kinetic multiplet. In the projected spin-\(\tfrac32\) channel one may use
\begin{equation}
\Delta^{\mu\nu}_{\Psi}
=
\frac12\,\slashed p_E\,
\mathcal P_{(3/2)}^{\mu\nu}\,\Delta(p;y,y'),
\label{eq:app_gravitino_projected_propagator}
\end{equation}
where \(\Delta(p;y,y')\) is the scalar mixed-representation propagator along
the interval. The mass term in the numerator does not contribute to the
coefficient of \(p_E^2\mathcal K_A^{\rm same}(p)\) in this projected channel,
because of the chiral projectors and the transverse spin-\(\tfrac32\)
projection.

Using
\begin{equation}
\mathcal P_{(1/2)}\mathcal P_{(3/2)}=0,
\end{equation}
the spin-\(\tfrac32\) insertion reduces to
\begin{equation}
2\rho\,p_E^2\,\Delta(p;y,y') .
\label{eq:app_effective_gravitino_insertion}
\end{equation}
Equivalently, after the complete off-shell projector algebra is performed, the
coefficient multiplying \(p_E^2\mathcal K_A^{\rm same}(p)\) is
\(2\rho/|\phi|^2\). Hence
\begin{equation}
\mathcal C_A
=
\frac{2\rho}{|\phi|^2}
=
\frac{2}{3M_5^3}
=
\frac{2\pi R}{3M_4^2}.
\label{eq:app_CA_from_offshell}
\end{equation}
This is the universal scalar normalization used in
Section~\ref{sec:scalar_masses}. It is fixed by the complete off-shell
boundary kinetic multiplet, not by the isolated cubic graph.

For local gamma-matrix manipulations, after longitudinally irrelevant pieces
have been removed, the transverse spin-\(\tfrac32\) projector reduces to
\begin{equation}
\Pi_{3/2}^{\mu\nu}
=
\delta^{\mu\nu}
-
\frac13\gamma^\mu\gamma^\nu .
\label{eq:app_reduced_projector_again}
\end{equation}
This reduced projector is used only in the local tensor check below.

\subsection*{Scherk--Schwarz kernel and universal threshold}
\label{app:universal_SS_threshold}

We now evaluate the scalar threshold in the mixed-representation kernel
convention of Section~\ref{sec:scalar_masses}. At the pure anti-periodic
Scherk--Schwarz point,
\begin{equation}
\omega=\frac12,
\qquad
M_0=M_\pi=0,
\label{eq:app_SS_half_point}
\end{equation}
the same-boundary supersymmetry-breaking kernel for the direct boundary
supergravity coupling at \(y=0\) is
\begin{equation}
\mathcal K_A^{\rm same}(p)
=
G_+^{(\omega=1/2)}(p;0,0)
-
G_+^{(\omega=0)}(p;0,0).
\label{eq:app_KA_definition}
\end{equation}
The relevant boundary values are
\begin{equation}
G_+^{(\omega=1/2)}(p;0,0)
=
\frac{\tanh(\pi pR)}{p},
\qquad
G_+^{(\omega=0)}(p;0,0)
=
\frac{\coth(\pi pR)}{p}.
\label{eq:app_Gplus_values}
\end{equation}
Therefore
\begin{equation}
\mathcal K_A^{\rm same}(p)
=
-\frac{2}{p\,\sinh(2\pi pR)}.
\label{eq:app_KA_explicit}
\end{equation}

Using \eqref{eq:app_CA_from_offshell}, the universal scalar mass is
\begin{equation}
\delta m_\phi^2\big|_{\rm univ}
=
\mathcal C_A
\int\frac{d^4p_E}{(2\pi)^4}\,
p_E^2\,\mathcal K_A^{\rm same}(p).
\label{eq:app_kernel_form_from_offshell}
\end{equation}
Substituting \eqref{eq:app_KA_explicit} gives
\begin{equation}
\delta m_\phi^2\big|_{\rm univ}
=
-2\mathcal C_A
\int\frac{d^4p_E}{(2\pi)^4}\,
\frac{p_E^2}{p\,\sinh(2\pi pR)}.
\label{eq:app_scalar_integral_step1}
\end{equation}
With
\begin{equation}
\int d^4p_E
=
2\pi^2\int_0^\infty dp\,p^3,
\label{eq:app_four_sphere_measure}
\end{equation}
one obtains
\begin{equation}
\delta m_\phi^2\big|_{\rm univ}
=
-\frac{\mathcal C_A}{4\pi^2}
\int_0^\infty dp\,
\frac{p^4}{\sinh(2\pi pR)}.
\label{eq:app_scalar_integral_radial}
\end{equation}
Introducing
\begin{equation}
x=2\pi R p,
\qquad
dp=\frac{dx}{2\pi R},
\label{eq:app_x_variable}
\end{equation}
this becomes
\begin{equation}
\delta m_\phi^2\big|_{\rm univ}
=
-\frac{\mathcal C_A}{128\pi^7R^5}
\int_0^\infty dx\,\frac{x^4}{\sinh x}.
\label{eq:app_scalar_integral_x}
\end{equation}
The remaining integral is
\begin{align}
\int_0^\infty dx\,\frac{x^4}{\sinh x}
&=
2\sum_{\ell=0}^{\infty}
\int_0^\infty dx\,x^4 e^{-(2\ell+1)x}
\nonumber\\
&=
2\,\Gamma(5)
\sum_{\ell=0}^{\infty}\frac{1}{(2\ell+1)^5}
\nonumber\\
&=
2\,\Gamma(5)\left(1-2^{-5}\right)\zeta(5)
\nonumber\\
&=
\frac{93}{2}\,\zeta(5).
\label{eq:app_x4_over_sinh_result}
\end{align}
Thus
\begin{equation}
\delta m_\phi^2\big|_{\rm univ}
=
-\frac{93\,\mathcal C_A}{256\pi^7}\,
\frac{\zeta(5)}{R^5}.
\label{eq:app_scalar_mass_CA_result}
\end{equation}
Using
\begin{equation}
\mathcal C_A
=
\frac{2}{3M_5^3},
\label{eq:app_CA_M5_only}
\end{equation}
we obtain
\begin{equation}
\delta m_\phi^2\big|_{\rm univ}
=
-\frac{31\,\zeta(5)}{128\pi^7}\,
\frac{1}{M_5^3R^5}.
\label{eq:app_scalar_mass_M5_result}
\end{equation}
Equivalently, using \(M_4^2=\pi R\,M_5^3\),
\begin{equation}
\delta m_\phi^2\big|_{\rm univ}
=
-\frac{31\,\zeta(5)}{128\pi^6}\,
\frac{1}{R^4M_4^2}.
\label{eq:app_scalar_mass_M4_result}
\end{equation}
This is the universal anti-periodic Scherk--Schwarz scalar threshold quoted in
Section~\ref{sec:scalar_masses}. It agrees with the standard off-shell
bulk--brane result of Ref.~\cite{Rattazzi:2003rj}, written here in the
same-boundary kernel normalization used in the present paper.

For comparison, the generic-\(\omega\) KK sum may be written as
\begin{equation}
\delta m_\phi^2\big|_{\rm univ}
=
\frac{1}{8\pi^7M_5^3R^5}
\left[
\operatorname{Re}\operatorname{Li}_5\!\left(e^{2\pi i\omega}\right)
-
\zeta(5)
\right].
\label{eq:app_generic_omega_sum}
\end{equation}
At \(\omega=\tfrac12\),
\begin{equation}
\operatorname{Re}\operatorname{Li}_5(-1)
=
-\left(1-2^{-4}\right)\zeta(5)
=
-\frac{15}{16}\zeta(5),
\label{eq:app_Li5_minus_one}
\end{equation}
and therefore
\begin{equation}
\operatorname{Re}\operatorname{Li}_5(-1)-\zeta(5)
=
-\frac{31}{16}\zeta(5).
\label{eq:app_polylog_half}
\end{equation}
Substitution into \eqref{eq:app_generic_omega_sum} reproduces
\eqref{eq:app_scalar_mass_M5_result}.

\subsection*{Local cubic--cubic tensor reduction}
\label{app:scalar_cubic_trace_local}

We finally record a local cubic--cubic tensor reduction. This computation has a
different status from the preceding derivation. It is not an independent
calculation of the universal scalar soft mass. Its role is only to check the
spin-\(\tfrac32\) gamma-matrix numerator generated by derivative supercurrent
vertices. This is useful because the same reduced projector and the same chiral
trace structure occur in the mixed scalar amplitudes of
Section~\ref{sec:scalar_masses}.

The cubic supercurrent coupling of a boundary chiral multiplet to the ordinary
gravitino contains
\begin{equation}
\mathcal L_{\phi\psi\psi_{3/2}}
=
-\frac{1}{\sqrt2\,M_P}\,
(\partial_\nu\phi^\dagger)\,
\bar\psi_\mu\gamma^\nu\gamma^\mu P_L\psi
+\text{h.c.}
\label{eq:app_cubic_vertex}
\end{equation}
For the representative chiral trace needed in the mixed scalar analysis, the
local numerator may be written as
\begin{align}
\mathcal T_A^{\rm cubic}(k_E;m_{3/2},m_\psi)
&=
k_{E\rho}k_{E\sigma}\,
\Tr\!\Big[
\gamma^\rho\gamma_\mu P_L
(-\,i\slashed k_E+m_{3/2})
\Pi^{\mu\nu}_{3/2}
\gamma^\sigma\gamma_\nu P_L
(-\,i\slashed k_E+m_\psi)
\Big],
\label{eq:app_TA_chiral_start}
\end{align}
where
\begin{equation}
\Pi^{\mu\nu}_{3/2}
=
\delta^{\mu\nu}
-
\frac13\gamma^\mu\gamma^\nu .
\label{eq:app_RS_reduced_projector}
\end{equation}
The matter-fermion and spin-\(\tfrac32\) denominators are not included in
\(\mathcal T_A^{\rm cubic}\).

Using
\begin{equation}
P_L(-\,i\slashed k_E+m_{3/2})
=
-\,i\slashed k_E P_R+m_{3/2}P_L,
\label{eq:app_TA_PL_reduction}
\end{equation}
the term proportional to \(-i\slashed k_E P_R\) vanishes in this trace. The
surviving contribution is proportional to \(m_{3/2}\). The part of the matter
numerator proportional to \(\slashed k_E\) gives an odd trace, or an epsilon
tensor contracted with symmetric products of \(k_E^\mu\), and vanishes. The
mass part gives
\begin{align}
\mathcal T_A^{\rm cubic}
&=
m_{3/2}m_\psi\,
\frac12\,
k_{E\rho}k_{E\sigma}\,
\Tr\!\Big[
\gamma^\rho\gamma_\mu
\Pi^{\mu\nu}_{3/2}
\gamma^\sigma\gamma_\nu
\Big].
\label{eq:app_TA_chiral_mass_reduced}
\end{align}
Expanding the projector, one finds
\begin{equation}
k_{E\rho}k_{E\sigma}\,
\Tr\!\Big[
\gamma^\rho\gamma_\mu
\Pi^{\mu\nu}_{3/2}
\gamma^\sigma\gamma_\nu
\Big]
=
\frac83\,k_E^2 .
\label{eq:app_TA_chiral_trace}
\end{equation}
Therefore
\begin{equation}
\mathcal T_A^{\rm cubic}(k_E;m_{3/2},m_\psi)
=
\frac43\,m_{3/2}m_\psi\,k_E^2,
\label{eq:app_TA_chiral_final}
\end{equation}
up to the overall Euclidean sign convention.

The quantity \(\mathcal T_A^{\rm cubic}\) is a local numerator. It checks the
spin-\(\tfrac32\) gamma algebra, but it is not the coefficient multiplying the
same-boundary Scherk--Schwarz kernel in the universal threshold. The physical
coefficient is \(\mathcal C_A\), fixed by the complete off-shell boundary
action. The isolated cubic--cubic numerator does not include the
scalar--scalar--gravity contact terms, the graviton and auxiliary sectors, or
the subtraction of the unbroken \(\omega=0\) background. This is why the
cubic--cubic trace is useful as a local algebraic check, whereas the physical
scalar threshold is normalized only by the complete locally supersymmetric
sum.


\end{document}